\documentclass[amsfonts,aps]{revtex4}

\begin{document}

\title{On the problem of the relation between 
phason elasticity and phason dynamics
in quasicrystals} 

\author{Gerrit Coddens}

\address{Laboratoire des Solides Irradi\'es, 
Ecole Polytechnique,\\
F-91128-Palaiseau CEDEX, France}  

\date{today}   


\widetext

\begin{abstract}
It has recently been claimed that 
the dynamics
of long-wavelength phason fluctuations 
has been observed in 
{\em i}-AlPdMn quasicrystals.\cite{Francoual} 
We will show that 
the data reported 
call for a more detailed development of 
the elasticity theory
of Jari\'c and Nelsson\cite{Jaric} 
in order to determine the
nature of small phonon-like atomic displacements
with a symmetry that follows the phason 
elastic constants.
We also show that a simple model with a 
single diffusing tile is sufficient 
to produce a signal
that (1) is situated at a ``satellite position'' 
at a distance $q$ from each Bragg peak, that (2) 
has an intensity that 
scales with the intensity of the corresponding Bragg peak,  
(3) falls off as $1/q^{2}$
and (4) has a time decay constant that is proportional to $1/Dq^{2}$.
It is thus superfluous to call for a picture of ``phason
waves'' in order to explain such data, especially as
such ``waves'' violate many physical principles.

\end{abstract}
\maketitle

{\section{Introduction}}

It has recently been claimed that 
the dynamics
of long-wavelength phason fluctuations 
has been observed in 
{\em i}-AlPdMn quasicrystals.\cite{Francoual} 
These claims were based on the observation 
of very  long
relaxation times in the so-called speckle 
patterns 
of the diffuse scattering measured 
with coherent X-rays.
We report here the results of a number of 
investigations and model calculations
on coherent scattering signals for phason
dynamics, that show that 
there are problems with these claims.
Let us resume some important points of our paper:

(1) There is no proof that the (kinetics of the) diffuse scattering 
observed in these data is due to tile disorder produced by atomic jumps.
Alternative interpretations for the  data are possible
and have not been ruled out.
Chemical disorder and a field of small atomic displacements
that are not phason jumps can lead to very similar signals.

(2) Atomic jumps are in principle not correlated over long distances,
especially when they participate in atomic self-diffusion.
Experimental data never code correlations between
atomic jumps. They correspond to correlations between atomic positions.
Most importantly, the diffusion of a single tile already shares many
features with the signal reported, while
it does not involve any correlations whatsoever.

(3) Identifying the diffuse scattering with jumps is in 
contradiction with quasielastic neutron scattering data 
which clearly demonstrate
that the number of jumps increases with T, while the diffuse scattering
shows exactly the opposite behaviour. 

(4) Introducing a concept of a phason as a 
displacive mode of atomic jumps in a QC is problematic.
An ondulating cut (``sine wave'') does not
produce  a physical wave in physical space.
The displacement field defined by
the ``sine wave'' is non-periodic, 
and the concept of such a mode introduces 
notions that are unphysical, such as spooky long-distance  
correlations between isolated phason jumps.
The ``sine wave'' is in this respect not any better
than a rigid translation of the cut.

(5) One has confused coherent scattering signals
with evidence for collective effects,
Fourier components with  physical waves, the Fourier transform
of the QC with the Fourier transform of a field of atomic jumps.

(6) One has overinterpreted Lubensky's statement\cite{Lubensky} that
phasons are diffusive. 

(7) The available data  are not specific enough to warrant claims
that they would present evidence for the random tiling model.

(8) Contrary to statements
by several authors, the terminology ``phason jumps'' is by all standards
a correct terminology.

(9) The authors  analyze
the $Q$-dependence of the fluctuating
signal as though it would be 
a {\em static structure factor}, while this can never 
be correct as the dynamical signal 
they have measured is governed 
by an intrinsically different  {\em dynamical structure factor}.
More precisely their photon correlation spectroscopy 
measurements probe the dynamics in terms
 of the intermediate scattering function 
$I({\mathbf{Q}},t)$. That at a given value of ${\mathbf{Q}}$ this 
intermediate scattering function is of the form
to $I({\mathbf{Q}})\,e^{-t/{\tau_0}}$, where $I({\mathbf{Q}})$ more or less
follows the trends of the static structure factor $S({\mathbf{Q}})$, 
does not at all warrant that the interpretation and the 
calculation of  $S({\mathbf{Q}})$ 
could be applied as  valid procedures
for the calculation and the interpretation of $I({\mathbf{Q}})$.

Several of these problems are quite subtle and require a detailed discussion.
The task is rendered difficult by interferences between the problems
listed above. 
The paper is organized as follows. In the second Section we prove
that the ``sine wave'' proposed by the authors is not a proper wave,
and that it is not periodic. We take the occasion to show that the interpretation
of the data is not unique. In the third Section we show why the coherence
of the data does not imply that there would be correlations between atomic jumps.
In the fourth Section we explain why the neutron data rule out
the scenario of tile flips proposed by the authors. In Section V we discuss
a number of problems for the understanding of the elasticity theory.
With respect to point (9)  we show  in Section VI what a correct methodology of approach 
to coherent quasielastic scattering signals
from phason dynamics should be. We discuss several models and
derive here an {\em extremely important result}:
A simple model with a single diffusing tile is sufficient 
to produce a signal
 that (1) is situated at ``satellite positions'' 
at a distance $q$ from each Bragg peak, that (2) has an intensity that 
scales with the intensity of the corresponding Bragg peak,  (3) falls off as $1/q^{2}$
and (4) has a time decay constant that is proportional to $1/Dq^{2}$.
In Section VII we reexamine the evidence presented in favour of the random
tiling interpretation. In Section VIII we discuss the validity
of the claim that ``phason jumps'' would not be a correct terminology.
And in Section IX we conclude.\\

{\section{Periodic Delusions}}

{\subsection{The function $u_{\perp} \sin(q_{\parallel}\, x_{\parallel})$ is not a wave}

Various authors\cite{Francoual,Janssen} have proposed ``phason waves'' of the type
$u_{\perp} \sin(q_{\parallel}\, x_{\parallel})$,  with wavelength 
$\lambda = 2\pi/q_{\parallel}$ along parallel space $E_{\parallel}$, and  polarisation
with an amplitude $u_{\perp}$along perpendicular space $E_{\perp}$.
But it is an obvious mathematical fact that this does not define a periodic 
displacement wave in parallel space $E_{\parallel}$.
To see and convince oneself of this, it suffices to inspect the displacement pattern 
the ``sine wave'' brings about on the drawing 
that defines the cut-and-projection algorithm in 
$E = E_{\parallel} \oplus E_{\perp} = {\mathbb{R}}^{2}$
that generates the Fibonacci sequence.
The very misleading notation 
as a perfect sine wave $u_{\perp} \sin(q_{\parallel}\, x_{\parallel})$
creates the illusion that
there is a periodicity by definition. Certainly, 
the values this function takes
in $E_{\perp}$ are periodic.
But this is not the end of the story, as we have to
inspect how this is translated by the algorithm
 into a displacement field in $E_{\parallel}$.
With respect to this issue, the value the sine function takes
in perpendicular space is not the only variable that comes
into play. What also comes into play at a given value of $x_{\parallel}\in E_{\parallel}$
is the position of the atomic surface ${\cal{W}}$ at $(x_{\parallel},x_{\perp})$ 
 with respect to the cut.
This position will define the minimum value $t(x_{\parallel})$ 
of perpendicular space ondulation
it takes to induce a jump.
Let us call the end points of ${\cal{W}}$, $A$ and $B$.
We take $A$ to be the point with the higher
$x_{\perp}$-value. The minimum value $t(x_{\parallel})$ is 
the distance between the point
where parallel space cuts the atomic surface and one of the endpoints
$A$ or $B$ of that atomic surface.  It is the comparison
between $t(x_{\parallel})$ and $u_{\perp} \sin(q_{\parallel}\, x_{\parallel})$,
rather than the value $u_{\perp} \sin(q_{\parallel}\, x_{\parallel})$ alone
that will determine if there is a jump or otherwise.
The function $t(x_{\parallel})$ is quasiperiodic.
Let us note the disordered QC generated by the sine wave 
$u_{\perp} \sin(q_{\parallel}\, x_{\parallel})$ as ${\mathbb{QC}}^{*}$,
and the perfect QC as ${\mathbb{QC}}$. To give ${\mathbb{QC}}^{*}$ as a
whole a periodic description through 
a superspace embedding one needs to lift it to dimension 3.
This is obvious from the fact that the Fourier spectrum
of ${\mathbb{QC}}^{*}$
is spanned by three reciprocal lattice vectors, viz. the two reciprocal
lattice vectors that span the Fourier spectrum of the Fibonacci lattice ${\mathbb{QC}}$
and the vector $q_{\parallel}$. Note that when we calculate the
Fourier spectrum of ${\mathbb{QC}}^{*}$ to better approximations, by taking
in higher order terms of the Taylor expansion that is used in the calculation (see below),
the harmonics with wave vectors $2q_{\parallel}$, $3q_{\parallel}$, etc...
will enter, such that the statement that $q_{\parallel}$ is a third generator
of the Fourier spectrum is perfectly accurate (see below).

The Fourier spectrum we are talking about here, is  the one of
 the modified QC, i.e. ${\mathbb{QC}}^{*}$, which is
 of course different from the Fourier spectrum of the displacement
 field ${\mathbb{D}}$ of phason jumps. 
 The Fourier spectrum of the jump field ${\mathbb{D}}$
itself is not an experimentally measured quantity.
The field ${\mathbb{D}}$ is not at all periodic, as we show now
on the example of the Fibonacci chain.

The cut in the superspace description of the Fibonacci chain
can pass either through zero or through two endpoints of atomic surfaces.
In fact, if there is at least one, it means that the cut is exactly
at the position of inducing a jump. The other end point of that jump
will then also be an endpoint of an atomic surface, viz. of that
atomic surface one is swapping to.
When the  endpoint
of the first atomic surface is of the type A, then the endpoint on the second 
atomic surface will be of the type B, because all this together defines a single jump.
There will be no  endpoint of any other atomic surface on the cut,
because else the slope of the cut would be rational.
This shows that the number of atomic surface endpoints can 
only be $0$ or $2$, and that when 
there are two of them, they are separated by a jump distance, because
they define a single jump and there can be only one jump at a time in the cut.
It is therefore possible to take a point $O$ on the cut, such that, to the right
of it, the cut does not contain a single endpoint of an atomic surface.
Let us take $O$ as the origin of the $x_{\parallel}$ coordinates.

Choose an arbitrary large positive integer number $K \in {\mathbb{N}}$.
Consider now the interval $[0,K\lambda]$ on parallel space.
Consider the atomic positions $x_{j}$ in this interval.
In each point $x_{j}$, it will take a minimal {\em non-zero} translation  
$t_{j} \neq 0$ of the cut along $E_{\perp}$
to induce an atomic jump, because, by our choice of $[0,K\lambda]$,
we know that in $x_{j}$ the cut does not pass 
through the end point
of an atomic surface. For all the atomic positions $x_{j}$ within 
$[0,K\lambda]$ we can define such a minimal non-zero translation $t_{j}$. 
The finite set of 
strictly positive numbers
$|t_{j}|, j \ge 0$ does not
contain $0$, such that it has a strictly positive minimum value. 
Call $a$ this minimum, and take $0 < |u_{\perp}| < a$.
Then $u_{\perp} \sin(q_{\parallel} \,x_{\parallel})$ will not induce a 
single atomic jump over the whole interval $[0,K\lambda]$.
By taking increasing values of $K$  we can make 
this situation arbitrarily bad. In fact, $a$ functions as a density parameter
for the jump field.
Nevertheless, the ``sine wave'' does define a non-zero displacement
field, because the Fourier spectrum of ${\mathbb{QC}}^{*}$ is different from
the Fourier spectrum of ${\mathbb{QC}}$.

This argument clearly indicates where the logical error lies in the reasoning
of  people who think that $ u_{\perp} \sin(q_{\parallel} \,x_{\parallel})$
would define a periodic displacement field. They have focused on the
blind spot of the formal appearance of the ``sine wave'',
without thinking about the way it cuts through the quasiperiodic pattern
of the atomic surfaces.

It is obvious from the previous that $\lambda$ has not the meaning of a wavelength 
for the displacement field ${\mathbb{D}}$ (see also below).
The value $a$ is a measure for the density of jumps
on parallel space. Of course, the density 
is inversely proportional
to the average distance between two next neighbour jumps.
As the jumps all involve the same jump 
distance $(\tau -1)/\sqrt{2+\tau}$, 
this value defines the maxima of
the displacement field defined by the sine wave.
The value of this maximum amplitude 
$(\tau -1)/\sqrt{2+\tau}$ cannot be changed
by varying $u_{\perp}$.
Hence we see that varying the amplitude of the ``sine wave''
changes the average distance between the maxima of the displacement field,
rather than its amplitude.
That is certainly not a wave: When we change the 
amplitude of a true wave,
it is exactly the opposite that happens, viz.
the distance between its maxima  remains the same, while the 
amplitude of its maxima varies!
In fact, $u_{\perp}$ does not act as an amplitude
of the displacement field. It rather defines its 
density (or the average
distance between displacements). 
The fact that the maximum amplitude of the physical 
displacement field is forcedly
 $(\tau -1)/\sqrt{2+\tau}$ is already sufficient to provide  
 us with a handwaving argument for the fact that
$u_{\perp}$ acts as a density parameter.
It is obvious that when $u_{\perp}$ increases,
the effects must become stronger. As the only two possible local values
for the displacement field are $0$ and  $(\tau -1)/\sqrt{2+\tau}$,
$u_{\perp}$ only can act on the number of jumps, not on the 
maximum amplitude of the field.

Our construction shows that $\lambda$ is not a period
of the displacement field. In fact, if X is the first point to the
right of O where we have a jump, then at position $Y: |YX| = \lambda$ to the left
of X, there  is no jump, as it should be if $\lambda$ were a period.
In fact, in general there will even be no atomic position at
$Y$, because in general $\lambda$ will not correspond to a distance between
two points of the Fibonacci sequence. 
The displacement field has no periodicity at all.
In fact, restricting the {\em domain} of a periodic function of ${\mathbb{R}}$
to the aperiodic dicrete set of QC lattice points ${\mathbb{QC}}$ already
destroys the periodicity {\em stricto sensu}, but the situation
is rendered worse by restricting also its {\em range} in ${\mathbb{R}}$
to a set that only contains two values $0$ and  $(\tau -1)/\sqrt{2+\tau}$.
The function is even no longer quasiperiodic with a two-dimensional ${\mathbb{Z}}$-module
because its periodic embedding requires a superspace dimension of 3 rather than 2.
It can thus {\em a priori} not be claimed on the basis of an argument
of (wrongly) assumed periodicity that its Fourier transform 
would have non-zero components
at $Q_{B} + q_{\parallel}$, where $Q_{B}$ defines a Bragg peak of the QC, and 
$q_{\parallel}$ is defined by the relation
$2\pi/\lambda = q_{\parallel}$.
But we shall see below that this is nevertheless true.
It is also {\em a priori} not obvious that the Fourier decomposition 
of the ``sine wave'' would yield $Q_{B} + q_{\parallel}$-values 
that all have the same
value of $q_{\parallel}$. But again, this is true.
 
To save the periodicity, one might argue that
one made indeed an error but that this just consisted in expressing
one's ideas through the wrong formula. 
One might argue that a displacement wave of jumps of the type
$u_{\parallel} \sin(q_{\parallel} \,x_{\parallel})$ surely can exist.
But this is also badly wrong: It is impossible to define any sine wave based
on a unique jump distance on the Fibonacci chain, because it would just
contradict quasiperiodicity (see our argument
about restricting the domains of periodic functions). 
As we already stated, $\lambda$ will only 
in very exceptional cases correspond to a distance between
two atomic positions of the Fibonacci sequence.
Hence the positions $P$ were the atomic jumps $PQ$ would have to take
origin will in general fall somewhere in between the atomic positions
of the quasicrystal. Moreover, the intensity of the diffuse pattern
will not scale with $Q_{\perp,B}^{2}$, when the polarization is along
$u_{\parallel}$.

Finally, we must make a remark about the fact that our approach has been 
one-dimensional. We can see that our objection can be avoided by
taking the direction of the wave vector along the {\em periodic} direction
of a ``two-dimensional'' quasicrystal
(like a decagonal phase). In that case the ``sine wave''
does not run into the same conceptual difficulties.
We may note that this reminds us of the electron 
microscopy observations
of a ``large phason jump'' by Edagawa et al.\cite{Edagawa}, 
but these authors remain cautious
about the interpretation of their data in terms 
of a ``sine wave''\cite{Ames2}.
Secondly, domain walls have been observed in icosahedral
phases by electron microscopy, with a high density of tile flips.
Only a few orientations of such domain walls will
be energetically allowed, and it would also require
a lot of energy to move them as a whole. The Fourier transform
of such a plane is a line.  
This represents then a correlation in the atomic positions where
the flips occur, in the sense that they are all
located in the same plane. But the $q_{\parallel}$ vectors 
that one could associate with the lines mentioned would 
in the ``sine wave'' picture correspond
to a correlation in the direction normal to the plane,
not within it.
The domain wall can also not be of the ``sine wave'' 
form proposed by the
authors, as the displacement field defined by 
a ``sine wave'' is not two-dimensional.
And within the plane, relaxing phason jumps will 
not be any more correlated
than along a Fibonacci chain (see also below).

{\em In conclusion: There does not exist any true, 
periodic sine wave of atomic jumps 
in physical space on a QC.}\\

{\subsection{Calculation of the diffraction pattern of the disordered QC}}

We must criticize the way a single Fourier component 
in the diffuse scattering pattern
is given physical meaning in reference \cite{Francoual}.
Let us first point out that a Fourier componenent of 
the diffuse scattering
corresponds to a density wave. 
In the theory of Jaric' and Nelsson,\cite{Jaric} the basic
formalism is indeed based on density waves.
These density waves are not restricted to
the discrete set of atomic positions of the QC
as the authors of reference \cite{Francoual} propose in their interpretation,
but they are defined on the whole continuum of parallel space, where they are
allowed to cancel mutually, etc... In fact this mutual
canceling is what happens most of the time because the QC
is a discrete density.
The diffraction diagram as a whole, with the Bragg peaks and the diffuse 
scattering, describes thus a priori a field of {\em atomic positions}, not a field
of {\em atomic jumps}.

Let us calculate the diffraction pattern of a Fibonacci chain,
whose corresponding two-dimensional ``supercrystal'' 
has been modulated (in superspace) by the ``sine 
wave'' $u_{\perp} \sin(q_{\parallel}\, x_{\parallel})$.
After applying the ``sine wave'' the (non-decorated) two-dimensional lattice
will be a truly modulated crystal, whose two-dimensional diffraction
pattern will be given by Bragg peaks  $\delta({\mathbf{Q}} - {\mathbf{Q}}_{B})$ and
satellites $\delta({\mathbf{Q}}-({\mathbf{Q}}_{B} \pm {\mathbf{q}}_{\parallel}))$.
Let us note the window function of the strip method as ${\cal{W}}$.
The points of this modulated lattice that fall
into the strip ${\cal{W}} \times E_{\parallel}$ project to the 
atomic postions of the disordered QC.
Alternatively, we can decorate the atomic positions of the modulated ``supercrystral''
with atomic surfaces ${\cal{W}}$ and cut with $E_{\parallel}$.
What we have to calculate is:

\begin{equation} \label{Eq1}
\int\,
e^{\imath {\mathbf{Q\cdot r}}}\,
\sum_{(M,N)\in {\mathbb{Z}}^{2}}\,
\delta({\mathbf{r - r}}_{M,N} - {\mathbf{u}}_{M,N})\, 
* {\cal{W}}({\mathbf{r}})\,d{\mathbf{r}} =
(\int\,e^{\imath {\mathbf{Q\cdot r}}}\,{\cal{W}}({\mathbf{r}})
\,d{\mathbf{r}})\,\times\,
(\sum_{(M,N)\in {\mathbb{Z}}^{2}}\,
e^{\imath {\mathbf{Q\cdot}}({\mathbf{r}}_{M,N} + {\mathbf{u}}_{M,N})}\,),
\end{equation}

\noindent where ${\mathbf{u}}_{M,N} = 
{\mathbf{u}}_{\perp} \,
\sin({\mathbf{q}}_{\parallel}{\mathbf{\cdot r}}_{M,N})$,
and where we have indexed the nodes of the ``supercrystal'' by $(M,N)$.
When $u_{\perp}$ is small we can develop 
$e^{\imath{\mathbf{Q\cdot u}}_{\perp}\,
\sin({\mathbf{q}}_{\parallel}{\mathbf{\cdot r}}_{M,N})}$ 
in a Taylor expansion, such that 
we obtain for the second term of
the product in the righthand side of Eq. (\ref{Eq1}):

\begin{equation}\label{Eq2}
\sum_{(M,N)\in {\mathbb{Z}}^{2}}\,
(\, e^{\imath {\mathbf{Q}}{\mathbf{\cdot}}{\mathbf{r}}_{M,N}}
+ ({\frac{{\mathbf{Q\cdot u}}_{\perp}}{2}})\,
( e^{\imath ({\mathbf{Q+q}}_{\parallel}) {\mathbf{\cdot}}{\mathbf{r}}_{M,N}}
- e^{\imath ({\mathbf{Q-q}}_{\parallel}) {\mathbf{\cdot}}{\mathbf{r}}_{M,N}})\,).
\end{equation}

\noindent The first exponential will lead to the set of Bragg peaks,
the second and third exponentials will lead to Bragg peaks shifted
by  $\pm {\mathbf{q}}_{\parallel}$ and further weigthed with 
$ {\frac{1}{4}} (Q_{\perp,B} u_{\perp})^{2}$.
We see thus that there will be satellites.

The situation is completely analogous to what we observe
with a phonon modulation on a periodic crystal.
E.g. a  longitudinal phonon displacement field $u \sin(qx)$ 
on a periodic one-dimensional
crystal will introduce satellites at $\pm q$ with respect to each Bragg
peak $Q_{B}$. These satellites correspond to density waves with wavelength
$2\pi/(q+Q_{B})$. The function $u \sin(qx)$ is defined over the whole of
${\mathbb{R}}$. By using it to define a displacement wave,
we restrict it to the discrete set of the lattice nodes of the crystal.
As a corollary its Fourier transform changes from
$\delta(Q\pm q)$ for the function whose domain is ${\mathbb{R}}$,
to the infinite set of Dirac measures at $Q_{B} \pm q$.
In other words, the price to pay for restricting a continuous function
to a lattice, is that one must convolute its Fourier transform
with the reciprocal lattice. We see that $2\pi/(Q_{B}+q)$ 
corresponds to a density
wave, not to a period, and that we need a whole set of them
in order to render the intensity in between the atomic
positions of the phonon-modulated crystal zero. 
The result for the QC follows the same philosphy, with
a more complicated reciprocal lattice.

It may look surprizing and counterintuitive that we find that the 
position of these satellites for the QC
is independent from the amplitude $u_{\perp}$, while we have shown
that for small enough amplitudes there will be much longer
distances between the jumps than given by $2\pi/{q}_{\parallel}$.
But this is nevertheless correct (see below).
This formalism corresponds to the paradigm used 
in reference \cite{Francoual}.
It corresponds to the idea of Equations [42] and [43] in 
the paper by Janssen et al.\\

{\subsection{Criticism of the wave approach}}

There are very serious problems with the approach outlined above:\\

{\subsubsection{Topological problems}} 

A problem with the application of the elasticity theory 
to phason dynamics is that
Lubensky's theory is first derived
for the  example of a structure obtained as a union of two
{\em incommensurately modulated sublattices} (see his Figure 3.1).
As such it creates the impression that he would have avoided the major pitfall
of obtaining the excitations from a cut through higher-dimensional
dynamics, and the reader thankfully relaxes vigilance.
But later on Lubensky lifts the description of the QC 
to higher-dimensional space. 
The danger of the idea of a cut through
higher-dimensional dynamics thus surfaces again,
in less alarming appearances
 by presenting it the other way
around. Eventually we find people talking 
about notions  unifying phasons and phonons as 
being waves that only differ by having different
polarization vectors. 
Between Lubensky's Figure 3.1 and this final formulation
 there is a big leap
with many gaps in the argument. 

The phonon dynamics of the Fibonacci 
chain can in general not be obtained
by just making a cut through the phonon dynamics of a 
two-dimensional lattice,
as is suggested by the Equation [42] in the paper of Janssen et al.
The phonon problem of the Fibonacci is an unsolved problem
of an horrendous difficulty.
This has several origins: 

(a) The topology of the problem
is different from that of the phonon problem
on an unbounded lattice. Neighbour interactions with 
atomic surfaces that are out of the acceptance
window, i.e. out of the superspace strip, are just cut away. This gives rise
to boundary conditions.  This is the  discrete
analogon of a wave equation on a domain $D$. 
In the example of the Fibonacci chain with only first
neighbour interactions, the boundary problem is 
quite extreme, as
every atomic position is on the boundary  $\partial D$ of 
the domain $D$ of the wave equation, such that $D$ has
an empty interior. 
In fact in ${\mathbb{Z}}^{2}$
each lattice node has 4  first neighbours, while on the Fibonacci chain
each point only has 2 neighbours. This
can be formulated as a condition 
that is the discrete analogon  of
a boundary condition of the  type ${\mathbf{e}}_{n}\cdot \nabla \psi = 0$,
 on $\partial D$ for the wave equation 
$\Delta \psi = 1/c^{2} \partial^{2} \psi/\partial t^{2}$;
here ${\mathbf{e}}_{n}$ is the normal to the boundary.
But in the discrete problem ${\mathbf{e}}_{n}$ does not vary smoothly.
 It takes different orientations,
depending on the question if its first neigbour environment
corresponds to $L\cdot L$, $L\cdot S$ 
or $S\cdot L$ (where $L$ and $S$ are the interatomic distances).

(b) Using  another approach, the two-dimensional lattice must 
be considered to
 be decorated with
atomic surfaces, which implies that there is an
infinite number of ``atoms'' $s$ in the unit cell $(M,N)$.
The dynamical matrix that defines  all the interactions 
between pairs of atoms in first-neighbour cells
$(M,N,s_{1}), (M+1,N,s_{2})$, etc..,
is infinite.

Postulating wavelike solutions as is done e.g. in Eq. [43] 
of the paper of Janssen {\em et al.}
is, at least in general, very wrong. There is not the slightest proof 
that this is reasonable in general.
Rigorous studies of the phonon problem on the Fibonacci chain through
the transfer matrix method seem to indicate that such an {\em Ansatz}
is just wrong. This in turn raises the question if there are phason ``modes''
(in the usual sense of a wave)
all together. 

A QC is not a continuum
but a discrete lattice. The discrete approach prevails
thus over the continuum approach. The continuum approach
is obtained by taking the long-wavelength limit of the discrete problem.
When we propose thus a continuous phonon mode,
it must be validated by checking if it makes sense by taking the limit of
the discrete solutions. 
For continuous phonon modes this makes sense.
For phasons the situation is all together different:

(a) There are a number of famous papers by Levitov and others\cite{Levitov} that show
that in  general phasons in QCs crystals are not continuous.

(b) In contrast with what happens with the phonon modes, the displacement field
of the postulated phason wave is not at all sinusoidal in parallel space.
When we make the discrete graph $u(x_{\parallel})$ {\em vs.} $x_{\parallel}$
of a long-wavelength phononlike 
sine wave diplacement field 
 on the Fibonacci chain, and we look
at this graph  from very far, 
such that the atomic positions seem to
fill a continuum, the picture of the sine wave will become clearly visible.
It is this that gives sense to the continuum limit for phonons.
When we do the same for the displacement field of the phasonlike ``sine wave'',
we will not be able to make sense of it: In certain regions the 
graph will appear to consist of two horizontal 
lines $u(x_{\parallel})= (\tau - 1)/{\sqrt{2+\tau}}$,
and $u(x_{\parallel}) = 0$,
as though it would correspond to a two-valued  ``function''.
The two lines will have  grey shading , rather than being just black.
In other regions, it will appear to be just $u(x_{\parallel}) = 0$ alone.
It is thus no longer possible to keep telling accurately what is
going on in the QC when we move out to inspect it from further away with
a coarsened resolution. Taking the ``long-wavelength limit'' ceases to 
be a self-evident concept or method,
because (1) one cannot define limits for discontinuous functions,
and (2) the displacement field cannot be described or regularized as the restriction
of a continuous wave to the QC.

(c) The postulated phason modes lead to notions of long distance coherence
between atomic jumps that are completely unphysical.
This will be developed below: The closer the jumps are in time,
the farther away they must be in space! Also, in all experimental studies
of atomic self-diffusion, the paradigm and the evidence 
is that atomic jumps are not correlated
over long distances. 

(d) We may add to this that the infinite lattice allows also
for modes with an exponential decay in space.
These are excluded in a real  crystal
because these exponential waves diverge, in one direction or another,
but in the QC, divergence along the
$x_{\perp}$ direction is harmless, due to the restriction to the strip.
Hence, if the wavelike solutions were correct, they would lack generality.\\

{\subsubsection{Incompleteness of the data}} 

While in a crystal 
the spectral reponse for a phonon modulation 
 is rigorously the same at every Bragg
peak, in a QC the intensity
is different at every Bragg peak. It is thus no longer
representative to study a single Fourier component at $Q_{B}\pm q_{\parallel}$ 
for a given $q_{\parallel}$-value
at a choosen Bragg peak defined by $Q_{B}$. One has a
 whole set of satellites, with their
respective intensities. It would be normal to check experimentally
that these relative intensities at $Q_{B}\pm q_{\parallel}$ 
 for various Bragg peaks
are compatible with the calculation
of the picture of a superspace phonon.
It would also be normal to check experimentally if the same $q_{\parallel}$-value
at different Bragg peaks $Q_{B}$  yields identical relaxation times.\\

{\subsection{Solution of a paradox}}

We must now treat the apparent contradiction that
$q_{\parallel}$ does not correspond to  a period of 
the displacement field ${\mathbb{D}}$,
while it appears in the diffraction spectrum.

{\subsubsection{An important distinction}} 

Let us first note 
that we must make a difference between 
${\mathbb{QC}}^{*}$, which is
the whole disordered quasicrystal, i.e. a set of 
{\em atomic positions}, 
and the displacement field ${\mathbb{D}}$, which 
is a set of {\em atomic jumps} located
at certain atomic positions, and which can have a 
very large average distance
between the points where it is non-zero.\\

{\subsubsection{The disordered quasicrystal}}

It may look paradoxycal that the distance between 
successive jumps in the displacement field
is much larger than the wavelength of the sine wave, 
while the wave vector $q_{\parallel}$
shows up in the Fourier spectrum of the disordered QC. 
It is nevertheless true.
This is actually no more paradoxical than that the 
reciprocal lattice vectors of the the ${\mathbb{Z}}$-module
of the superspace embedding (like $2\pi/L$ and 
$2\pi/S$ in the Fibonacci chain)
appear (as generators) in the diffraction pattern of the QC, 
while they are not periods.
In fact, we may consider the superspace lattice, 
decorated with its atomic surfaces
as modulated by the sine wave 
$u_{\perp} \sin(q_{\parallel}\, x_{\parallel})$.
Following the method of Dewolff, Janssen and 
Janner,\cite{superspace} we can 
approach such a modulation by lifting the reciprocal lattice
to a three dimensional periodic lattice.
This allows us to see that the Fourier module
will now have the additional generator $q_{\parallel}$.
And this does not depend on the amplitude $u_{\perp}$.
We can also see that the whole spectrum of harmonics
of  $q_{\parallel}$ can in principle occur. It is just that they
have too weak intensities when they are not observed. E.g. when 
$u_{\perp}$ becomes larger or if the modulation is not sinusoidal, 
they may show up. (E.g. if in our derivation of Eq. (\ref{Eq2}) one
develops the exponential to second order, the harmonics $2q_{\parallel}$
come into play). This can perhaps
even be used to find out how we can describe the system
without violating the conservation of atoms when the modulation is strong.
We can see from this very clearly that $\lambda$ is not a period,
just like the interatomic distances $L$ and $S$ in the 
Fibonacci chain are not periods.\\

{\subsubsection{The displacement field}}
Let us now check how it is with the paradox for
the proper displacement field ${\mathbb{D}}$. 
The Fourier spectrum of
the displacement field is {\em not} the difference of the Fourier
spectra of ${\mathbb{QC}}^{*}$ and ${\mathbb{QC}}$; 
for a displacement from $x_{1}$ to
$x_{2}$, $e^{\imath Q(x_{2} - x_{1})} \neq 
e^{\imath Qx_{2}} - e^{\imath Qx_{1}}$, also not in first order.
But let us inspect what we obtain when we subtract the
Fourier transforms of ${\mathbb{QC}}^{*}$ and ${\mathbb{QC}}$. It would yield
the Fourier transform of a set of dipoles: There is a negative
Dirac measure at each starting point $x_{1}$ of a jump,
and a positive Dirac measure at each end point $x_{2}$ of a jump.
Hence we have ${\cal{F}}[\,\sum (\,\delta(x-x_{2}) - \delta(x-x_{1})\,)\,]$.
What we would need to calculate is 
${\cal{F}}[\,\sum \, (\tau -1) 
\,{\mathbf{e}}_{\parallel}\,\delta(x-x_{1})\,/{\sqrt{2+\tau}}]$.
In the superspace description the difference spectrum gives rise
to just a subtraction of the densities of the atomic surfaces.
At one end of an atomic surface the subtraction yields 
a positive density, in the region
where the atomic surfaces do not overlap,
at the other end a negative density. When the jump occurs to the right
it is the upper end of a subtracted atomic surface 
where the positive density occurs,
and the lower end of the neighbouring  subtracted atomic 
surface where the negative 
density occurs. We certainly need these pieces of density to
make a calculation of the Fourier transform of the displacement
field, but we need a method to select only the negative pieces,
and weighting it with $\pm  
(\tau -1)  \, {\mathbf{e}}_{\parallel}/\sqrt{2+\tau}$ 
where the sign depends
on the question if the jump occurs to the left or to the right.

To simpfly the problem it is better to consider
the two displacement fields ${\mathbb{D}}_{1}$,
defined by $u_{\perp} (1 + \sin(q_{\parallel}\, x_{\parallel})\,)$,
and ${\mathbb{D}}_{2}$
defined by $-u_{\perp}$. The first one will only yield jumps to the right.
The second one only jumps to the left.
Subtracting the Fourier transform of the perfect QC from the Fourier transform
of the QC modulated by ${\mathbb{D}}_{1}$, will correspond
to the Fourier transform ${\cal{F}}_{1} = {\cal{F}}(S_{1})$ of 
a set of dipoles $S_{1}$, 
that are all oriented
to the right. Call the set of origins
of the jump vectors of ${\mathbb{D}}_{1}$, $B_{1}$,
and a dipole oriented to the right $d$. 
Then ${\cal{F}}_{1} = {\cal{F}}(B_{1} * d) = 
{\cal{F}}(B_{1}) {\cal{F}}(d)$,
from which ${\cal{F}}(B_{1})$ can be calculated.
(There is a problem with this, in that ${\cal{F}}(d) 
= e^{\imath Q (\tau -1)/\sqrt{2+\tau} } - 1$
contains zeros, but these do not occur on the satellites).
Similarly we can calculate the Fourier transform of
the set ${\cal{F}}(B_{2})$ of origins of jump 
vectors defined by ${\mathbb{D}}_{2}$.
If we take ${\cal{F}}(d)$ to deconvolute, the set $B_{2}$ will be
negatively weighted. With this weighting
 $ (\tau -1) \, (\,{\cal{F}}(B_{1}) + 
 {\cal{F}}(B_{2})\,)/\sqrt{2+\tau}$, will then be the Fourier transform
of the displacement field ${\mathbb{D}}$. 
When in a given point both a jump to the right 
occurs within in ${\mathbb{D}}_{1}$
and a jump to the left within ${\mathbb{D}}_{2}$, then the two jumps
will conveniently add up to zero. In all other points, the jumps
will occur with the proper weighting.
This point hinges of course on
the fact that the amplitude of $u_{\perp}$ must be  small enough.
If the amplitude of the sine wave is too large, there will be successive
jumps of a same atom, and the jump distances will not all be  $(\tau -1)/\sqrt{2+\tau}$.
In conclusion, also for ${\mathbb{D}}$
the solution of the paradox is  that $q_{\parallel}$  
does indeed occur in the Fourier spectrum.\\

{\subsubsection{Conclusion}}

We see that doubling the amplitude $u_{\perp}$ will 
{\em in first order} double the intensity
 of the satellites both in the Fourier spectrum of ${\mathbb{QC}}^{*}$,
and in the Fourier spectrum of the displacement
field ${\mathbb{D}}$. But of course, this does not imply that
${\mathbb{QC}}^{*}$ or ${\mathbb{D}}$ would be periodic, or 
that ${\mathbb{D}}$ would
be physically meaninful (see below).\\

{\subsection{Clearly distinguishing the various fields}}

Let us resume the situation.
A single Fourier component at some value $Q_{B} + q_{\parallel}$
in the diffuse scattering is a  continuous density wave
and as such cannot be the Fourier transform of a field of phason jumps.
Even a set of Fourier components $Q_{B} + q_{\parallel}$
with the same $q_{\parallel}$-values does not correspond to  a
periodic wave of jumps: 

(a) (If we include also the Bragg peaks,) it just corresponds
to a disordered quasicrystal ${\mathbb{QC}}^{*}$, not 
to the field of jumps that would permit
to go from the perfect quasicrystal ${\mathbb{QC}}$ to ${\mathbb{QC}}^{*}$.
(Of course the ``difference'' between the pristine 
${\mathbb{QC}}$ and the disordered
${\mathbb{QC}}^{*}$ corresponds to a field of jumps).

(b) We have proved that when a ``sine wave''  transforms 
the perfect ${\mathbb{QC}}$ into 
the disordered ${\mathbb{QC}}^{*}$ with such a $q_{\parallel}$-based
 set of Fourier components with wave vectors $Q_{B} + q_{\parallel}$, then this
 does not imply at all that the transformation from the perfect
 ${\mathbb{QC}}$ to the disordered ${\mathbb{QC}}^{*}$ would be be based on
a {\em periodic}  wave of atomic jumps at positions of ${\mathbb{QC}}$.\\

Thus in a diffuse scattering pattern
that uniquely corresponds to the disorder generated by tile flips
(e.g. in the Monte Carlo simulations of Tang et al.\cite{Tang,Shaw}),
the intensity at $Q_{B} + q_{\parallel}$ can never have physical meaning, as it
corresponds to a continuous sine wave by definition. 
The diffuse scattering intensity $Q_{B} + q_{\parallel}$
in the disordered-tiling model is only a Fourier component, without any
real physical meaning. It corresponds to a density wave for the 
disordered quasicrystal ${\mathbb{QC}}^{*}$.
To obtain a physical meaning we must combine 
all Fourier components $Q_{B} + q_{\parallel}$ at all Bragg peaks. 
And in a given point of space,
these Fourier components will most of the time just cancel mutually
to yield zero density, as the atomic positions are a discrete set.
By combining all contributions with wave vectors
$Q_{B} \pm q_{\parallel}$, corresponding to a single $q_{\parallel}$
linearly with their appropriate intensities,
we may reconstruct the whole disordered tiling 
${\mathbb{QC}}^{*}$, produced by the ``sine wave''
with wavelength $2\pi/q_{\parallel}$ from 
the perfect quasicrystal ${\mathbb{QC}}$.

We see thus that we have the following scheme:

\begin{eqnarray*} \label{scheme}
\begin{array}{ccc}
Bragg~peaks~\{\,Q_{M,N} \parallel (M,N) \in {\mathbb{Z}}^{2}\,\} & 
\begin{array}{c}
    ~\\
    \longrightarrow\\
   \pm \, q_{\parallel}\\
       \end{array} & 
Bragg~peaks~\{\, Q_{M,N}\pm q_{\parallel},0 \parallel (M,N) 
\in {\mathbb{Z}}^{2}\,\}\\
\downarrow {\small{{\cal{F}}}} & ~ & \downarrow {\small{{\cal{F}}}}\\
perfect~{\mathbb{QC}} & \begin{array}{c}
    ~\\
    \longrightarrow\\
    ~aperiodic~field~of~phason~jumps~{\mathbb{D}}\\
    \end{array} & disordered~{\mathbb{QC}}^{*}\\
\downarrow lift  & ~ & \downarrow lift \\    
 {\cal{W}} * {\mathbb{Z}}^{2} & \begin{array}{c}
    ~\\
    \longrightarrow\\
    periodic~modulation~wave ~q_{\parallel}\\
    \end{array} &   {\cal{W}} * modulated~ {\mathbb{Z}}^{2}\\        
\end{array}
\end{eqnarray*}

\noindent Here only the last line in the diagram corresponds to
superspace quantities. The two other lines refer to parallel spaces:
The first line to  reciprocal space, the second line to direct space.

It is important to realize that the periodic modulation wave 
is only the Fourier transform of $q_{\parallel}$ in the superspace
sense, and that the Fourier transform of $q_{\parallel}$
in $E_{\parallel}$ has no physical meaning, e.g. it does not correspond
to the
aperiodic field of phason jumps ${\mathbb{D}}$. The 
periodicity of the modulation in superspace
does not imply that the field of phason jumps ${\mathbb{D}}$ would be periodic.
This field is definitely aperiodic. The field of the phason jumps 
is not directly measured, and its Fourier transform is not 
a single value $\delta(Q - (Q_{B} \pm q_{\parallel}))$.
The only quantities that can be directly measured are the ones on
the first line of the diagram.\\

{\subsection{Further criticism of the ``sine wave'' interpretation}}

The way
the authors analyse the isolated diffuse intensities at $Q_{B} + q_{\parallel}$
by drawing the relaxation time as a function of $q_{\parallel}$
to prove that it would correspond to a diffusive mode of jumps
is approximate and incomplete. As we have seen it 
corresponds to a physically meaningless
density wave.
To obtain meaningful jumps in the case of a sine wave modulation
one would have to study a whole  set of  $Q_{B} + q_{\parallel}$-values 
simultaneously,
as only a whole set will yield the disordered QC,
created by a field of jumps. 
There is an experimental
problem that we do not know exactly how the corresponding 
amplitudes are to be combined . This is the very same problem
as for  reconstructing the perfect QC from the intensities of its Bragg peaks.
Even if we leave this practical problem aside, there remain
conceptual problems.\\

{\subsubsection{The postulated diplacement field is unphysical}}

We have seen that when $u_{\perp}$ is sufficiently
small, $q_{\parallel}$ can correspond to a field that induces very few
atomic jumps that are very far away from each other,
with no jumps at all in between. If we imagine $u_{\perp}$
as built up from infinitesimal contributions
$\delta u_{\perp}$, we can have the sine wave
continuously growing from $0$ amplitude to the small amplitude
$u_{\perp}$. In a given large patch of the QC, we will see then
the tile flips occuring consecutively, one by one, and the consective
atomic jumps will be separated
by very large distances. The closer the phason jumps
are in time, the further away they will be in space!
That does not evoke an image of
a coherent process with correlated jumps! For correlated jumps,
we would like to see them not too far one from another,
such that they can interact by a force. Of course,
we can repeat the same argument for any increase of
$u_{\perp}$ from a finite reasonable value $u_{\perp}^{(1)}$
to another finite reasonable value $u_{\perp}^{(2)}$.
We have not proved such an argument rigorously for the case $0$ is not
one of these two values, but we doubt
that anyone would claim the opposite to be true.
Now what one sees in a speckle experiment
at a single value of $Q_{B}+q_{\parallel}$ is exactly 
a  change of intensity (in the form of an exponential decay)
between two
values close to a main value,
that is interpreted to correspond
to such a change of amplitude of the ``sine wave''.
This does not correspond to any normal
interpretation of the concept of a {\em mode}.\\

{\subsubsection{Lack of uniqueness}} 

There exist other modulations than of the phonon type,
e.g. compositional modulation, or magnetic modulation. They
can be calculated by exactly the same formalism as for a displacive
modulation and 
will yield satellites at the very same positions in reciprocal space.
The information is thus {\em unspecific}.
We cannot claim without further justification that we
have a displacement modulation. 

When a specific interpretation for an observation is claimed,
and when the interpretation is {\em a priori} not unique,
like it is the case with satellites in a Fourier spectrum,
the choice of the specific interpretation must be motivated.
It can happen that the motivation given for the claim
is not sufficient to make away with all possible ambiguities.
This means that the charge of proof for the claim has not been
met appropriately and that there are loopholes in the justification. 
When this is the case, this will in general give rise
to objections.
A good way to point out that the uniqueness
of an interpretation has not been proved,
is giving the possibility of a counter example.
 It is pointless and not appropriate to reply to the suggestion of
 such possibilities
that one would have to prove them, since that would amount to
a reversal of the charge of proof.
E.g. when one wants to point out a loophole in a proof of a
mathematical theorem, this does not imply that one would
have to prove that the {\em theorem} is wrong.
Pointing out a possible loophole shows that the
proof is incomplete, and as such that 
the {\em proof} of the theorem is wrong.
In an inductive science like physics, there
is no parade against infinite scepticism.
Therefore, the objections must remain reasonable.
All possible {\em reasonable} objections
must be duly incorporated in the discussion of the interpretation of the
data, and it must be clarified how bad they could be.
We want to raise the reasonable objection that the interpretation
of the diffuse scattering data is in bad lack of proof of
uniqueness. The complicated structure of the present paper,
with its many subdivisions in sections, subsections, etc...
is probably enough to show the profound confusion
this has produced.\\

{\em 1. Chemical Disorder.}
The authors have speculated several times\cite{Ames2,Krakow} about the possibility
that chemical disorder could be a kind of phason defect.
In reference \cite{Ames2} it is e.g. stated
that de Boissieu 
``wonders if the perp-space Debye-Waller factor component
is just another way of accomodating chemical disorder of atoms''.
It is also stated
that ``it is in agreement with the diffuse scattering''.
It is thus obvious that this possibility
must be ruled out, before one can claim that
the modulation is displacive.
As pointed out above, chemical modulation can be treated by 
exactly the same formalism as displacive modulation.
In that case, it is the complicated decoration of the atomic surface 
that interacts with the ``sine wave''. The change of the type of atom
is then possible by varying the amplitude of the ``sine wave'', and
the variation can occur without necessarily implying a phason jump.
In other words, in this case, the ``reason'' for the change of type of atom
is not given by the superspace description {\em in se}.
We may finally add that in AlMnPd, a strong component of diffuse scattering
has been observed at small angles\cite{Bellissent}, that is flat with
$Q$ and has an intensity that
corresponds to what one would obtain from a calculation assuming
complete chemical disorder. This seems to indicate that the presence of at least
some chemical
disorder has to be taken as a very serious possibility.
Introducing a chemical modulation ``sine wave'' can be done
without any reference to elasticity. This implies that when the data
are due to chemical disorder, they are irrelevant for
the {\em elastic} stability issues raised by the random tiling model, as Widom's
theory is based on displacive modulation.\\

{\em 2. An alternative type of displacive modulation.}
But even if we stick to a displacement
modulation, there is an alternative. Imagine a QC ${\mathbb{QC}}^{(\alpha)}$ 
obtained by replacing
the atomic surfaces  ${\cal{W}}$ of ${\mathbb{QC}}$ by other 
atomic surfaces ${\cal{W}}^{(\alpha)}$ that are tilted
by a small angle $\alpha$ with respect to ${\cal{W}}$, and whose lengths
are adjusted such as to keep the condition of conservation of
atoms satisfied.\cite{note}
The whole derivation of the calculation of the
Bragg peaks of ${\mathbb{QC}}^{(\alpha)}$ remains the same.
Hence, the positions of the Bragg peaks are the the same
in ${\mathbb{QC}}^{(\alpha)}$ and ${\mathbb{QC}}$. 
Only the intensities of the Bragg peaks
are changed. The Fourier transform of the tilted atomic surface
varies along the direction of this atomic surface,
but does not vary along the direction
that is perpendicular to it, such that
the value of $Q_{\perp}$ of the Bragg peak that occurs
in the expressions  
of the intensities $\sin (Q_{\perp}W/2)/(Q_{\perp}/2)$, 
must be replaced by 
$Q_{\perp}\cos\alpha - Q_{\parallel} \sin\alpha$. 
As the length $W$
of the atomic
surface also enters the expressions, it must also be properly
modified to the new value $W/\cos\alpha$. 
The final result is $\sin[\,(Q_{\perp} - Q_{\parallel} 
\tan\alpha)\,{W\over{2}}\,]/[\,(Q_{\perp}\cos\alpha 
- Q_{\parallel} \sin\alpha)/2\,]$.
 
Again we modulate the positions of 
the undecorated ``supercrystal''
with a sine wave $u_{\perp} 
\sin(q_{\parallel}\, x_{\parallel})$.
Note that this is slightly different from 
cutting the tilted atomic surfaces
with the ``sine wave'', because in the 
latter, the value
where the amplitude of the ``sine wave'' 
must be calculated 
would be slightly shifted along
$E_{\parallel}$.
Call the resulting QC ${\mathbb{QC}}^{(\alpha) *}$.
As rigorously the same kind of calculation
comes into play for the satellites as for the 
Bragg peaks, we see  that 
also the satellites of ${\mathbb{QC}}^{(\alpha) *}$
will be in exactly the same positions as in 
${\mathbb{QC}}^{*}$ and 
that only their intensities will be different.
Note that, in the end, the factor 
${\mathbf{Q\cdot u}}_{\perp}$ can only lead
to a ${\mathbf{Q}}_{\perp}$-dependence, 
due to the scalar product
with ${\mathbf{u}}_{\perp}$.
There is a small variation in the intensities of
the satellites in that they have now to be calculated
as $\sin[\,(Q_{\perp} - (Q_{\parallel}+q_{\parallel}) 
\tan\alpha)\,{W\over{2}}\,]/[\,(Q_{\perp}\cos\alpha - 
(Q_{\parallel}+q_{\parallel}) \sin\alpha)/2\,]$.
As $q_{\parallel}$ is small this will
deviate from the value at the Bragg peak
by a term of the order $W \alpha q_{\parallel}$,
which is certainly second order with respect to
$({\mathbf{Q\cdot u}}_{\perp})^{2}/4$,
such that the intensity of the diffuse
scattering remains uniquely dependent
on this parameter (The real effect
of variation of the diffuse 
scattering intensity with $Q_{\parallel}$ 
lies in the variation in the intensity
of the Bragg peak with which it scales).

Hence, there is no qualitative difference between 
the diffraction diagrams
of ${\mathbb{QC}}^{(\alpha) *}$ and ${\mathbb{QC}}^{*}$.
It may disturb at first sight that the modulation
produces displacements along $E_{\parallel}$,
but this not conceptually different from the 
situation with phason jumps.
It is thus a possible alternative.

This alternative is more physical.
In ${\mathbb{QC}}^{(\alpha) *}$ there will 
be small atomic displacements in every
atom of the QC, even when $u_{\perp}$ is very small.
Nobody will find it difficult to believe from
such a picture that the atoms are elastically  
coupled from neighbour to neighbour.
There is thus no postulate of spooky 
correlations over large
distances between two isolated jumps, 
while the whole intermediate 
region remains unaffected, as in ${\mathbb{QC}}^{*}$.
There is no such problem that the closer two jumps
are following each other in time, the further 
away they must be in space.
Moreover, when $u_{\perp}$ is very small, these 
small atomic displacements
will outnumber the very scarce atomic jumps.

An essential point is that the real-space 
atomic diplacements created
by the fluctuation of the cut are much larger
and not harmonic in the model of the authors.
An atomic jump can only occur in an double-well potential,
and the function that describes such a potential is at least
of the fourth degree. The potential is thus not harmonic.
We think that our alternative leads to a less unphysical
interpretation of the data, without a real necessity to 
invoke tile flips  as the basic ingredient.
The displacement fields are just like classical
phonon displacement fields, except for the fact that
they are parameterized by other, perpendicular 
space coordinates.
We also want to stress that the procedure of tilting
the atomic surfaces is only a first-order approximation
to illustrate the idea. In reality, we should
introduce a kind of devil's stair case
in order to account for far-away changes
of configurations. Furthermore, we should modify
the atomic surfaces in such a way that it does
preserve the symmetry. Duneau\cite{Duneau} has shown 
that in an admittedly wrong, polynomial
approach for icosahedral symmetry, the minimal 
degree of the polynomials
that 
comply with this
condition is three.
The idea of modulating the atomic surfaces
finds its confirmation in numerical calculations
of the dynamics, where it has to be introduced 
in order to relax
the initial system.
The idea of modulating the atomic surfaces is also
present in a work of Steurer,\cite{Steurer} who calls
it the IMS setting (as opposed to the QC setting). 

It remains to explain how the diffuse 
scattering resulting
from our alternative model could decrease 
when the temperature
is raised, but there are many possibilities to do this.
Widom's instability\cite{Widom}, with tile-flip phason 
elasticity replaced
by a phason elasticity based on small atomic
 displacements,
would already to the job. 
But it is even 
not necessary
to claim that the QC would not be stable.
A mere softening of the elastic constants would do.
There are many examples
known of elastic constants or phonon modes 
that  soften in a given
temperature range, 
without triggering any phase transition
at all. 

It is even not necessary to invoke a softening
of the elastic constants. It could just be
that the system acquires supplementary
possibilities to reduce the strain.
One possibility is e.g. that thermal 
vacancies contribute
to the relaxation of  the observed strain fields.
It is quite plausible that their number becomes 
significant
at the temperatures where the fluctuations observed 
by the authors set in.  Also fast phason hopping 
between two positions
could help in relaxing strain fields.

We may add a note about the fact that the diffuse scattering 
seems to follow predominantly the phason elastic constants.
The diffuse scattering reveals ``frozen'' phasons rather than ``frozen''
phonons.
Perhaps one thinks this is surprizing, and perhaps
one thinks it must be 
full of highly significant information.
 But a frozen phonon would stipulate an amplitude for an atomic
displacement field in a certain place, 
just on the basis of its position coordinate,
without any hindsight on its environment. The phonon wave could be
just totally out of phase with respect to the reality of
the local environment. In sharp contrast with this,
the phason variable
is a parameter that refers to the local environment.
In order to know the environment of a point $x$
in a crystal, it suffices to take the non-integer part
of $x/a$, where a is the lattice parameter.
Hence $x$ is a good quantity to define a phonon 
displacement field in a crystal, as it contains 
the necessary information about
the local environment.
In a QC, $x_{\parallel}$  is not a good parameter to define 
a displacement field in this respect.
The parameter $x_{\perp}$ is much better suited to define
the local environment.
We think that this indicates that a displacement field
in a QC can only reasonably expected to be of the frozen phason type.\\

{\em 3. Other possibilities?}
We cannot pretend that the list of possible alternatives has
been exhausted with these two counter examples. 
The whole problem with the interpretation of the diffuse scattering
is that it is hopelessly difficult, due to the large amount
of diffuse scattering scenarios in general, combined 
with the complexity of the structure
and of quasiperiodicity, more specifically. It is for this reason
that it is impossible to propose with certainty 
an alternative interpretation of the diffuse scattering data,
and that we were obliged in the previous lines
to limit ourselves to suggesting 
the existence of alternative possibilities.
This caution on our behalf should
not be misrepresented  as a weakness of
our viewpoint with respect to the viewpoint
that has been published, since that would
amount to a reversal of the charge of proof.
It rather illustrates the weakness 
of the published viewpoint in that it has taken a
non justified shortcut to the necessary caution.
We refer the reader to Section VI where we
develop a completely different approach and 
show how a jump model with a single diffusing tile
reproduces several characteristics of the data,
while there is no modulation of the
superspace lattice whatsoever.

We should also warn the reader against a play of words.
As the authors have rightly pointed out,
the word ``phason'' has been used with many different meanings.
This was also pointed out by Janssen et al \cite{Janssen}
and, much earlier, in reference \cite{Europhysics}.
Therefore it is all the more surprizing
that the authors rely on the mere fact that 
their data can be described in terms of ``phason elasticity''
to identify them without any further discussion
with phason jumps, as though this would be self-evident.
Clearly concepts cannot be identified
on the mere basis that they are homonyms. 
The two examples of chemical disorder and small 
non-phason-jump small atomic displacements
clearly may have physical realizations 
that could be described by a ``wave'' with perpendicular-space
polarization, even if
they both can have also other physical realizations that contain
a parallel-space component in the polarization.
\\

{\subsubsection{A possible ambiguity}}

There is not a single $q_{\parallel}$-value
but a whole distribution of them. It is obvious that
there is an ambiguity  that has not been addressed:
 The perpendicular-space amplitude $h_{\perp}$ of the ondulation
is not unambiguously defined: At each atomic position $h_{\perp}$ can
be varied at will as long as it does not lead to a swap of atomic surface.
In between the atomic positions $h_{\perp}$ can also be varied.
The physical situation in $E_{\parallel}$ allows 
thus for more than one decomposition in ``sinusoidal phason waves'',
it i.e. in sets $Q_{B} + q_{\parallel}$. 
This is perhaps connected to the fact that $Q_{B} + q_{\parallel}$ can also
be written as $Q_{B}^{*} + q^{*}_{\parallel}$, where $Q_{B}^{*}$ is another Bragg peak.
The information at
$q_{\parallel}$ is thus not unambiguous. It will contain a strong contribution
from a strong Bragg peak and other, weaker contributions.
We can reduce the importance of the latter problem by
making the speckle measurements in the vicinity of a strong Bragg peak $Q_{B}$,
such that the decomposition $Q_{B} + q_{\parallel}$ will dominate over all
other ones.\\

{\subsubsection{How is a non-sinusoidal ondulation of the cut Fourier-decomposed? }}

While the Fourier transform of a sine-wave
modulation will produce a set of  $Q_{B} + q_{\parallel}$-values
and corresponding intensities that allow to reconstruct
a field of pure phason jumps, it is not granted that
the Fourier decomposition of the whole ondulation of the cut 
can be kept stepwise,
putting each  atomic jump as a whole into one set of 
$Q_{B} + q_{\parallel}$-values that reconstruct  a sine wave 
that will generate this jump. Two jumps in succession
of the same atom, that do not result in a zero displacement,
will not forcedly lead to a Fourier decomposition
that contains pure phason jumps in the
$Q_{B} + q_{\parallel}$-sets. We see that this raises
the question of how large the amplitude of the sine wave can
be without ceasing to be sufficiently small for 
the validity of the approximation
that is claimed to justify the formalism.
In our development above,  we have more or less acted
as though the amplitude $u_{\perp}$ of the ``sine wave'' were less than
the amplitude of ${\cal{W}}$, in order to remain within
the limit of small amplitudes, and to stay away from problems
like violating the criterium of the conservation of the number of atoms.
This is not an important restriction, as what we want to deal
with in general are small variations of the amplitude,
and the only thing that really matters is that the
total ondulation of the cut conserves the number of atoms. 
In any case, it may thus well be that
a set of measured $Q_{B} + q_{\parallel}$-values with its intensities
defines fractional jumps.
What will remain of the interpretation of the diffusion constant analysis
in terms of a ``sine wave'' of jumps if this is the case?
The data from  $Q_{B} + q_{\parallel}$-sets may well reflect
small atomic displacements that are not whole phason jumps. 
A field of such
small atomic displacements cannot be distinguished from a field
where the displacements are not the result of only phason jumps.
This underlines once more our argument that it cannot possibly
be claimed on the basis of the data, that what one observes
is due to disorder induced by tile flips.\\

{\subsection{Summary}}

In conclusion of this Section, we have shown that the interpretation of
an intensity at $Q_{B} + q_{\parallel}$ in terms of a {\em periodic}
displacement field with wavelength 
$2\pi/q_{\parallel}$ of correlated jumps is wrong and unphysical. 
It contains  several tacit, misleading 
conjectures, whose truth is just taken for granted, while they
are obviously fallacies.\\

{\section{Coherent Delusions}}

{\subsection{Important Warning}}

Coherent scattering signals collect 
contributions from all 
particles of the system that have a non-zero 
coherent scattering amplitude.
It is thus a many-particle signal.
Despite all possible folk lore, this should 
not be overinterpreted.
That the scattering is coherent does not prove that it 
corresponds to
a ``collective'' signal or to a signal 
of some correlations,
as the latter implies that the particles 
are no longer 
independent and  move in a concerted, 
correlated fashion, 
due to some coupling or interaction, as is 
e.g. the case for phonons. That can very well
be the case, but it is not a necessity.
The oppositions independent vs. 
correlated,
coherent vs. incoherent, single-particle vs. 
many-particle
cannot be amalgamated.

(1) Coherent scattering with a certain 
structure can also
occur in a system wherein the dynamics of all 
particles are 
totally independent.
A clear example of this are the Monte 
Carlo simulations of 
Tang et al. \cite{Tang,Shaw} The tile flips are here completely
random and independent, but they lead to a clear
coherent signal that is actually very similar in its 
reciprocal-space properties
to the one observed by the authors (It is only the temperature
dependence that permits to invalidate the pristine random tiling 
model used in this simulation). 
The same conclusions have been reached by Naumis et al.\cite{Naumis}
They stated that in the long-wavelength limit
one cannot draw conclusions as to the presence of correlations,
because also totally uncorrelated jumps will give rise
to a coherent signal.

(2) Coherent scattering signals can be obtained
from the dynamics of a single particle, provided
it is a coherent scatterer. This is 
independent by definition.

(3) Conversely, many-particle systems can 
give rise to incoherent
signals, even if the dynamics are 
strongly correlated.
It just suffices that the particles 
are incoherent scatterers. 

The reader who wants to come to terms with these issues
is refered to reference\cite{Coddens}.
We will discuss how one calculates coherent scattering signals below.\\

{\subsection{Atomic jumps in self-diffusion problems are not correlated}}

It is generally admitted that phason jumps can give rise to
long-range atomic self-diffusion.
The idea has been pushed to the extreme in the model
of Kalugin and Katz \cite{KaluginKatz}, which predicted
extremely fast diffusion, but no evidence has been found
in favour of this prediction. What has been observed
is  extremely fast phason hopping.
We think that there is no absolute proof for the possibility
that phason jumps lead to diffusion, due to the simultaneous
presence of vacancy diffusion. But this is nevertheless
generally believed to be true, because it is hard to see
an obstacle against it.
In any case, Francoual et al.
make an analysis of their data in terms of a macroscopic diffusion constant.
But in all atomic self-diffusion studies, the prevailing paradigm
is that atomic jumps are not correlated over large distances.
This adds up to our previous arguments that postulating
correlations between atomic jumps over long distances
is unphysical.

Such an objection cannot be rebutted by
stating that it would be a mere speculation.
First of all, the objection just formulates the 
prevailing paradigm. In this respect it is
the postulate of correlations between the jumps that 
appears as a speculation. Secondly,
it would consist in a reversal of the charge of proof.\\

{\subsection{Atomic jumps in order-disorder transitions}}

The objection can also not be rebutted by invoking an order-disorder transition.
First of all, invoking an order-disorder transition
completely changes the context. It is difficult to see
how an order-disorder transition could tell us something
about atomic jumps in a self-diffusion process.
Secondly, atomic jumps do not become correlated in an order-disorder transition.
Once again, it is only the atomic positions, not the atomic jumps,
 that become correlated in
an order-disorder transition.
What one can e.g. do to illustrate this point in the situation
close to the  order-disorder phase transition
is considering a periodic lattice of asymmetric double-well potentialss.
E.g. the  minima of the left wells lie much lower than the minima
of the right wells. In each double well, an atom
is allowed to jump. When the temperature rises the difference between the
two wells
becomes smaller, and at high temperature the wells 
are completely symmetric, while at the transition temperature
they are highly asymmetric.
At all temperatures the jumps
in the different double-wells  are stochastic,
but when the temperature decreases, the jumping atoms
will have a much longer residence time in the
lower left well minima than in the higher right well minima.
While at high temperatures the residence times will be equal, because
the two wells are equally deep. The asymmetry will develop
when the temperature is decreased.
In the asymmetrical double-well the jump rates left-right and right-left 
are different, and their ratio is governed by a
Boltzmann factor. 
Such models have been used to describe the phenomenology of order-disorder transitions.
This clearly shows that the jumps remain independent,
while it is the most probable positions of the atoms (within the left well minimum
in between the jumps) 
 that become correlated.\\

{\subsection{Not every reciprocal lattice vector is a wavelength}}

In general,
there is no guarantee for the attribution
of a wavelength to the quantity
$q_{\parallel}$ as the authors do. 
The authors believe that the QC is not stable
and that its zero temperature ground state is a crystal.
The low-temperature phase is not observed
because the kinetics become frozen, such that the phase is not reached.
They want to interpret $q_{\parallel}$
as a wavelength associated with the phase transition towards the 
low-temperature
periodic, crystalline phase.
It may well be that the random tiling scenario
predicts a diffuse scattering as observed by Francoual et al.
But the ``wave lengths'' of this scenario
are in superspace, and do not have the real meaning of
a wavelength in physical space, as we have already shown.
We can also give some physical reasons within the context
used by the authors.

As there is no conclusive interpretation
of the diffuse scattering, considerations
about the nature of the {\em ad hoc} stipulated 
purely hypothetical transition 
(order-disorder, first or second order)
cannot play a role at this stage.
Consider thus the clear analogon
of a (second-order) antiferromagnetic 
phase transition.
 
At approaching
the N\'eel temperature from above, larger and larger 
antiferromagnetically ordered
domains (or clusters) will occur that 
will take longer and longer
times $\tau_{0}$ to decay. This will show 
up as diffuse scattering
intensity centered at the antiferromagnetic 
Bragg position, e.g. at
${\mathbf{Q}} = [{1\over{2}},0,0] {2\pi\over{a}}$ of 
the future low-temperature phase,
where $a$ is the lattice parameter 
of the high-temperature phase.
The intensity at ${\mathbf{Q}} + {\mathbf{q}}$ will
have a characteristic decay time $\tau_{0}$, which 
is a measure of how long
an antiferromagnetically ordered cluster of 
size $2\pi/q$ will persist in time
without being disrupted by the spin flip dynamics.
We see that it is $2\pi/Q$ rather than $2\pi/q$
that characterizes the wavelength $2a$
of the spin wave that is being built up. The quantity
$2\pi/q$ is not a long wavelength of some spin wave, 
but an instantaneous domain size, a coherence length
of the short wavelength spin wave. 
The time $\tau_{0}$ is
not characteristic of the spin flips 
themselves (which are local),
but of the absence of spin flips within 
a domain of size 
$2\pi/q$. These domain sizes increase 
when the spin flip
dynamics slow down on approaching $T_{N}$.

In this discussion, we use the 
phase transition
only to illustrate a possibility 
of an interpretation.
In the context outlined above, 
this possibility will remain valid 
in its general ideas,
even if there is no phase transition at stake at all:
$q$ refers to a domain size, rather than 
to a wavelength.
By analogy, we see that it does not go without
justification in 
the quasicrystal,
to associate $2\pi/q_{\parallel}$ with some 
hypothetical long wavelength 
phason wave. Moreover, if there were 
some wavelength $\lambda$
in the phason dynamics,  diffuse 
scattering should eventually build maxima
at {\em new} Bragg peak
positions at ${\mathbf{Q}}$ with  
$Q=2\pi/\lambda$,
or at satellites positions at 
${\mathbf{Q}}= {\mathbf{Q_{B} +q}}$
rather than remaining smeared out over 
a continuum of positions ${\mathbf{Q_{B} +q}}$
 in a distribution centered on 
${\mathbf{Q_{B}}}$, with
$q=2\pi/\lambda$ (if it is the signature
of the mechanism behind the transition).
As the diffuse scattering in QCs remains centered at 
the Bragg peaks of the high temperature regime
and its maxima do not define new 
Bragg or satellite positions,
the {\em ad hoc} interpretation in 
terms of critical scattering
announcing a phase transition that 
would not be 
reached due to the slowing
down of the phason dynamics, lacks proof.
Note that the slowing down of the spin flips
is what triggers the antiferromagnetic 
phase transition
rather than impeding it!\\

{\subsection{Experimental data never contain direct information about 
correlations between atomic jumps}}

We have seen above that the whole discussion
about correlations between the jumps originates in a confusion
that  has been made between
correlations between {\em atomic positions} and correlations between
{\em atomic jumps}. We will see below that this
confusion has been amplified by overinterpreting
the coherence of the signal.
The confusion manifests itself already in the apparent paradox
that $\lambda$ is not a period of the displacement field,
while $2\pi/\lambda$ occurs in the Fourier spectrum
of the disordered quasicrystal. 
Correlations between atomic jumps can just
not be directly observed in an experiment.
We would like to stress this point further on the basis
of the Van Hove formalism for coherent neutron scattering.

The Van Hove formalism says that what you measure in coherent neutron scattering
is the Fourier transform of the probability of having a particle $j$
at position $x$ at time $t$ if the same or another particle 
$k$ (whereby $k=j$ is thus allowed)
was present at position $0$ at time $0$. Hence, what we deal with in a measurement are 
correlations between positions
and not correlations between jumps.
The very same formalism applies, {\em mutatis mutandis}, for X ray intensities.\\

{\subsection{Conflicting time scales}}

We have seen that a Fourier component at a single value of
$Q_{B} + q_{\parallel}$ corresponds to a density wave for the
disordered QC, and that a set of them with the same value of
$q_{\parallel}$ would
define a displacement field that is not periodic.
The time constant for the relaxation
of this field is claimed to be of the order
of minutes, on the basis of the speckle experiments.
Let us first of all point out that it is difficult
to claim that two phason jumps separated by a long distance,
could by correlated in time at such a time scale of the order of minutes.
In the mean time, the corresponding atoms will have jumped independently
a huge number (at least of the order $10^{12}$) of times, 
as the relaxation time for individual
jumps measured by neutron scattering and M\"ossbauer spectroscopy
lies in the range from  a few picoseconds to a few nanoseconds.
What permits us to say that  phason jump number $n$ at a given position $x$
will be the one phason flip that correlates through
the ``sine wave'' with a flip, at position $x'$,  hundreds of Angstroms away, 
a few minutes earlier?
Of course it can be argued that one has to filter out the fast components.
It has been  stated\cite{Ames2}
that "In this case of the long-wavelength phason fluctuation,
individual phason flips, which are local rearrangements of underlying tiles
or atomic jumps, correlate at long distance because 
the same Fourier mode is responsible
for them".
We would like to point out that this contains a kind
of reversal of the causality that is by no means 
warranted by the information content of the data.

To point out why, it is sufficient to refer to the pristine random
tiling model Monte Carlo calculations by Tang et al.\cite{Tang,Shaw}  
This model will yield the very same
type of diffuse scattering around the Bragg peaks, as
in a model wherein the kind of correlation evoked above is stipulated.
This point was also made by Naumis et al.\cite{Naumis} (see above).
 And this happens 
despite the fact that all the jumps are totally uncorrelated by definition:
What one does in the model is to flip tiles at random
by Monte Carlo. The only experimentally detectable qualitative difference 
between the model simulated by
 Tang et al. \cite{Tang,Shaw} and Widom's model\cite{Widom} is 
 the T dependence (which is opposite). 
The model of Tang et al. \cite{Tang,Shaw} will yield similar 
time decays of the speckle,
because any wavelength that might have been built up in the 
fluctuation pattern
is entirely random and must therefore decay away again with time.
Of course, these chance Fourier components are also building 
up with time 
following the very same time constants.
They come and go.
In the model of Tang et al.,\cite{Tang,Shaw} there is no 
correlation whatsoever between the jumps,
but of course the random configurations of the system will yield the
Fourier components with their time decay. These are chance Fourier components
of random correlations. The authors invert the logical order of cause and effect 
by attributing the origin of the correlation to this Fourier component, such that
it would be the Fourier component that causes the fluctuations.
In reality the fluctuations are random, and the Fourier analysis of
these fluctuations yields the chance Fourier component.
There is not the slightest element in the experimental data
that permits to justify this kind of reversal of cause and effect
that is present in the claims of the authors.
Such a reversal might be necessary to make the 
conceptual change from the model
of Tang et al.\cite{Tang,Shaw} to a  reading of Widom's 
model\cite{Widom} in terms of tile flips (rather
than small atomic displacements),
but there is nothing in the experimental data that can be claimed to be
evidence for it.

Certainly, in the model of Tang et al.\cite{Tang,Shaw} it can occur that
locally, a tile flip only becomes possible 
after another one,
and one could build a chain of such possibilities
over a long distance. But many other tile 
flips  disrupt
this chain all the time at a very high rate, as explained above.
The origin of the coherence of the signal is not the presence of such
marginal chains.
What is overinterpreted in the data
is the fact that the coherence of the signal is not due to
correlations between atomic jumps, but to correlations
between atomic positions.
In fact, the totally uncorrelated jumps
in this model will nevertheless give rise
to strongly structured coherent signals, but
these are due to the correlations between the atomic positions.
The paradox is thus, once gain, due to a confusion
between correlations of atomic positions and correlations of jumps.

Altough it is not relevant for the data,
its is perhaps interesting to end this subsection on a remark
about random correlations between jumps.
For the atomic jumps themselves one should
not confuse the {\em constraints} of the 
random tiling
model with some correlations or 
lack of independence
in the tile flips. 
The jumps are totally uncorrelated within the 
given set of constraints dictated by the 
random tiling model.
To give an analogon: For two completely 
independent walkers
in a city where all streets run only eighter 
North-South or East-West,
like New York (without Broadway), we might 
find a mysterious correlation
in that they are found to walk always only
in mutually perpendicular or parallel directions.
This is not a mysterious correlation between the
two independent walkers, but a constraint imposed by
the city map of New York.
Hence, yes it is true that the directions of the jumps
in the random tiling model are strongly ``correlated'',
but nevertheless the jumps are totally independent by 
the very construction of the Monte Carlo simulation.

{\section{The Temperature Dependence}}

{\subsection {What it is all about}}

The diffuse scattering data cannot correspond to disorder 
generated by phason jumps.
Up to now we have analyzed the diffuse scattering in the assumption
that it would be due to tile flip disorder.
We have already pointed out alternative possibilities, e.g. chemical disorder.
Here we show that the data simply cannot correspond
to  quasicrystal disorder that would be created
by tile flips.

The quasielastic neutron scattering signal that
corresponds to tile flips has been studied in great detail.
The intensities of the quasielastic signals in these 
data show that the number of phason jumps increases when 
the temperature is raised. This is model-independent
factual information. If the
diffuse scattering observed by the authors
were to correspond to tile flip kinetics
it should thus follow the same temperature behaviour
as in the neutron data. In reality it follows 
the opposite one: the diffuse scattering diminishes when 
the temperature is raised.
This contradiction is unassailable.
Perhaps, this requires a more detailed discussion.\\

{\subsection{Coherent and incoherent scattering in many-particle problems}}

{\subsubsection{Method of Calculation}}

First of all the reader should check 
in Reference \cite{Coddens} how a coherent quasielastic signal
is calculated. Such a  coherent quasielastic signal
is the temporal Fourier transform of the speckle signal whose
intensity decays exponentially with time.
We cannot possibly reproduce the whole development
of that paper. Perhaps it suffices to say that one first
has to map out the whole set of configurations ${\cal{C}}_{\mu}$
that the system can take. Each configuration ${\cal{C}}_{\mu}$ 
can be represented
as a point in a higher-dimensional space. When an atomic jump
with relaxation time $\tau_{0}$ takes the system from configuration
${\cal{C}}_{1}$ to configuration ${\cal{C}}_{2}$, we connect
the corresponding points by a line that we label with $\tau_{0}$.
This way the whole dynamics can be mapped in the terms of a
graph in higher-dimensional space. We can then say
that the whole system is an abstract, single particle that diffuses
through first-neighbour jumps on the higher-dimensional graph.
This high-dimensional diffusion problem
can then be expressed in terms of a set of coupled
linear differential equations with a jump matrix,
just as the normal, less abstract diffusion problem of
a single particle in physical space.

We may note that not only coherent scattering,
but also incoherent scattering for a many-particle system can  be calculated 
following this method. Contrary to folk lore,
this leads to a more complicated and structured neutron signal than
coherent scattering.  In coherent scattering,
it is not relevant where the individual particles are within a configuration.
All configurations obtained by particle exchange operations
are in the same equivalence class.
All that counts in coherent scattering is the configuration. 
But, in incoherent scattering one must
keep track of a single particle.
In incoherent scattering,
two identical configurations, where an individual particle
finds itself in different positions, are therefore  no longer equivalent.
The configuration space to be considered is therefore
much larger in the case of incoherent scattering.
The reader can verify this on the calculation of the model
in reference \cite{Coddens}. 

There is still one step missing if we want to compare this 
with the speckle data, which are obtained with X-rays. 
In general the scattering formalisms for various techniques
are quite analogous. They can be just transcribed
{\em mutatis mutandis} from one probing particle
to another one. But it is only with the advent
of synchrotron sources that the study of dynamics has been
opened to X-ray studies.
In fact, X rays do not have the necessary energy resolution
(which lies at the very best in the $eV$ range)
to separate even the broadest quasielastic scattering 
(which occurs in the $\mu eV$ or $meV$ range)
from purely elastic scattering.
The diffuse scattering observed is thus an energy-integrated 
whole of all quasielastic
and elastic signals. That means that one has to
add up all quasielastic form factors from the calculation described above
in order to obtain the diffuse scattering signal.
A diffuse signal at a paricular $Q$-value will in general
arise from many contributions with different relaxation times.
 But in the case of the speckle data,
there is a time dependence, such that the separation between 
the quasielastic scattering
and the elastic scattering is obtained in the time domain.
It also appears that, at least in a first approximation,
at a given $Q$-value, one measures only one, dominant
value of  the  many relaxation times that are probably present. \\

{\subsection{Applying it to quasicrystals}}
 
If one were able to diagonalize
the huge jump matrix that describes the whole of the phason 
dynamics, along the methods of calculation descibed above, 
one would find a (very large) number 
of characteristic times,
each leading to a Lorentzian signal with a dynamical 
structure factor. Such a jump matrix exists both
for the coherent and for the incoherent scattering case.
There is a coherent and an incoherent signal even for the case of
completely independent jumps.
Such a theoretical treatment is beyond reach however,
due to the sheer size and complexity of configuration space.
The observed neutron scattering data correspond to the sum
of coherent and incoherent scattering (contrary to 
the statements of reference \cite{Francoual}
that the neutron scattering 
signals would be incoherent).
 The long-time signals attributed to
phason dynamics in reference \cite{Francoual} would just correspond to 
some of these Lorentzians,
with very long relaxation times.

These long relaxation times are not elementary
parameters of the jump model, but functions 
of more elementary
jump times. These functions 
pop up as the inverses of the eigenvalues 
$\lambda_{j} = f_{j}(\tau_{1}, \cdots \tau_{n})$ 
of the jump matrix
that has been defined in terms of the few more
elementary jump times, $\tau_{1}, \cdots \tau_{n}$, 
which are much faster. The 
Q-dependence of the intensities 
of all Lorentzians $\Lambda_{j}$ is given by
the corresponding structure factors. 
Any temperature dependence enters into the model
through the temperature dependence
of the elementary jump times in terms of 
activation energies.
Through the functional dependence 
$\lambda_{j} = f_{j}(\tau_{1}, \cdots \tau_{n})$ evoked,
the temperature dependence of all Lorentzians
is thus dictated by the temperature dependence
of the elementary jump times.
According to crude criteria such as 
increasing or decreasing 
of jump times or intensities, the long time dynamics
should thus have the same temperature dependence
as the short time dynamics, and any person who wants
to formulate a claim that they could show 
opposite behaviour
will have to work very hard to gain 
credibility for it.

But the experimental observation 
(from quasi-elastic neutron scattering)
that the intensity,
rather than the width
of the fast signals increases with temperature 
is unusual.
In a first approach, one might argue that 
it could be due to the finite energy resolution
of the neutron scattering experiments: 
At low temperatures
the dynamics are too slow to be resolved and appear
as elastic. At higher temperatures they 
become resolved
leading to the illusion that the intensity 
of the elastic peak
decreases and the intensity of the 
quasielastic signals
increases accordingly. But this is not what 
happens:
The width of the quasielastic signals 
cannot be detected 
to change with temperature,
while their intensities change drastically 
in a way that
cannot be attributed to some broadening.
One needs to introduce an assistance scanario 
to explain this
very unusual behaviour.

But what one observes in the diffuse scattering 
and has been called the effect of an inverse Debye-Waller factor,
is that the intensities (i.e. structure factors)
of the very slow signals {\em decrease} while 
the temperature is raised,
{\em with the intensities being transfered to 
the Bragg peaks!}
If there were no assistance scenario in the jump dynamics,
the intensity and its Q-dependences for a given
Lorentzian $\Lambda_{j}$ would remain 
the same at all temperatures.
In fact, the structure factor of the 
Lorentzian $\Lambda_{j}$
associated with 
$\lambda_{j}$ is not changed by
a speeding up of the dynamics 
(allowance made for the phonon
Debye-waller factor).
At the very best, its intensity at a given Q-value
would appear to be associated with a 
{\em faster} relaxation time,
through the functional relationship
$\lambda_{j} = f_{j}(\tau_{1}, \cdots \tau_{n})$.
If one wants to change $f_{j}$ or the 
structure factors, rather than just
$\tau_{1}, \cdots \tau_{n}$, one has to 
introduce special assumptions.
The observation that the quasielastic 
neutron scattering intensity
increases with temperature forced us to
introduce such an assumption in the 
form of an assistance
scenario, as even the elastic 
intensity is determined by
the jump model: It corresponds to the 
eigenvalue $0$ of the
jump matrix, and hence its intensity or 
structure factor should normally
not change with temperature. In other words:
Allowing for the effect of the Debye-Waller 
factor due to the phonons,
the ratios of the
various structure factors, including the elastic one, 
should have remained the same.
In the assistance scenario, the elastic 
intensity decreases
with temperature, because one introduces 
long-lived excited
states, whose population is governed by 
a Boltzmann factor with a large
activation energy.
What kind of most extraordinary
{\em ad hoc} assumptions 
would have to be introduced into the 
jump model in order to 
obtain a {\em decreasing} diffuse 
intensity as observed by the authors,
that could be attributed to phason jumps
despite the fact that the 
quasielastic intensity 
corresponding to fast phason 
jumps has been observed
to {\em increase} by neutron 
scattering?

To resume the situation: The number of tiles that
flip increases when the temperature is raised.
Therefore, the intensity of the signal that 
should betray the presence
of these tile flips, e.g. off-Bragg-peak 
diffuse scattering
claimed to correspond to the structural 
disorder produced by the flips,
should also increase when the temperature 
is raised.
Such a temperature dependence runs 
contrary to what the authors observed.

The authors \cite{Francoual} then decided to proceed by
adopting an ``alternative
random tiling model'', proposed by Widom, wherein
the diffuse scattering intensity decreases
when the temperature is raised.
This is certainly important for comparing different
versions of the random tiling model,
that differ in their predictions about the temperature
dependence of the data, but
it sidetracks the attention with respect
to the important issue we want to address here:
Such a change of model does not change a iota to the 
fact that the diffuse scattering
intensity cannot be attributed to 
structural disorder
produced by tile flips as it has 
the wrong temperature
behaviour. It follows that  the diffuse scattering,
which they call the ``phason fluctuations''
of the alternative random tiling model,
must be dissociated from tiling disorder,
i.e. the ``phason fluctuations'' are 
not tile flip kinetics.

This point cannot be rebutted by arguing that this would not be
contradictory because 
it is only the pristine random tiling model,
as used in the simulations of Tang et al.\cite{Tang,Shaw} that would
ask for an intensity that increases with rising temperature,
while Widom's more elaborate model\cite{Widom}
can also incorporate other temperature dependences.
That would be a totally pointless misrepresentation of the issues.
{\em We are not opposing the temperature dependence of the data to the predictions
of the random tiling model} (although it is interesting to point out
that the temperature dependence discards the pristine model 
in favour of Widom's model\cite{Widom})
{\em we are opposing the temperature dependence to the 
information given by quasielastic
neutron scattering data!}\\

{\section{What is phason elasticity on the microscopic level?}}

By formulating their claims in a 
macroscopic language
of phason elasticity, the authors\cite{Francoual} created a 
language barrier:
Nobody understands
how the macroscopic phason elasticity 
is supposed to relate to the 
microscopic-level tile flips, if it does at all. 
And therefore, nobody knows how 
the previously mentioned 
microscopic objections about the temperature dependence
should be translated across this 
language barrier.
 In Widom's paper\cite{Widom}
it is stated that on lowering the temperature
the QC moves away from the ideal 
random tiling conditions
and that this drives an elastic instability.
It is not told how this should be described 
on the microscopic level. The elastic instability
could e.g. correspond to a distortion
of the tiles rather than to their mere flips.
Widom's paper\cite{Widom} talks about an inverse 
Debye-Waller effect
on the {\em elastic} intensity.\\

{\subsection{Debye-Waller factors}}

What the authors call the inverse Debye-Waller effect
is the fact that when the temperature increases
the diffuse scattering decreases, and the intensity
is transfered to the Bragg peak, such that simultaneously the intensity
of the Bragg peak increases.

A true Debye-Waller diminishes all the elastic intensity
at the profit of the inelastic intensity. There is a sum rule between the
elastic and the inelastic intensity.
The Bragg peaks and the diffuse
scattering intensities are {\em both elastic} intensities,
such that they should be both affected the same way.
What is the solution of this apparent contradiction?

What Widom calls the Debye-Waller factor in his paper\cite{Widom}
is not the true Debye-Waller factor, but a theoretical
auxiliary quantity that consists on integrating on only
the very long-wavelength modes (i.e. on only an infinitesimal
domain of $Q$-vectors). It thus excludes the whole
phonon density of states (as e.g. measured by Suck), except
that infinitesimal part. Similarly, it excludes all the phason
dynamics  measured by TOF neutron-scattering, except an infinitesimal part.
Within this long-wavelenth approximation Widom then calculates
his auxiliary quantity that indeed applies to the transfer of
intensity between the Bragg peak which is elastic and the
diffuse scattering which becomes inelastic in this approach.
This leads to the interesting possibility
 that the true Debye-Waller factor, obtained by
including the rest of the full $Q$-range, could
again invert the tendencies described by Widom,\cite{Widom} i.e.
some Bragg peaks could show the inverse effect, while
others could show the normal effect.\\

{\subsection{Diffusion constants}}

We may  note that 
a linear relationship 
$\tau_{0} \propto 1/q^{2}$ certainly  does not prove 
that the speckle data are produced by a diffusion mechanism.
The relevant parameter for the diffuse
scattering is $1/q^{2}$, such that
the first terms of any Taylor expansion
for the underlying physics will lead
to $\tau_{0} = C_{1} + C_{2} (1/q^{2})$.
The authors
have previously considered the relationship 
$\tau_{0} = C_{1} + C_{2} (1/q^{2})$
and this resulted in a much better fit
of their data.
Such a relationship is totally unspecific.
In general, fits of the type
$\Gamma= 1/\tau_{0} = \Gamma_{0} + D q^{2}$ 
(where $\Gamma_{0} \neq 0$ is an indication
for confined motion)
are used to analyze data when we already
know that they are produced by a diffusion mechanism,
not to present the data as supplementary 
evidence that a diffusion 
mechanism would be at work.

But let us admit that the analysis of the authors
in terms of a diffusion constant is correct.
We want to point out that
the use that the authors make of Lubensky's
statement\cite{Lubensky} that ``phason modes are diffusive''
is  misleading, 
as it creates the impression
that the data analysis would contain  
proof for their interpretations, while this is not true.

E.g. Huang scattering in standard crystals is 
traditionally
described as a ``frozen phonon''. (This has nothing to do
with  phonon dynamics:
The kinetics of frozen phonons will lead to 
(very narrow) quasielastic scattering, while
dynamical phonons correspond in general to 
non-zero frequencies).
{\em The kinetics of  frozen phonons shall
be ``diffusive'' in many instances, although phonons are
not qualified as diffusive by Lubensky.}\cite{Lubensky}
That the kinetics of the diffuse scattering is
of the relaxational or even of 
the ``diffusive'' type, in the interpretation Francoual et al.
want to give to it is thus unspecific.

There is a conceptual contradiction between
wanting to describe a phenomenon as diffusive
and taking simultaneously a ``sine wave'' {\em Ansatz} as the authors do.
The ``sine wave'' makes a decay that occurs simultaneously over
the whole of space, while a diffusive phenomenon
should progressively spread out in space. Diffusive
``sine waves'' are a {\em contradictio in terminis}
(see also below).

We may finally note that time scales of the order of minutes
put the interpretation of the data beyond any possible cross-checking
with other spectroscopic techniques.
This is a dangerous situation that demands for extreme caution. 
It is preferable to have a situation
where we can follow the physics from time window to time window,
with different techniques that have overlapping time scales
if we want to link two observations in vastly different
time domains to a same phenomenon as the authors do,
by identifying diffuse scattering on the minute time scale
with phason dynamics on the pico-second time scale.\\

{\subsection{Phason elasticity}}

It might be misleading that 
theoretical physicists also use
the term elasticity for situations where 
there are no 
interactions between atoms. E.g. in the 
simulations of
Tang et al.,\cite{Tang,Shaw} the tile flips are totally independent,
but still a ``phason elasticity'' can be 
defined to refer to
an ``entropic restoring force''.
If this is not appreciated properly, it 
can lead 
to more confusion in terms of ``elastic wave'' 
pictures that do not apply.\\

{\section{Delusions about correlations}}

{\subsection{Introduction}}

{\subsubsection{Conficting paradigms}}

The ``phason wave'' is claimed to be collective and diffusive, 
to consist of correlated jumps
and to have a sine or cosine profile.
There are serious problems with the physics of this claim:

A first, and major problem is that nothing is 
done to clearly explain to the reader
what this is supposed to mean, e.g. by an 
unambiguous definition or by a detailed
description in terms of
a number of figures showing the diplacement 
field and how the diffusion 
process is supposed to proceed step by step
on a QC tiling. Instead of that,
the claim is put forward as though its contents 
would be conceptually trivial and self-evident.
But this is not at all the case, although it does not 
transpire on a first reading.
It is only when one tries to make sense of the claims, 
that one gradually
discovers the problems, and ends up being trapped 
within the frustrating task of
trying to figure out a reasonable mental representation
of what is going on. 

What the authors propose is not
an explanation but a collection 
of {\em ad hoc} assumptions, which are
conceptually contradictory
while taken from two {\em conflicting} incompatible paradigms,
viz. the phonon model and the jump diffusion model.

{\em Diffusion and jump dynamics}. In a diffusion process or in a jump model 
particles are seen 
to move backwards and forwards
stochastically. With time the positions of the diffusing 
particle become more and more
remote from the starting position. After long times,
the  distance from the initial position is 
seen to be proportional 
with the square root of the elapsed time, 
and the displacement vector
has an isotropic distribution.
The diffusing particles all move independently:
There is no such thing as coherent diffusion at 
900 K in a metallic alloy. 
{\em A priori}, the only correlation that does
exist between the diffusing particles, is that 
two of them cannot occupy a same lattice site
simultaneously.  In some cases, there can be a 
local correlation between atomic jumps,
in the sense that the jump of one particle can be 
necessary in order to enable 
the jump of another particle. We have acknowledged for 
this possibility already in our very first
papers.
The diffusion equation contains the {\em first} partial time derivative of
a site-occupation probability field.
That the  Lorentzian {\em width in energy} 
of the quasielastic signal 
is given by $\Delta \hbar\omega = D q^{2}$ 
is a result that can be derived for 
incoherent scattering in the case of
single-particle diffusion.
And then $q$ does not refer to a distance to some nearby Bragg peak. 
It is a unique single q-value to be taken with 
respect to $Q=0$.
That this result could be extrapolated to the problem of
several diffusing particles is a {\em guess}, 
that needs to be proved.
Even for the case of the diffusion of two 
particles on a periodic lattice, 
the mathematically rigorous
calculation is very difficult, because the 
``fermionic'' condition
that two particles cannot be on the same site breaks the
translational invariance that allows one to 
diagonalize the jump matrix in
the single-particle problem. 
Of course one can take the approximation that 
with only two particles the concentration
is so diluted that one
can  neglect without
any practical incidence
the probability that the particles
become so close  that one would have to take 
care of the ``fermionic'' condition.
But it is not obvious at all that the general  
phason problem on the Fibonacci chain can 
be treated within such a limit
of infinite dilution, as all particles are allowed to jump.
It must also be stressed that in the single-particle problem
there are no replicas of the intensity
around every Bragg peak.

{\em Phonons and lattice dynamics}. In a phonon 
problem there is no stochastic back and 
forwards wandering process of the
particles. The phonon propagates linearly and 
deterministically  
on a straight line in a well defined direction 
with a well determined
wave vector. 
The wave equation contains the {\em second} 
partial time derivative 
of a displacement field.
The {\em position in energy} of the inelastic signal 
is defined for a set of 
${\mathbf{Q+q}}$-values at a distance ${\mathbf{q}}$
from the Bragg peaks ${\mathbf{Q}}$, i.e. the 
spectrum contains replicas
around every Bragg peak.

We may note that in their simplest formulation 
on a periodic lattice,
both dynamical models lead to a set of
coupled partial differential equations, that 
can be expressed in matrix form,
and that (e.g. when only first-neighbour interactions
or jumps to first-neighbour sites are considered) 
these matrices  are identical
apart from a pre-factor. Therefore they lead to 
identical wavelike eigenvectors,
due to the translational invariance of the 
periodic lattice (This is just Bloch's theorem).
But this is not the end of the story as these 
eigenvectors have to be
fed into two very different formalisms.
The fact that the time derivative
is of the first order in the diffusion problem, 
while it
is of the second order in the phonon problem, 
leads to an essential difference:
In the phonon problem we end up with 
oscillating time behaviour
and propagating waves.
In the diffusion problem there are no waves:
The dynamics of the diffusion problem lead to 
relaxation with
an exponentially decaying time behaviour. 
The eigenvectors couple also to different formalisms 
for the calculation of the
$Q$-dependence of the signals. One of the important 
consequences
of this is that the phonon problem leads to a 
$Q$-dependence with
replicas of the inelastic signal at every Bragg peak, 
while the diffusion 
problem is 
{\em a priori} exepected to be 
devoid of such replicas, despite the 
fact that the eigenvectors
of the two matrix formulations are the same.

Already at this stage it transpires  that 
phononlike wave behaviour 
is not what one would call the hallmark
of diffusion. 
There is therefore no way that it could be serious
 practice to introduce a wave {\em Ansatz}
for a diffusion process, without any discussion 
or justification.
Despite these very elementary basic principles, 
the authors  cherry-pick in each of the 
two conflicting paradigms
those aspects that can be claimed to confirm 
their views, and ignore the
ones that manifestly contradict them.
The authors do not touch upon the conflicting nature of 
the two paradigms,
and shortcut 
the obligation to prove that 
such assumptions are not mutually exclusive,
by showing that they can be derived from a 
single contradiction-free model.
Of course, the isolated components of these claims, 
taken one by one, are physically sound,  such that
there is a kind of ``d\'ej\`a vu''
that surrounds them and makes
them  look plausible on 
a first superficial impression.
But the dangerous other side of the medal 
is that such a methodology can be
like assembling  griffins,  unicorns, sphinxes, centaurs, cyclopes
or flying horses from parts of existing animals.
E.g. polarization vectors  are only defined for propagating waves,
not for diffusion phenomena, and equations for atomic jumps 
are not derived from elasticity considerations.

{\em Statics vs. dynamics}. The claims also blend concepts 
into the dynamics that are 
static rather than dynamical.
The alleged dynamical phason wave is 
a ''sine wave'' that in reality has only 
meaning in a static, structural sense.
The calculations presented are calculations of a 
(component of a) static structure factor.
It is meaningless to think of dynamics
being accurately described by multiplying a static structure factor
with a time dependence in terms of an exponential decay.
A dynamical signal is not governed by an expression
 of the type $S({\mathbf{Q}}) \,e^{-t/\tau_{0}}$,
where $S({\mathbf{Q}})$ is an {\em elastic structure factor}. 
The $Q$-dependence  
of a dynamical signal follows a 
{\em dynamical structure factor}.
The fact that a polarization vector along perpendicular
space is attributed to
this  purely static description very suggestively
 creates the impression that it is a dynamical concept,
which it is not. 
The fact that the dynamical phason wave
is presented as having a polarization along perpendicular space
also clearly shows that the dynamical concept that one wants to
introduce is being thought of
as being obtained from a cut through higher-dimensional
dynamics, beit that this is easily denied,
by drawing one's attention to the static
presentation. It is very hard to spot that the latter is 
concerned with a purely
static object that
does not correspond to dynamics.
But the calculation takes exception with the 
established prescription for the
calculation of a dynamical structure factor. 
We came across the consequences hereof already in Section I,
as they reveal themselves through the fact
that the time decay of the sine wave introduces completely unphysical
correlations over arbitrarily large distances, that are alien to real 
diffusive phenomena (see  I.F.1).
In fact, the {\em Darstellung} of a static ''sine wave'' that exponentially 
fades away from the structure
has nothing to do with a dynamical wave,
neither of the propagating nor of the diffusive type.

A static structure factor is built up from weighted expressions of the 
 type ${\cal{F}}({\cal{C}}){\cal{F}}^{*}({\cal{C}})$,
 where ${\cal{C}}$ is a configuration of the system.
 A dynamical structure factor is built up from weighted expressions of the 
 type ${\cal{F}}({\cal{C}}_{2}){\cal{F}}^{*}({\cal{C}}_{1})$,
 where ${\cal{C}}_{1}$ and ${\cal{C}}_{2}$ are different (initial and final)
 configurations of the system. 
  A process wherein the initial and the final state are identical,
 as in ${\cal{F}}({\cal{C}}){\cal{F}}^{*}({\cal{C}})$, does 
 - {\em by definition} - not correspond to dynamics.
 (Of course it could be argued  
 that there could be a constant $\alpha \neq 1$
 such that ${\cal{F}}({\cal{C}}_{2}) = \alpha {\cal{F}}({\cal{C}}_{1})$,
resulting in a structure factor that is only apparently static.
Perhaps this looks as a convincing paradigm.
But  as we already discussed the thereby postulated long-distance correlations
are unphysical. How compelling the evidence is
for an interpretation in terms of such  
a paradigm will be discussed in more detail in subsubsection VI.6.).
 
 We will discuss a first series of models  of 
many-particle 
diffusion where it can be shown rigorously
that  there is no wavelike spatial
dependence of the signal. 
It can be shown that for sufficiently small amplitudes,
the sine wave proposed is incompatible with any long-range
diffusion (see below). This is reminiscent 
of the threshold in the Katz-Kalugin
scenario (which was however derived under very 
different assumptions).
It is generally believed that in the 
temperature range
of the data of the authors, the Katz-Kalugin 
percolation transition has not yet
occurred. The claim of the authors of a diffusive phason wave
is formulated without any mention
of this kind of limitation, which clearly reveals 
the misleading character 
of its presentation that looks totally devoid of
any possible problem.
Moreover, for 
any kind of deformation of the cut that is bounded by this amplitude,
the Q-dependence of the signal is smooth (see below), 
while the sine wave paradigm predicts discrete 
contributions at
positions ${\mathbf{Q+q}}$, again without any 
mention of a possible limitation.

We will also discuss a model of many-particle 
diffusion that leads
to an intensity pattern that shares four basic 
features with the
diffuse scattering, viz. (1) there is a replica 
at every Bragg peak, (2) the intensity scales with
the intensity of the Bragg peak,
(3) it falls of as $1/q^{2}$ 
where $q$ is the distance from the Bragg peak,
and (4) the relaxation has a
time decay constant that is proportional
to $1/Dq^{2}$. (It is probably not true that the 
intensity could be modeled within the 
framework of the theory
of Jari\'c and Nelsson as being purely 
due to phason elasticity only,
but this can hardly be a point as 
it is totally obvious that we are dealing
here with phason dynamics. Perhaps, we should 
learn from this that  ``phason elasticity'' 
is not the 1-1 criterium
to decide if a signal is due to phasons that it
is claimed to be).
But in this model the QC configurations that produce 
this intensity pattern do
not at all correspond to a sine wave ondulation of the 
cut, and $q$ has 
no meaning of a wavelength whatsoever. In fact, 
in this model the set 
of QC configurations that comes into play
is exactly the same one for every $q$-value! 
This shows the arbitrariness of associating
a wavelength with the quantity $q$ and of 
imposing the modulation wave picture on
the data. This picture has thrived on 
the basis of the difficulty
of the calculations that are needed to 
question its validity.\\

{\subsubsection{Problems with elasticity}}

A second kind of difficulties that are  
passed under silence
concerns the generalisation of the 
elasticity theory to superspace.
It is true that shifts of the cut can lead 
to configurations that have very similar
energies. However, there is no low-energy 
path that connects
these configurations: The energy levels in 
the two wells
of the double-well potential may indeed have  
very close values,
but they are separated by a huge potential 
wall in between them, at least
a few eV. In other words, 
as we already pointed out in the previous subsubsection, extrapolating a 
sine wave concept
from a context of thermodynamics to a concept of dynamics
is not justified. The energy scale of dynamics 
is in the meV or sub-meV
region, not in the eV region.
In fact, the neutron data suggest that e.g. a vacancy 
has to come along to assist the jump process 
(The assistance energy
has been measured to be 0.6 eV in AlCuFe). 
It is absurd to postulate the existence of a 
whole choreography of vacancies (or whatever other assisting processes)
as would be needed to produce the long-wavelength 
correlations between jumps
that are suggested by the authors.

It may be reminded at this point that elasticity only 
plays a second-stage r\^ole in the calculation of the static structure factor.
In a first stage calculating the static structure factor is 
concerned with a matter of density
 waves. These waves are parametrized through the
 six-dimensional Fourier variables $q_{j}$. 
 The diffuse scattering profile
 must to first approximation be a second-degree polynomial 
 in the variables $q_{j}/q^{2}$. To account for the symmetry
 constraints it is normal to analyse
 these intensities in terms of the symmetry-adapted second-rank tensors.
 The second-degree polynomials  
 follow from symmetry even without calling upon elasticity
 considerations. Of course, the intensity profiles will comply with
 these symmetry considerations.
 Elasticity  comes in when we want to take into account the
 proper weighting of the contributions by the Boltzmann factors,
 for which an estimate of the energies involved is needed.
 This elasticity argument is thermodynamical, not dynamical.
 Hence what the diffuse-scattering formalism calculates are not {\em dynamical waves}
 (with a dynamical structure factor), 
 but a Fourier decomposition in 
 (physically meaningless) density waves
 of {\em static} modulation {\em fields} 
 (with static structure factors), which
are being {\em thermodynamically} weighted by calling in 
 elasticity considerations to estimate the relevant energies.
 The fluctuation
of a ``wave'' (or more precisely of a Fourier component, 
see subsubsection VI.6 below) 
is not the same thing as a wave
of fluctuation:
 That the thermodynamics fluctuate (and that the
 diffuse scattering allows for a Fourier decomposition) 
 is by no means a sufficient ground
 for postulating the existence of {\em dynamical waves} with a
 dispersion relation,
 and a calculation of these fluctuations  in 
 terms of a static structure factor
 as $S({\mathbf{Q}}) \,e^{-t/\tau_{0}}$
does not correspond to an accepted formalism for a 
calculation of the dynamics. 

Obviously other mechanisms for diffuse scattering will be subject
to the same symmetry considerations. What will distinguish 
an interpretation of the data
based on such mechanisms from an elastic interpretation will
be the values of the weighting factors  of the various components.
These weighting factors are elastic constants in the elastic
interpretation, but have a different meaning in another
interpretation.
A weighting factor that 
is different from its value calculated on the basis of  
an elasticity {\em Ansatz} may indicate that
the origin of the diffuse scattering is not elastic.
It is then not appropriate to call such weighting factors
elastic constants. (If an instability is observed,
it is then also not elastic. E.g. 
if the QC needs a very fine tuning of
its chemical composition in order to be stable, 
a small off-stoichiometry
of the sample may induce a small chemical instability. 
One could pass this way through the vicinity of a  
chemical instability
within the phase diagram by varying the temperature, 
as was suggested by Gruschko. Such possibilities 
cannot be interpreted within the conceptual framework 
of a softening of the elastic constants).

The concerns about high energy barriers 
suggest that we might be in a regime of pinned
phason dynamics, where the hydrodynamical 
theory does not apply,
as discussed by Lubensky himself. 
In fact the whole present approach of the diffuse scattering
in terms of a theory of linear phason elasticity 
hinges on the assumption that
we are not in the regime where phasons are pinned.
The deep problem that this assumption is not at all proved,
is systematically being ignored and eclipsed behind a 
triumphant presentation
of the claims. 
The fact that phason jumps are observed 
and that they
require a high assistance energy is a strong indication that the QC
is still in the pinned regime. 
It was in fact mentioned as a possibility by Lubensky that 
 pinning only disappears at the melting temperature.
  If Lubensky's approach is complete, it should
 be able to explain why phasons are propagating and not diffusive in
 an incommensurately modulated structure like biphenyl
 while they are diffusive
 in a QC. In both cases the ingredients are Landau theory
 and a broken symmetry
 from a higher-dimensional description. The difference
 is of course that the atomic surfaces are not continuous in QCs,
 but this is not an ingredient that enters into the
 continuum theory.

The picture of an elastic wave with a polarisation 
along perpendicular space
{\em misleads} in that the displacements it 
really brings about are along parallel space,
not along perpendicular space.
Let us consider an atomic surface that is just 
about to induce a jump.
This atomic surface is centered on a node of 
the superspace lattice.
When a superspace phonon jiggles the nodes of 
the superspace lattice
along the direction of perpendicular space, 
one gets the impression
that the nodes just can make tiny very low-energy 
vibrations along perpendicular space
in a local harmonic potential well, while in 
reality the real motion induced
is a large jump over the barrier, that requires 
a large assistance energy.
In stead of making small displacements within a 
minimum of a potential 
well within the harmonic regime,
the jumping atom has to go over the maximum of 
the potential in
a highly non-harmonic regime!
In other words, it is not correct to treat the 
displacements of the atomic surfaces
as equivalent to displacements of the nodes of 
the same amplitude, 
although nothing transpires about this in
a calculation of the diffraction pattern of the 
QC thas is modulated by a sine wave
in superspace.
With a superspace phonon, the energies involved 
at the intersection
of the atomic surface with the cut and at the 
node of the atomic surface
are totally different. One can thus not think 
of an atomic surface
vibrating as whole with a given energy given 
by the superspace phonon,
because such a picture does not differentiate 
the huge energy differences that
come into play between different positions on 
the atomic surface. 
These energy differences also jeopardize the 
idea of a wavelength.
In fact, in a real sine wave picture 
all points where the amplitude have the same value 
are conceptually energetically equivalent,
while in the QC the corresponding local environments 
can be energetically 
totally different.

The term in the elastic tensor that is at stake is not
one that tells you how one atom that moves along perpendicular
space moves in response (along parallel space) 
to another atom that moves with a similar
polarisation along perpendicular space. The real 
motion is along parallel space.
In other words, if we want to use a picture 
of superspace  elasticity,
we must couple the atomic surfaces between 
each other uniquely along
parallel space, and the coupling (or the 
potential energy) 
will vary along the atomic surface. 
With  a sine wave representation  there 
is  no way how a derivative
of the potential energy with respect to 
the perpendicular space displacement can be defined.
The intermediate infinitesimal
displacements needed to define such a 
derivative just do not exist:
The displacement can only take two values: jump or no-jump.
A use of this type  of the concept of 
phason elasticity is therefore flawed.
A jumping atom at the maximum in the double-well potential 
is not at all within the harmonic regime that is assumed 
when one defines elastic constants.

{\subsection{Rigorous models}}

{\subsubsection{Model 1. The limit of small amplitudes}}

{\em{Conceptual problems with very small amplitudes}}.
Let us first develop a model wherein the 
amplitude of the distortion of the cut
is bounded, i.e. $(\exists a)(\forall x_{\parallel})
( u_{\perp} (x_{\parallel}) < a)$. 
It is well known that an atomic surface can be subdived
into disjoint parts, each one of which describes 
one precise local
environment. For the Fibonacci chain, the central 
part of the atomic surface
describes a local configuration $L\cdot L$. This 
part has a total length of
$W_{0} = (\tau - 1)/\sqrt(2+\tau)$, while the total 
atomic surface has a length 
$W=(\tau + 1)/\sqrt(2+\tau)$.
If we take a maximum amplitude of $(W-W_{0})/2$, 
the atoms that
are in a first-neighbour environment $L\cdot L$ 
cannot jump. This subdivides
the whole Fibonbacci chain in segments $LSL$ and 
$LSLSL$, between atoms that cannot jump,
while within the segments jumps are possible. 
Within such a segment
there can be correlations between the atomic 
jumps in the sense that e.g. in $LLSSL$, 
the atom between
the two short distances will only be able to 
jump if another jump
will have exchanged one of the two letters $S$ for an $L$.
But there can be no correlations across the 
boundaries of the segments:
In other words, the dynamics on the segments 
are mutually independent.
The absence of such correlations over longer 
distances corresponds to the
 normal standard paradigm, that cannot be just 
 ignored or questioned
arbitrarily: Taking exception
with it by postulating a possibility 
of additional correlations over long distances
must be considered as a most unusual, awkward procedure.
In a first approximation, we thus assume that we 
cannot have correlations across
a local configuration $L\cdot L$.

It is already clear from this, that with the maximum amplitude choosen
there can be no diffusion whatsoever, because all atomic motion remains
confined within the segments. We will show also that the 
quasielastic signal that corresponds to such a model can be calculated
and has a smooth $Q$-dependence, such that it cannot possibly
correspond to the sine wave interpretation.
Nonetheless, the  sine wave interpretation has 
been postulated to be meaningful and straightforward by the authors,
without any mention of a possible reserve about 
the generality of its validity, e.g. 
by warning that the amplitudes should not be taken too small.

As we will see, it is  easy to map out the configuration spaces
for the jump dynamics on the isolated segments,
and due to the independence of the segments,
it follows that the configuration space for the whole
Fibonacci chain factorizes into a product of such
segment configuration spaces.\\

{\em{Larger small amplitudes.}}
 One can relax the amplitude condition by going 
 to the $\tau^{m}$ times inflated tiling,
 with $m$ even,
and choose it such that only the atoms that are at a position
$LL$ in the inflated tiling are not allowed to 
jump (When $m$ is odd,  $L\cdot L$ of
the inflated tiling corresponds to $LS\cdot LS$, 
such that the sectioning
does not occur in an environment $L\cdot L$).
These atoms correspond to a $\tau^{m}$ times 
deflated domain on the atomic surface,
such that a larger amplitude 
$(W- (\tau-1)^{m}W_{0})/2$ for the cut distortion can be accepted.
The atomic motion will now be confined to 
$\tau^{m}$ times larger segments.
This shows that we have to go to the limit 
of an amplitude that is equal to the whole length $W/2$  
before we can have unbounded, long-range diffusion.
This raises the question: How large can the 
amplitudes become, before
we start to see second-order satellites in the 
calculation of the sine wave modulation
calculated above?\\

{\em{Calculation of the Dynamics for Isolated 
Segments LSLSL and LSL.}}
As we already pointed out, the first step in 
the calculation of a coherent quasielastic signal
is mapping out the configuration space.
For the isolated segment $LSL$ the configuration 
space consists just of
$SLL - LSL - LLS$, where the hyphens indicate 
that the configurations are ``connected''
in the sense that a jump with relaxation time 
$\tau_{0}$ permits to go (both ways) 
from one configuration to the other.

For the isolated segment $LSLSL$, the configuration space
corresponds to all permutations of the letters in 
the five-letter word $LSLSL$
that lead to real changes. We can map it onto the problem
of two ``pseudoparticles'' $S$ that diffuse on five 
pseudosites (the letters of the word
$LSLSL$), where the letters $L$ would code an 
``empty pseudosite''. 
These two pseudoparticles $S$ (1) cannot be 
simultaneously on the same site,
and (2) they cannot leapfrog. If these two 
conditions were not
necessary, the configuration space (with 
its connectivity) would be just a square lattice
of $5 \times 5$ nodes $(j_{1},j_{2}) \in 
({\mathbb{Z}} \cap [1,5])^{2}$, where
$j_{1}$ and $j_{2}$  label the respective positions of 
the two pseudoparticles. 
The jump matrix in configuration space would map then onto
the jump matrix of the diffusion problem of an abstract particle
 (the system) that diffuses
on this square lattice (the configuration space of 
our problem) between first-neighbour
positions.
The two conditions (1) and (2)
correspond to exluding the line $j_{1} = j_{2}$, 
from this configuration space,
 and removing also one of the two triangular
sections produced by this line, e.g. 
the triangular part $j_{1} > j_{2}$.

The problem of the pristine square lattice 
without fermionic 
constraint (which is thus is not sectioned by
$j_{1} = j_{2}$) can 
be solved rigorously and algebraically
by using translational invariance.
Its 25 eigenvectors ${\mathbf{v}}^{(k_{1},k_{2})}$ 
are each defined by their 25
components
 $v^{(k_{1},k_{2})}_{j_{1},j_{2}} = 
 \cos({\pi\over{10}}(k_{1}-1)(2j_{1}-1)) 
\cos ({\pi\over{10}} (k_{2}-1)(2j_{2}-1))$. 
This means that
each eigenvector has 25 components that we 
label by the compound
indices $(j_{1},j_{2})$, where $j_{1},j_{2}$ 
are integers within $[1,5]$. 
There are 25 eigenvectors
that we label  by the compound indxices $(k_{1},k_{2})$,
where $k_{1},k_{2}$ are integers within $[1,5]$.
This is possible because the eigenvectors 
are the tensor products 
of the five  five-dimensional eigenvectors 
${\mathbf{v}}^{(k)}$ for the 
single-pseudoparticle diffusion 
problem, with components given by
$v^{(k)}_{j} = \cos({\pi\over{10}}((k-1)(2j-1)) )$.
  The corresponding eigenvalues for the 
  pristine square-lattice problem are
$ - {4\over{\tau_{0}}} (\sin^{2} ({\pi\over{20}}(k_{1}-1)) + 
\sin^{2} ({\pi\over{20}}(k_{2}-1)))$, i.e. 
sums of two eigenvalues of the 
single-pseudoparticle problem.

But when the fermionic condition has to be taken into account
and the symmetry is broken
by the cut $j_{1}=j_{2}$, the simple calculation is no longer possible.
In fact, the eigenvalues have no longer any relation whatsoever
with those of the problem on the square lattice.\\

{\em{Mathematical Formalism for Larger Isolated Segments.}}
Going to higher levels of inflation $m$, with $m$ even, 
we will have $F_{n}$ letters $S$
and $F_{n+1}$ letters $L$ (for the larger segment), 
such that the problem can be mapped
onto the diffusion problem of $F_{n}$ particles on 
$F_{n+2}$ sites.
Here $n = m + 3$ (e.g. for the $\tau^{2}$ inflated 
tiling, $m=2$ and we have
 $F_{5} = 5$ particles on $F_{7} = 13$ sites for the 
 larger segment).
The fact that the particles cannot be on a same site 
and cannot leapfrog
implies that the number of configurations 
(on the larger segment) is 
$\left ( \begin{tabular}{c}
$F_{n+2}$\\
$F_{n}$\\
\end{tabular}\right )$.
On the smaller segment it is
$\left ( \begin{tabular}{c}
$F_{n+1}$\\
$F_{n-1}$\\
\end{tabular}\right )$.
E.g. when $m=0$, we have $n=3$ for the segments 
$LSLSL$, with
$\left ( \begin{tabular}{c}
$F_{5}$\\
$F_{3}$\\
\end{tabular}\right ) = \left ( \begin{tabular}{c}
$5$\\
$2$\\
\end{tabular}\right ) = 10$ configurations, while 
for the segments $LSL$
we have 
$\left ( \begin{tabular}{c}
$F_{4}$\\
$F_{2}$\\
\end{tabular}\right ) = \left ( \begin{tabular}{c}
$3$\\
$1$\\
\end{tabular}\right ) = 3$ configurations.

The $F_{n}$ particles will have coordinates 
$(j_{1}, j_{2} \cdots j_{F_{n}}) \in
([1,F_{n+2}]\cap {\mathbb{Z}})^{F_{n}}$, which is a 
$F_{n}$-dimensional lattice bounded by a
hypercube of edge length $F_{n+2} - 1$. The condition 
that the particles
cannot be on the same sites defines now forbidden
hyperplanes $j_{1} = j_{2}, j_{2} = j_{3}, \cdots 
j_{F_{n}-1} = j_{F_{n}}$,
which are sectioning the hypercubic lattice. Again, 
the diffusion problem on 
the hypercubic lattice
could be easily solved in general by using translational invariance,
but, again, the  cuts by the hyperplanes break the symmetry and
spoil any possibility of a general algebraic solution.

In any case, generally spoken, one will have to allow
for segmentation lengths that are quite large before
one will obtain very small non-zero eigenvalues.
Finally, it may be noted it is not impossible that the
condition $u_{\perp}(x_{\perp}) \le (W- (\tau-1)^{m}W_{0})/2$,
might remove more configurations from the hypercubic lattice
than we described in terms of its sectioning with the hyperplanes
$j_{1} = j_{2}, j_{2} = j_{3}, \cdots 
j_{F_{n}-1} = j_{F_{n}}$ above. But this is not important
for the basic principle of the arguments we will develop below.
The sizes of the configuration spaces we have considered above are
then  upper limits for the sizes we can expect. The numbers $n_{\nu}$ below have
then to be replaced by smaller numbers.\\

{\em{Mathematical Formalism for a whole chain of 
isolated segments.}}
The configuration space for the whole Fibonacci chain
is a Cartesian product of the configuration spaces for
segments of $F_{n}$ letters  $S$ and $F_{n+1}$ 
letters $L$. The eigenvalues
for the whole chain are the sums of the eigenvalues 
for each segment,
and the eigenvectors are the tensor products of the 
eigenvectors of the segments.
The fact that we cannot give the eigenvectors 
and eigenvalues
for the problems on the isolated segments explicitly, 
does not impede one to
develop the calculation further and to
reach general conclusions for the dynamics of 
the whole Fibonacci 
chain based on such segments. 
The further calculation is analogous to the
one in \cite{myPRB}, which shows clearly that the
coherent quasielastic signal has a smooth 
$Q$-dependence.
Most importantly, the signal is not composed 
of more or less similar 
replicas around every Bragg, i.e. it does not show the  
salient hallmarks of a convolution product 
that are present in the data.
In fact, the eigenvalues that are sums of eigenvalues 
of several single-segment
problems remain silent.

We give here a shortened account of the calculation. 
The reader can compare it 
with the calculation in \cite{myPRB}. This should
in principle permit him to understand the whole 
argument, even if at first sight
it might look intimidating.
Let us note a configuration on the Fibonacci chain as 
${\cal{C}}_{\{ j_{\nu} \}} =
{\cal{C}}_{\cdots j_{1} j_{2} \cdots j_{\nu} \cdots}$,
which means that in segment $\nu \in {\mathbb{Z}}$  
the configuration is $c_{j_{\nu}}$.
As pointed out above, ${j_{\nu}}$ can take here any 
value from $1$ to $n_{\nu} = \left ( \begin{tabular}{c}
$F_{n+2}$\\
$F_{n}$\\
\end{tabular}\right )$, for the larger segment,
and from $1$ to $n_{\nu} = \left ( \begin{tabular}{c}
$F_{n+1}$\\
$F_{n-1}$\\
\end{tabular}\right )$, for the smaller segment. 
Here $n -3 = m$ is 
the (necessarily even) inflation
level.

The Fourier transform of the  configuration 
on the Fibonacci chain is:

\begin{equation}
{\cal{F}}({\cal{C}}_{\{ j_{\nu} \}}) = 
\sum_{\nu \in {\mathbb{Z}}}\,
{\tilde{{\cal{F}}}}(c_{j_{\nu}}) e^{\imath Q x_{\nu}},
\end{equation}

\noindent where $x_{\nu}$ is the left boundary of the
segment $\nu$, and ${\tilde{{\cal{F}}}}(c_{j_{\nu}})$ 
is the Fourier transform of the 
configuration on the segment calculated in relative 
coordinates $x-x_{\nu}$.
The eigenvectors ${\mathbf{v}}^{ \{ k_{\nu} \}} = 
{\mathbf{v}}^{(\cdots k_{1} k_{2}\cdots k_{\nu} \cdots)}$ 
are tensor products

\begin{equation}
{\mathbf{v}}^{ \{ k_{\nu} \}} = \bigotimes_{\nu \in 
{\mathbb{Z}}}\, {\mathbf{v}}^{(k_{\nu})},
\end{equation}

\noindent such that their components are of the form
$v^{ \{ k_{\nu} \} }_{ \{ j_{\nu} \} }$ $=$
$v^{(\cdots k_{1} k_{2}\cdots k_{\nu} \cdots)}_{\cdots j_{1} j_{2} 
\cdots j_{\nu} \cdots}$
$=$ $\cdots v^{(k_{1})}_{j_{1}} v^{(k_{2})}_{j_{2}} \cdots 
v^{(k_{\nu})}_{j_{\nu}} \cdots$.
Note that $k_{\nu}$ are here just labels for the eigenvectors, 
and that the eigenvectors
are not at all assumed to be of the translationally invariant type
$v^{(k_{\nu})}_{j_{\nu}} = 
\cos({\pi\over{2N_{\nu}}}((k_{\nu}-1)(2j_{\nu}-1)) )$, where
$N_{\nu}$ would take the value $F_{n+1}$ or $F_{n}$ depending 
on the type of segment.
However it is assumed that by definition the label 
$k_{\nu} = 1$ corresponds to the eigenvalue $0$
and that the corresponding eigenvector is then 
$[1,1,\cdots 1]^{\top}$, which will e.g. be the case
if all jump times in the dynamics on a single segment 
are the same. (Actually it is sufficient
that all jumps take place in symmetrical double-well potentials).
We must calculate 

\begin{equation}
\sum_{\{ j_{\nu} \}}\, {\cal{F}}({\cal{C}}_{\{ j_{\nu} \}})\,
v^{ \{ k_{\nu} \} }_{ \{ j_{\nu} \} } =
\sum_{\{ j_{\nu} \}}\,(\sum_{w \in {\mathbb{Z}}}\,
{\tilde{{\cal{F}}}}(c_{j_{w}}) e^{\imath Q x_{w}}\,)\,
(\,\prod_{\nu} \,v^{(k_{\nu})}_{j_{\nu}}\,).
\end{equation}

\noindent This is a sum over $w$. The term $w$ is given by

\begin{equation}
\prod_{\nu \neq w}\, (\sum_{j_{\nu}}\,v^{(k_{\nu})}_{j_{\nu}})\,
(\sum_{j_{w}}\,{\tilde{{\cal{F}}}}(c_{j_{w}}) \, 
e^{\imath Q x_{w}} \, v^{(k_{w})}_{j_{w}}\,)
= (\,\prod_{\nu \neq w}\, \sqrt{n_{\nu}}\, \delta_{k_{\nu}\,1}\,)\,
(\sum_{j_{w}}\,{\tilde{{\cal{F}}}}(c_{j_{w}})\, 
e^{\imath Q x_{w}}\, v^{(k_{w})}_{j_{w}} \,).
\end{equation}

\noindent When $k_{\nu} = 1$ on segment $\nu$, the eigenvalue is $0$
and the normalized eigenvector has $n_{l}$ identical components 
$1/\sqrt{ n_{l}}$ on the segment, 
which explains
the prefactor $\sqrt{n_{\nu}}$ of $\delta_{k_{\nu}\,1}$. 
We do not have to care too
much about these prefactors, since after squaring amplitudes, 
and weigthing all
configurations to the same initial probability, they will 
disappear from the final result.

The sum over $w$ is only non-zero in two cases. The first case is
that $(\forall \nu \in {\mathbb{Z}}) (k_{\nu} = 1)$, 
which leads to the eigenvalue
$\lambda = 0$, with the corresponding eigenvector whose components
are all equal to $1/\prod_{\nu}\,\sqrt{n_{\nu}}$. 
Leaving apart this normalization factor, as it will 
eventually disappear anyway,
we obtain  the result:

\begin{equation}
\sum_{w} \sum_{j_{w}}\,{\tilde{{\cal{F}}}}(c_{j_{w}}) 
\,e^{\imath Q x_{w}},
\end{equation}

\noindent which is just the sum of the Fourier 
transforms of all random tiling configurations.
The second case is that $(\exists ! w\in 
{\mathbb{Z}}) (k_{w} \neq 1) \& 
(\forall \nu\neq w ) (k_{\nu} = 1)$. Then the 
eigenvalue will be $\lambda_{w}$
and the result will reduce to just one term:

\begin{equation}
\sum_{j_{w}}\,{\tilde{{\cal{F}}}}(c_{j_{w}}) \,
e^{\imath Q x_{w}}\,v^{(k_{w})}_{j_{w}}.
\end{equation}

Apart from the phase factor, $e^{\imath Q x_{w}}$, 
this is exactly the expression 
for the analogous dynamical problem for just the isolated segment $w$.
Carrying the calculation through until the end we 
must take the squares of the amplitudes.
We get then an elastic term (a constant in the time domain,
or $\delta(\omega)$ in the energy domain), that is 
weighted with the (averaged) diffraction intensity
of the random tiling model, and 
$\left ( \begin{tabular}{c}
$F_{n+2}$\\
$F_{n}$\\
\end{tabular}\right ) + \left ( \begin{tabular}{c}
$F_{n+1}$\\
$F_{n-1}$\\
\end{tabular}\right ) - 2$
quasielastic terms $e^{-\lambda_{w}t}$ (in the time 
domain) that are weighted with 
structure factors that have exactly the same $Q$-dependence as 
in the dynamical jump models for the small and the 
large single segments.
We can understand the result also in an intuitive way, 
arguing that
the segment dynamics are independent.
\\

{\em{Postulating correlations by brute force?}}
When $a < W/2$ one can of course argue, that the 
atomic positions in the various 
segments can be correlated
anyway in a sine wave fashion, but (1) this contradicts the accepted
paradigm of the {\em stochastic} nature of atomic jumps in alloys,
(2) reverses the charge of proof and (3) cannot cover up the fact 
that the authors completely missed this point.
The exceptional character of such a claim should at least
have been pointed out explicitly instead as presenting
it as a trivial and self-evident.

We may note that with large isolated fragments, we allow already for 
large distances of correlation, and that we have shown for these
cases that the signal lacks the hallmarks of a convolution product
in the form of replicas around each Bragg peak.
One could argue also that the correlations aimed at 
are of a different nature
than the ones we have explored. But this can only show how
it even has not been defined what the as self-evident 
presented correlations
are supposed to be. In our opinion postulating 
correlations of another 
type than we have allowed for 
(1) must be clearly defined, (2) properly justified 
to be physically
plausible, and (3) it must be proved  by a model calculation
that it leads indeed to the type of signal observed.\\
%
%

{\subsubsection{Model 2. The insolvable limit of large amplitudes 
and configuration space percolation}}

When $a > W/2$, the configuration space is no longer factorized.
We are then entitled to hope that small eigenvalues 
will enter the game.
But we may note that the relative proportions of letters $S$ and $L$
render it very doubtful that the eigenvectors and eigenvalues
could be reasonably approximated by the solution 
that one could propose for
a small number $m$ of particles on an infinite lattice,
viz. by considering that it is to a high degree 
of accuracy
the tensor product of $m$ eigenvectors for the 
single-particle problem.
For these reasons the {\em Ansatz} of the authors 
that the eigenvalues
would be of the type that leads within 
the energy domain to a Lorentian signal 
with a width $D Q^{2}$ lacks a proper justification.
In any case, this model case is of an 
unassailable difficulty.
The same can be said about models for two- and 
three-dimensional quasicrystals.


One can consider the Fibonacci chain as the
limit of a finite diffusion problem, whose size tends to infinity.
In the finite problem there will be $\beta$ letters $S$,
and $\xi - \beta$ letters $L$. The phason jumps dynamics can
then be formulated as the jump diffusion problem of
$\beta$ particles (the letters $S$) on $\xi$ sites 
(the total number of letters), as we already did above for models 1 and 2. 
(Alternatively, it can be formulated
as the jump diffusion problem of $\xi - \beta$ particles
(the letters $L$) on $\xi$ sites. The solutions will be the same). All 
one-dimensional diffusion problems with $\beta$ diffusing 
particles on $\xi$ sites (where $\xi \ge \beta$)
can be handled by using the irreducible representations 
of the permutation group $S_{\xi}$: 
Each of the $\xi -1$ generators 
{\small{$\left ( \begin{tabular}{cc}
$j$   & $j+1$\\
$j+1$ & $j$\\
\end{tabular}\right )$}}, (with $j \in [1,\xi-1]\cap{\mathbb{N}}$),
of the symmetric group $S_{\xi}$ interchanges 
the occupation of two adjacent sites $j$ and $j+1$.
Some of these interchanges lead to real meaningful physical changes, 
other ones do not,
because they lead to physically identical configurations.
All identical configurations build an equivalence class.
The question which permutations lead to identical configurations
decides in a further stage which ones of the ${\xi}!$ eigenvectors that
one can derive from the permutation group, remain silent in the 
final signal.
In fact, it can be shown that in order to obtain the 
solution of the physical problem 
it suffices to solve the problem over the permutation group, (i.e.
by considering all permutations to be different, despite the fact that some
of them correspond to configurations that are physically equivalent), 
and to sum afterwards over all configurations that are physically equivalent.
(The corresponding summing over the eigenvectors will in certain cases
lead to zero vectors, such that the result is no longer an eigenvector,
and the original eigenvalue ceases to be an eigenvalue).
With $\beta$ particles on $\xi$ sites we will have to sum over the
$\beta !$ permutations of identical particles and over the
$(\xi - \beta)!$ permutations of empty sites. In other words, 
each equivalence class
has cardinal $\beta ! (\xi -\beta)!$. (The equivalence of the
solutions for $\beta$ or $\xi - \beta$ diffusing 
particles mentioned above
 is due to the fact 
that they result in identical sums).
One of the steps in this proof consists in showing
that by summing the differential equations over these 
$\beta ! (\xi -\beta)!$
permutations we obtain the correct differential equations for the
equivalence classes.
The irreducible representations of the symmetric group $S_{\xi}$
are calculated on the basis of Young tableaus. 
For small values of $\xi$,
this embedding of the problem within the permutation symmetry 
group allows one 
to tackle the jump diffusion problems
on a case by case and representation by representation basis.
(Even in the simple case of the diffusion of a single particle,
the solutions are scattered over the various irreducible representations).
But the possible dimensions of the irreducible representation matrices 
grow with $\xi$, and it is not obvious to see relationships
that would permit us to link the calculations for $\xi$ with 
those for $\xi +1$. In other words,
it is hard to see  how one
could derive from this formalism 
a scheme that would permit us to find nice, general algebraic expressions for
the eigenvalues and eigenvectors in terms of $\beta$ and $\xi$
in a way that would be analogous to the expressions  
we were able to derive above (in the discussion of model 1)
in the presence of translational invariance, with its uniquely
one-dimensional irreducible representations. 
This hypothetical scheme 
would then have to serve as a platform for taking the limit
of the results when $\beta \rightarrow \infty $ and $\xi \rightarrow \infty $,
keeping the ratio $\beta/\xi$ constant.
The fact that the symmetry group that underlies the jump models
is the permutation group, and that the irreducible representations of
the permutation groups can reach arbitrarily high dimensions
thwarts the aspirations of finding simple, universal 
expressions for solutions of model 2.
\\

{\subsubsection{Model 3. Diffusing tiles}}

{\em{Introduction.}} Let us assume that on the Fibonacci 
chain a letter $S_{1}$ is allowed to diffuse, 
by processes $LS \leftrightarrow SL$
until it becomes the neighbour 
another letter $S_{2}$, in a situation $LS_{1}S_{2}L$. 
Then only two
possibilities are possible, {\em viz.}  backwards 
to $S_{1}LS_{2}L$
or forwards to $LS_{1}LS_{2}$. In the latter case we 
assume that $S_{2}$ takes over the active
r\^ole of $S_{1}$. Locally, the configuration space 
takes the topology of a linear chain:
$
\Leftrightarrow
\cdots SL.LSLSL.LS \cdots \Leftrightarrow
\cdots LS.LSLSL.LS \cdots \Leftrightarrow
\cdots L.LSSLSL.LS \cdots \Leftrightarrow
\cdots L.LSLSSL.LS \cdots \Leftrightarrow
\cdots L.LSLSLS.LS \cdots \Leftrightarrow
\cdots L.LSLSL.LSS \cdots \Leftrightarrow$. 
Moreover we see that
when we go from the left to the right in this sequence, the
Fibonacci chain is step by step translated to the left, 
as the segments 
are recovered (with a shift of one letter to the left), 
after the passage of
the active letter(s) $S$. We see that this is a 
many-particle jump model.
The model is diffusive. The single particles do not 
travel over long distances.
It is rather a pseudo-particle, the active letter $S$, 
that undergoes long-range diffusion.
From the viewpoint of this pseudo-particle, the model has all 
the characteristics of a single-particle model. 
If we allow only one tile or one letter $S$  to diffuse this way,
the model is perhaps not very realistic, but
 (1) it is the only case that is simple enough to 
 permit a rigorous calculation
and (2) it leads to remarkable results for the coherent 
scattering signals, from which important
lessons can be learned.
We can split this model into two submodels according 
to the philosophy of approach. 

{\em{Philosophy 1}}. In this philosophy we may assume that 
the error that initiated the motion of the active 
letter $S$ is at $-\infty$ , such that
there is no trace of the beginning of the process.
The sequence $
\Leftrightarrow
\cdots SL.LSLSL.LS \cdots \Leftrightarrow
\cdots LS.LSLSL.LS \cdots \Leftrightarrow
\cdots L.LSSLSL.LS \cdots \Leftrightarrow
\cdots L.LSLSSL.LS \cdots \Leftrightarrow
\cdots L.LSLSLS.LS \cdots \Leftrightarrow
\cdots L.LSLSL.LSS \cdots \Leftrightarrow$ mentioned above
is then unbounded, both to the left and to the right,
and the topology of configuration space is identical
to that of a linear chain.
The configurations ${\cal{C}}_{j}$ are  of the 
type $c_{1} S c_{2}$, 
where $c_{1} c_{2}$, is the full Fibonacci 
sequence, and $S$ is the active
letter $S$. 
The jump model is defined by
$(\forall j \in {\mathbb{Z}}) 
({\partial p_{j}\over{\partial t}}= 
{1\over{\tau_{0}}}(p_{j-1} - 2 p_{j} + p_{j+1})$
(where $p_{j}$ denotes the probability that
the QC is in the configuration ${\cal{C}}_{j}$).
The eigenvectors and eigenvalues of the ``infinite jump matrix''
are therefore straightforward
to calculate, and the Fourier transforms of the configurations also.
The eigenvectors ${\mathbf{v}}^{(q)}$ have components 
$(\forall j \in {\mathbb{Z}})
 (v_{j} = e^{\imath q (j-1)})$, where $q\in {\mathbb{R}}$. 
 The corresponding eigenvalues
are  $\lambda_{q} = - {4\over{\tau_{0}}} \sin^{2} {q\over{2}}$.
But as the assumption that there is no trace
of the beginning of the process is artificial, we will 
rather calculate the result for the
following model, that contains all the difficulties
on a larger scale.

{\em{Philosophy 2}}.
In this philosophy we  assume that 
we start from a perfect Fibonacci chain,
and initiate the dynamics with a first jump of a 
letter $S$, at some point $P$ after which we allow 
this letter $S$ to diffuse further. This initial
jump leaves a scar behind in the Fibonacci 
chain that permits to spot 
the point $P$ where the whole process started, 
e.g. by creating a local sequence $LLL$
that will not occur anywhere else in the chain. 
When the active letter
accidently gets back to the point $P$, this 
scar may be removed,
undoing all traces of the previous history. We 
recover the perfect Fibonacci chain and allow the process to restart
from a completely different place.
The configurations ${\cal{C}}$ are  of the 
type $c_{1} c_{2} S c_{3}$, 
where $c_{1} S c_{2} c_{3} $, is the full Fibonacci 
sequence, and $S$ is the active
letter $S$. 
In this case the topology of configuration space is  starlike.
A central point $O$ that corresponds to the 
perfect Fibonacci lattice is the origin of an infinity
of half linear chains. Each half linear chain is 
defined by an aggregate of two parameters,
viz. the starting point
and the direction (left or right) of diffusion of 
the active letter $S$.
We label the direction by $\epsilon \in \{ -1, 1 \} $, and the 
starting point by $m\in {\mathbb{Z}}$.
Let us call the perfect lattice the configuration ${\cal{C}}_{0}$.
All other configurations can then be labeled as ${\cal{C}}_{m,\epsilon,j}$,
where $(m,\epsilon)$ labels the half linear chain, and $j$ the 
position on the half linear chain
starting from ${\cal{C}}_{0}$.  
The Fourier transforms of all the intervening configurations on the 
Fibonacci chain are still easy to calculate. But the starlike 
topology of configuration space
implies that the infinite jump matrix
can no longer be diagonalized by using translational invariance. 
 In fact, the eigenvalue equations for the 
 probabilities $p_{m,\epsilon,j}$
 that the system is in the configuration 
 ${\cal{C}}_{m,\epsilon,j}$ are:\\
 
\begin{equation}
(\forall j \in {\mathbb{N}}\parallel j > 1)\,
(\forall m \in {\mathbb{Z}})\,(\forall \epsilon \in \{-1,1\})
  \,(\lambda p_{m,\epsilon,j} =
{1\over{\tau_{0}}}(p_{m,\epsilon,j-1} - 2 p_{m,\epsilon,j} + p_{m,\epsilon,j+1})),
\end{equation}

\begin{equation}
(\forall m \in {\mathbb{Z}})\,(\forall \epsilon \in \{-1,1\})
  \,(\lambda p_{m,\epsilon,1} =
{1\over{\tau_{0}}}(p_{0} - 2 p_{m,\epsilon,1} + p_{m,\epsilon,2})),
\end{equation}

\begin{equation}{\label{p0}}
\lambda p_{0} = \sum_{m \in {\mathbb{Z}}} \,\sum_{\epsilon \in \{-1,1\}}
\,(p_{m,\epsilon,1} - p_{0}).
\end{equation}

{\em Eigenvalues and eigenvectors of the model}. The diagonalization can  
be done rigorously by using the transfer
 matrix method on each of the half linear chains. 
 In fact, the first equation 
 that defines the eigenvalue problem of the jump matrix can be solved for
$p_{m,\epsilon,j+1}$ in terms of $p_{m,\epsilon,j}$ and 
$p_{m,\epsilon,j-1}$, such that in the end,
using this as a recurrence relation,
all $p_{m,\epsilon,j+1}$ can be expressed in terms of 
$p_{m,\epsilon,1}$ and $p_{0}$.
The recurrence equation is $p_{m,\epsilon,j+1} = (\lambda\tau_{0} + 2) 
p_{m,\epsilon,j} - p_{m,\epsilon,j-1}$.
The required expression is derived by solving the 
characteristic equation $\mu^{2} = (\lambda\tau_{0} + 2) \mu - 1$.
This equation has a double root ($\mu_{1}=\mu_{2}=1$) 
when $\lambda\tau_{0} = 0$.
The solution is then of the form $p_{m,\epsilon,j} = 
K_{m,\epsilon,1} + j K_{m,\epsilon,2}$. This remains only finite when
$j \rightarrow \infty$ when $K_{m,\epsilon,2}=0$, such that $(\forall j 
\in {\mathbb{N}}) (p_{m,\epsilon,j}= K_{m,\epsilon,1})$,
from which it follows that also $p_{0} = K_{m,\epsilon,1}$. 
Hence the unique eigenvector that corresponds
to $\lambda = 0$ takes the same constant value in all its components.
When the equation has two distinct roots, the solution is of the form
$p_{m,\epsilon,j} = K_{m,\epsilon,1}\mu_{1}^{j}+K_{m,\epsilon,2}\mu_{2}^{j}$,
 where $K_{m,\epsilon,1}$ and $K_{m,\epsilon,2}$ are determined
by $p_{0}$ and $p_{m,\epsilon,1}$. 
This leads to

\begin{equation}
p_{m,\epsilon,j} = p_{m,\epsilon,1} {\frac{\mu_{1}^{j} - 
\mu_{2}^{j}}{\mu_{1} - \mu_{2}}}
            - p_{0} {\frac{\mu_{1}^{j-1} -
             \mu_{2}^{j-1}}{\mu_{1} - \mu_{2}}}.
\end{equation}

\noindent As $\mu_{1}\mu_{2} = 1$, the solution will only remain finite 
when $j \rightarrow \infty$ if $| \mu_{1} | =  
| \mu_{2} | = 1$. By putting 
$\mu_{1} = e^{\imath q}$, with $q \in ]0,\pi[$ we obtain 
$\lambda\tau_{0} = -4 \sin^{2} {q\over{2}}$, and thus:

\begin{equation}{\label{eigen}}
p_{m,\epsilon,j} = p_{m,\epsilon,1} {\frac{\sin jq}{\sin q}} - p_{0} 
{\frac{\sin (j-1) q}{\sin q}}.
\end{equation}

\noindent The wole difficulty resumes then to finding 
the set of values $p_{0}, p_{m,\epsilon,1}$, satisfying
Eq. (\ref{p0}),
that defines a complete orthogonal set of eigenvectors 
for each value of $\lambda \ne 0$.
Note that we must take a maximal number of mutually 
independent solutions of Eq. (\ref{p0}),
and that these solutions must simultaneously lead 
to linear independent eigenvectors.
We will first act as though the number of quantities 
$p_{m,\epsilon,1}$ were a finite number $2N$, and then
generalize towards an infinite vector space. 
In other words, the star with an infinite number of half chains is taken 
as the limit of an $2N$-branched star, when
$N \rightarrow \infty$.
We will use a new type of labeling.
For $\epsilon=1$ we will label $(m,\epsilon)$ by $m$, and for
$\epsilon=-1$ we will note $(m,\epsilon)$ by $N+m$.
For convenience we will use an intermediary basis for the vectors
${\mathbf{p}} = (p_{0},p_{1,1},p_{2,1},\cdots,p_{m,1},
\cdots,p_{2N,1})$. 
Let us note the canonical basis as 
${\mathbf{g}}_{0}, {\mathbf{g}}_{1}, \cdots , 
{\mathbf{g}}_{m}, {\mathbf{g}}_{2N}.$

We define this intermediary basis as follows:

\begin{tabular}{lll}
${\mathbf{e}}_{0}$ & $=$ & $(1,0,0,0,\cdots,0)$\\
${\mathbf{e}}_{1}$ & $=$ & $({\mathbf{g}}_{1} - {\mathbf{g}}_{N+1})/\sqrt{2}$\\
${\mathbf{e}}_{2}$ & $=$ & $({\mathbf{g}}_{2} - {\mathbf{g}}_{N+2})/\sqrt{2}$\\
$\vdots$ & &\\
${\mathbf{e}}_{m}$ & $=$ & $({\mathbf{g}}_{m} - {\mathbf{g}}_{N+m})/\sqrt{2}$\\
$\vdots$ & &\\
${\mathbf{e}}_{N}$ & $=$ & $({\mathbf{g}}_{N} - {\mathbf{g}}_{2N})/\sqrt{2}$\\
${\mathbf{e}}_{N+1}$ & $=$ & $(0,1,1,1,\cdots,1)/\sqrt{2N} $\\
${\mathbf{e}}_{N+2}$ & $=$ & $(0,1,e^{\imath \psi},e^{\imath 2\psi},\cdots,
e^{\imath (r-1)\psi},\cdots, e^{\imath (N-1)\psi},
                                 1,e^{\imath \psi},e^{\imath 2\psi},\cdots,
                                 e^{\imath (r-1)\psi},\cdots, 
                                 e^{\imath (N-1)\psi})/\sqrt{2N} $\\
                     & $=$ & $ \sum_{r = 1}^{N}\, e^{\imath (r-1) \psi } 
                     \,({\mathbf{g}}_{r} + {\mathbf{g}}_{N+r})/\sqrt{2N}$\\                                 
$\vdots$ & &\\
${\mathbf{e}}_{N+m}$ & $=$ & $ \sum_{r = 1}^{N}\,e^{\imath (r-1) (m-1) \psi}\, 
                       ({\mathbf{g}}_{r} + {\mathbf{g}}_{N+r})/\sqrt{2N} $\\
$\vdots$ & &\\
${\mathbf{e}}_{2N}$  & $=$ & $ \sum_{r = 1}^{N}\,e^{\imath (r-1) (N-1) \psi}\, 
                     ({\mathbf{g}}_{r} + {\mathbf{g}}_{N+r})/\sqrt{2N} $\\
\end{tabular}

\noindent This is an orthonormal basis (We must 
take $\psi = 2\pi/N$ 
in the finite case). 
Let us call the components of ${\mathbf{p}}$
in this basis $\gamma_{m}$, such that ${\mathbf{p}} = \sum_{m}\, 
\gamma_{m} {\mathbf{e}}_{m}$.
A first eigenvector of our problem corresponds to 
$\lambda = 0$, 
and has the form
$(1,1,1,\cdots 1, 1) = {\mathbf{e}}_{0} + \sqrt{2N}\, 
{\mathbf{e}}_{N+1} $.

Assume first that $p_{0} = 0$. Then $\gamma_{0} = 
{\mathbf{p\cdot e}}_{0} 
= p_{0} = 0$.
Eq. (\ref{p0}) becomes then $0 = \sum_{m \ne 0}\,
\sum_{\epsilon \in \{-1,1\}} 
\,p_{m,\epsilon,1} = {\mathbf{p\cdot e}}_{N+1}\,
\sqrt{2N} = \gamma_{N+1}$.
The fact that $\gamma_{0} = \gamma_{N+1} = 0$ implies that 
the vectors ${\mathbf{p}}$ that satisfy
Eq. (\ref{p0}) all belong to the subspace
spanned by the vectors ${\mathbf{e}}_{1}$, 
${\mathbf{e}}_{2}$, ... ${\mathbf{e}}_{N}$, 
${\mathbf{e}}_{N+2}$, ${\mathbf{e}}_{N+3}$, ... 
${\mathbf{e}}_{2N}$.
In fact, each of these basis vectors alone satisfies 
the condition Eq. (\ref{p0}).
Note that the vectors ${\mathbf{p}}$ are not 
eigenvectors of our problem, but serve to 
define the eigenvectors,
by fixing the values of $p_{0}$ and $p_{m,\epsilon,1}$ 
that are needed to calculate all other components 
$p_{m,\epsilon,j}$ of an eigenvector
by using Eq. (\ref{eigen}). 

For a given non-zero eigenvalue $\lambda$ we have this 
way already found $2N-2$ linearly independent, and 
mutually orthogonal eigenvectors
with the choice $p_{0} = 0$. They are of the form 
${\mathbf{e}}_{m} \otimes 
(\sin q, \sin 2q, \cdots, \sin jq, \dots)$.
Let us now check what additional eigenvectors we can find
by supposing that $p_{0} \ne 0$. Eq. (\ref{p0}) becomes then
$(\sum_{m} \sum_{\epsilon \in \{-1,1\}}\, 
(p_{m,\epsilon,1} - p_{0}) ) - \lambda p_{0} = 0$. 
Now $p_{m,\epsilon,1} - p_{0}$ are the components
of the vector ${\mathbf{p}} - p_{0}\, 
{\mathbf{e}}_{0} - \sqrt{2N}\, p_{0}\, {\mathbf{e}}_{N+1}$.
Let us note this quantity as $\sum_{m}\, 
\gamma_{m} {\mathbf{e}}_{m}$. 
Then $\lambda p_{0} = \sum_{m,\epsilon} (p_{m,\epsilon,1} 
- p_{0}) = (1,1,1\cdots ){\mathbf{\cdot}}
({\mathbf{p}} - p_{0}\, {\mathbf{e}}_{0} - \sqrt{2N}\, 
p_{0}\, {\mathbf{e}}_{N+1}) =
({\mathbf{e}}_{0} + \sqrt{2N}\, {\mathbf{e}}_{N+1}) 
{\mathbf{\cdot}}({\mathbf{p}} - p_{0} 
{\mathbf{e}}_{0} - \sqrt{2N} p_{0} {\mathbf{e}}_{N+1}) = 
\gamma_{0} + \sqrt{2N} \gamma_{N+1}$.
Now ${\mathbf{e}}_{0}{\mathbf{\cdot p}} = p_{0}$, 
such that 
$\gamma_{0} = {\mathbf{e}}_{0} 
{\mathbf{\cdot}}({\mathbf{p}} - p_{0} \,
{\mathbf{e}}_{0} - \sqrt{2N}\, p_{0} \,
{\mathbf{e}}_{N+1}) = 0$. Hence
 $\gamma_{N+1} = \lambda p_{0}/\sqrt{2N}$, and 
 ${\mathbf{p}} - p_{0} {\mathbf{e}}_{0} - \sqrt{2N} 
 p_{0} {\mathbf{e}}_{N+1} = 
 (0,\gamma_{1},\gamma_{2},\gamma_{3},\cdots \gamma_{N}, 
 \lambda p_{0}/\sqrt{2N},
 \gamma_{N+2},\gamma_{N+3},\cdots \gamma_{2N})$, where
 the  $\gamma_{m}$ are completely free. 
In the intermediary basis
${\mathbf{p}}$ has thus coordinates $(p_{0},  
\gamma_{1}, \gamma_{2},
\cdots \gamma_{N}, p_{0}\,\sqrt{2N}\, [\,1 + \lambda/(2N)\,],
\gamma_{N+2},\gamma_{N+3},\cdots \gamma_{2N})$. 

If we construct additional eigenvectors with 
$(p_{0},  
\gamma_{1}, \gamma_{2},
\cdots \gamma_{N}, p_{0}\,\sqrt{2N}\, [\,1 + \lambda/(2N)\,],
\gamma_{N+2},\gamma_{N+3},\cdots \gamma_{2N})$,
 then 
they should be orthogonal
to the $2N-2$ ones we already derived. The 
Gramm-Schmidt procedure will therefore
reduce these additional eigenvectors to the ones 
that one can construct from 
$(p_{0}, 0,0,\cdots,0, (\,1 + \lambda/(2N)\,) \,
\sqrt{2N} p_{0},
0,0,\cdots,0)$.
As the eigenvectors are defined up to an arbitrary normalization
constant, we can take e.g. $p_{0} = 1$.
Developing this in the pristine canonical  basis, the 
component from ${\mathbf{e}}_{N+1}$ yields
$1 + \lambda/(2N)$ on all $p_{m,\epsilon,1}$. 
We can verify that with this choice of ${\mathbf{p}}$ for a given value
$\lambda$, Eq. (\ref{p0}) is indeed satisfied and yields 
the eigenvalue $\lambda$.
When $N\rightarrow \infty$ however, all $p_{m,\epsilon,1} 
\rightarrow 1$, which corresponds to
the eigenvector that goes with the eigenvalue
$\lambda = 0$. Hence, in the limit $N \rightarrow \infty$,  
only the eigenvalue $\lambda = 0$ can have an eigenvector
with the structure that corresponds to $p_{0} \ne 0$.
Hence the choices ${\mathbf{e}}_{1}$, ${\mathbf{e}}_{2}$, 
... for ${\mathbf{p}}$,
based on $p_{0} = 0$, lead to 
a maximal set of mutually independent
solutions for the constraint given by Eq. (\ref{p0}).
By the choice of the intermediary basis, whole lines (rather than 
half lines) are always treated at once,
and the solutions are eighter symmetrical ($s= 1$) or 
antisymmetrical ($s = -1$) in $j$.
This leads to eigenvectors 
${\mathbf{p}}^{(k,s,q)}$, with $k \in {\mathbb{Z}}, 
q \in ]0,\pi[, s \in \{-1,1\} $ given by:

\begin{equation}
{\mathbf{p}}^{(k,s,q)} = (1,{\frac{\sin 2 q} {\sin q}}, \cdots  \, 
{\frac{\sin j q} {\sin q}}, \cdots) \otimes {\mathbf{e}}_{k,s},
\end{equation}

\noindent where  the antisymmetrical eigenvectors 
${\mathbf{e}}_{k,1}$ 
are the generalisation 
of ${\mathbf{e}}_{k}$, with $0 < k\le N$ in the finite case,
while the symmetrical eigenvectors ${\mathbf{e}}_{k,-1}$ are 
the generalisation 
of ${\mathbf{e}}_{k}$, with $N+1 < k \le 2N$. If we take $j\in 
{\mathbb{Z}}$, we can write the components as: 

\begin{equation}{\label{Component1}}
(p^{(k,-,q)})_{m,j} = {\frac{\sin j q} {\sin q}}\,\delta_{k,m},
\end{equation} 

\begin{equation}{\label{Component2}}
(p^{(k,+,q)})_{m,j} = {\frac{\sin |j| q} {\sin q}} \, 
e^{\imath m \psi_{k}},
\end{equation}

\noindent which are mutually orthogonal by construction 
(and where the factor $1/\sin q$ can be dropped if $q\ne 0$. 
For $q=0$, we have $\lambda=0$ and the eigenvector is 
$(1,1,1,\cdots,1,1,\cdots )$. We can take $\psi_{k} \not\in \pi {\mathbb{Q}}$.
The use of $k$ is somewhat artificial
and serves only to label the eigenvectors, which are
defined by   $\psi_{k} \in ]-\pi,\pi[$. 

{\em Calculation of the coherent scattering signal}.
We now must apply the formalism of Reference \cite{Coddens}  
to calculate the structure factors that
go with each eigenvalue $\lambda$ and the time dependence 
$e^{-\lambda t}$. Let us note  that
in the limit $q \rightarrow 0$, $\lambda = -q^{2}/\tau_{0}$ 
to a first approximation, such that the width of
the Lorentzian obtained from the Fourier transform
of $e^{-\lambda t}$, will be of the type $Dq^{2}$.
For a given eigenvector ${\mathbf{v}}_{\lambda}$ that corresponds to
an eigenvalue $\lambda$, there will be a contribution $e^{-\lambda t}$
to the signal with a weight given by
$|\,\sum_{{\cal{C}}}\,{\cal{F}}({\cal{C}})\,
{\mathbf{v}}_{\lambda}({\cal{C}})\,|^{2}$, where the
components ${\mathbf{v}}_{\lambda}({\cal{C}})$ of the eigenvector
taken at the configuration ${\cal{C}}$, run through the whole starlike
graph of the configuration space. (It is thereby assumed, without loss of
generality that
all initial configurations have the same probability.
In fact, we should not forget that the infinite configuration space
is only the abstraction of a mathematical limit that is supposed 
to correspond to the
huge numbers of atoms that are
involved in a macroscopic solid. A real QC is always finite, 
and its configuration space also.
That the initial configurations all have the same probability
will be true for all real cases, i.e. for all stars with a finite number 
of branches whose length is also finite.
By taking the limit when the number and the length of 
the branches tend to infinity,
we will obtain a constant probability for the initial 
configurations all over the infinite star. Of course, one could also
try to assess the true energies involved in the 
configurations in order to calculate
their thermodynamical probabilities. But this corresponds to a 
level of precision and complexity 
that is far beyond the
scope and the aims of our monoatomic toy model).
We must take into account the occurrence of the modulus in $\sin |j| q$, and
the fact that sometimes $\psi$ is not a parameter, 
but in all calculations
the sum will contain two building bricks  of the type:

\begin{equation}{\label{structurefactor}}
F^{(k,q)} = \sum_{m\in{\mathbb{Z}}}\,\sum_{j\in{\mathbb{N}}}\,
{\cal{F}}({\cal{C}}_{m,j})\,e^{\imath m \psi}\,\sin j q.
\end{equation}

\noindent (The case when $\psi$ is not a parameter can be treated with the
same formalism by putting $\psi = 0$).

For this we must calculate the Fourier transforms 
${\cal{F}}({\cal{C}}_{m,j})$ of all configurations
${\cal{C}}_{m,j}$.
A configuration is defined by the starting point where the 
first flip away from
the perfect tiling occurred and the endpoint, i.e. the 
actual position of the
active, diffusing letter $S$.
We can describe these two points by decomposing the letter 
sequence of the perfect lattice 
${\cal{C}}_{0}$
as $c_{1}Sc_{2}c_{3}$, where $S$ denotes
the tile that is going to wander, and where the 
configuration reached ${\cal{C}}_{m,j}$
will be $c_{1}c_{2}Sc_{3}$. We  note the endpoint of 
$c_{1}$ as $P$ and the
point between $c_{2}$ and $c_{3}$ by $B$. We call the point 
between $S$ and $c_{2}$ in the perfect
tiling  $A$. (We have $x_{B} > x_{A}$). 
The global change from ${\cal{C}}_{0}$ to 
${\cal{C}}_{m,j}$
consists in:
(1) removing the atom at $P$, (2) moving all the atoms of $c_{2}$ 
over a distance of $\sigma = {1\over{\sqrt{2+\tau}}}$ to the left, such 
that $A$ ends up in $P$,
(3) putting the atom we removed in $P$ at the position 
$B$ where the atom
that initially occupied it has moved away over $\sigma$ 
to the left. 
(Here $\sigma$ stands for the length of the interval 
spanned by a letter $S$).
All other atoms in $c_{1}$ and $c_{3}$ remain unchanged.
The Fourier transform ${\cal{F}}({\cal{C}}_{m,j})$ of 
${\cal{C}}_{m,j}$ can thus be written as:

\begin{equation}{\label{generalform}}
{\cal{F}}({\mathbb{QC}}) + {\cal{F}}(\,{\mathbb{QC}}\,
\cdot\,(e^{-\imath Q\sigma} - 1)\,
\cdot\,\chi[AB] \,) + e^{\imath Qx_{B}} - e^{\imath Qx_{A}},
\end{equation}

\noindent where $\chi[AB]$ is the characteristic function 
of the interval $[AB]$, i.e. the function
that takes the value $1$ on the interval and the 
value $0$ elsewhere. When the tile diffuses to the left
$e^{-\imath Q\sigma}$ must be replaced by $e^{\imath Q\sigma}$,
while the definitions of $x_{B}$ and
$x_{A}$ must be maintained such that now $x_{B} < x_{A}$.
The points $A$ correspond to the index $m$, the 
points $B$ correspond to the index $j$.
We must now establish the relation that defines this 
correspondence.
There is a complication when we try to 
number the configurations, in that
it is not just a numbering of the letters of the Fibonacci sequence.
In fact, when we have $S^{*}LSLS \rightarrow 
LS^{*}SLS (= LSS^{*}LS) 
\rightarrow LSLS^{*}S (= LSLSS^{*})$, the active letter $S^{*}$
has jumped from the first to the fifth position in only
two moves. It is therefore better to count the 
configurations by using the
$\tau$-inflated quasicrystal. Let us note the 
letters of the non-inflated
QC by $\ell$ ans $s$, and those of the inflated QC by $L$ and $S$.
Then $L$ corresponds to $\ell s$ and $S$ corresponds to $\ell$.
The dynamical sequence evoked can then be rewritten as 
$s LL \rightarrow LsL \rightarrow LLs$,
where we reserve the use of the letter $s$ for 
the active letter $S^{*}$.
Hence the index $j$ can be established by counting the 
letters in the inflated tiling.
When a lattice point of the inflated tiling has 
(inflated) indices $(M,N)$,
then its $x$-coordinate is  $\tau (M+ \tau N)/
(\sqrt{2+\tau})$.
It corresponds to a number of letters  $M+N$.
In the superspace representation the point 
${\mathbf{r}} = (M,N)$ is $(M{\mathbf{e}}_{1} + 
N {\mathbf{e}}_{2})\tau$.
From this the number $M+N$ can be obtained as 
${\frac{1}{\tau}}\,({\mathbf{e}}_{1} + 
{\mathbf{e}}_{2}) \,{\mathbf{\cdot}} \,{\mathbf{r}}$.
It follows that $j = ({\mathbf{r}}_{B} - {\mathbf{r}}_{A})\,
{\mathbf{\cdot}}\, (1,1)/\tau$.
When the tile diffuses to the right, the starting points 
$A$ must be an endpoint of a letter 
$s$ of the non-inflated tiling,
hence the endpoint of a letter $L$ of the inflated tiling. 
They are thus counted by the letter $N$ and can be 
obtained from ${\mathbf{r}}_{A}\,{\mathbf{\cdot}}\, 
(0,1)/\tau$, provided 
we have well selected the points ${\mathbf{r}}_{A}$ to 
be endpoints of letters $L$.
When the tile diffuses to the left, the starting points 
$A$ must be the beginning point of a letter 
$s$ of the non-inflated tiling. They are therefore
also tied up with a letter $L$ of the inflated tiling,
counted by the letter $N$ and obtained
by $({\mathbf{r}}_{A}+ (1,0) \,)\,{\mathbf{\cdot}}
(0,1)/\tau$, provided 
we have well selected the points ${\mathbf{r}}_{A}$ to 
be endpoints of letters $L$.

This brings us to the selection of points. In the 
calculation of the Fourier
transform ${\cal{F}}({\cal{C}}_{m,j})$, all lattices points of 
the non-inflated
tiling enter into the calculation. They can be obtained by a 
cut method from
the superspace space presentation of the inflated tiling by 
decorating the nodes with
a composite atomic surface ${\cal{W}}_{0} = {\cal{W}}_{1} \cup 
{\cal{W}}_{2}$ that 
has two parts: a long one, ${\cal{W}}_{1}$ 
centered on the node, and a shorter
one, ${\cal{W}}_{2}$, at a distance $\sigma$ towards 
the left, that generates
the points of the non-inflated tiling that are not
present in the inflated tiling. 
${\cal{W}}_{1} = \{0\}\times [-{{2\tau + 1}\over{2\sqrt{\tau +2}}},
{{2\tau + 1}\over{2\sqrt{\tau +2}}}]$,
${\cal{W}}_{2} = \{-{1\over{\sqrt{2+\tau}}}\}\times 
[-{{2\tau + 1}\over{2\sqrt{\tau +2}}},
{1\over{2\sqrt{\tau +2}}}]$.
We call this composite atomic 
surface thus ${\cal{W}}_{0}$.
When the tile diffuses to the right, the points 
${\mathbf{r}}_{A}$ are generated by an atomic surface 
${\cal{W}}^{+}_{A} \subset {\cal{W}}_{1}$,
which is that part of ${\cal{W}}_{1}$ that is the projection of 
${\cal{W}}_{2}$ onto 
it: ${\cal{W}}^{+}_{A} = \{0\}\times 
[-{{2\tau + 1}\over{2\sqrt{\tau +2}}},
{1\over{2\sqrt{\tau +2}}}]$.
When the tile diffuses to the left,
the starting point $A$ must be the beginning point 
of a letter $s$ of the non-inflated tiling, hence it must be on
an atomic surface ${\cal{W}}^{-}_{A}={\cal{W}}_{2}$. 
Finally, the points ${\mathbf{r}}_{B}$ are generated by
an atomic surface ${\cal{W}}_{B} = {\cal{W}}_{1}$.
This way, each of the sets of points that we have to consider, 
viz. ${\mathbf{r}}$ for
${\cal{C}}_{m,j}$, ${\mathbf{r}}_{A}$ for $j$ and $m$, and  
${\mathbf{r}}_{B}$ for $j$,
build a different kind of QC, that we can note as 
${\mathbb{QC}}_{0}$, ${\mathbb{QC}}^{+}_{A}$, ${\mathbb{QC}}^{-}_{A}$
and ${\mathbb{QC}}_{B}$ respectivily. Each of these 
has its Fourier transform.
The positions of the Bragg peaks in these Fourier 
transforms will be the same for the four QCs,
but their weights (i.e. their amplitudes) will be 
different. 
They can be obtained from a calculation of the 
Fourier transforms 
$w_{0} = {\cal{F}}({\cal{W}}_{0}), w^{+}_{A} = 
{\cal{F}}({\cal{W}}^{+}_{A}), 
w^{-}_{A} = {\cal{F}}({\cal{W}}^{-}_{A}), w_{B} = 
{\cal{F}}({\cal{W}}_{B})$
of the corresponding atomic surfaces
in a standard fashion. 
We are now ready for 
the calculation of 
the stucture factor of Eq. (\ref{structurefactor}).

The Fourier transform of $\chi[AB]$ is 
$(e^{\imath Q_{\parallel}x_{B}} - 
e^{\imath Q_{\parallel}x_{A}})/iQ_{\parallel}$.
This has to be convoluted with the Fourier transform of the QC, 
which yields:

\begin{equation}
\sum_{{\mathbf{Q}}^{*}\in {\mathbb{L}}}\, 
w_{0}({\mathbf{Q}}^{*}{\mathbf{\cdot e}}_{\perp}) \, 
(e^{\imath ({\mathbf{Q-Q}}^{*})_{\parallel}{\mathbf{\cdot r}}_{B} }
- e^{\imath ({\mathbf{Q-Q}}^{*})_{\parallel}
{\mathbf{\cdot r}}_{A} })/i({\mathbf{Q-Q^{*}}}) 
{\mathbf{\cdot e}}_{\parallel},
\end{equation}

\noindent where ${\mathbf{Q}}^{*}$ stands for the ${\mathbf{Q}}$-values of 
the (two-dimensional) Bragg peaks 
and ${\mathbb{L}}$ is the set of these Bragg peaks. We have 
$\sin j q = (e^{ \imath q({\mathbf{r}}_{B} - {\mathbf{r}}_{A})\,
{\mathbf{\cdot}}\, (1,1)/\tau} - 
             e^{-\imath q({\mathbf{r}}_{B} - {\mathbf{r}}_{A})\,
             {\mathbf{\cdot}}\, (1,1)/\tau})/2\imath$,
while $e^{\imath m \psi_{k}} =  
e^{\imath ({\mathbf{r}}_{A}\,{\mathbf{\cdot}}\, (0,1)/\tau) \psi_{k}}$.
Let us first perform the sum over the configurations for
which $x_{A} < x_{B}$. Then $j \ge 0$.
For ${\mathbf{p}}^{(q,+,k)}$ we obtain:

\begin{equation}{\label{4terms}}
\sum_{{\mathbf{r}}_{A},{\mathbf{r}}_{B}}\, \sum_{{\mathbf{Q}}^{*}\in 
{\mathbb{L}}}\,
- \frac{w_{0}({\mathbf{Q}}^{*}{\mathbf{\cdot e}}_{\perp})} 
{2\,({\mathbf{Q-Q^{*}}}){\mathbf{\cdot e}}_{\parallel}}
(e^{\imath ({\mathbf{Q-Q}}^{*})_{\parallel}{\mathbf{\cdot r}}_{B} }
- e^{\imath ({\mathbf{Q-Q}}^{*})_{\parallel}
{\mathbf{\cdot r}}_{A} }) 
\,(e^{ \imath q({\mathbf{r}}_{B} - {\mathbf{r}}_{A})\,
{\mathbf{\cdot}}\, (1,1)/\tau} - 
             e^{-\imath q({\mathbf{r}}_{B} - {\mathbf{r}}_{A})\,
             {\mathbf{\cdot}}\, (1,1)/\tau})\,
             e^{\imath ({\mathbf{r}}_{A}\,{\mathbf{\cdot}}\,(0,1)/\tau) \psi_{k}},
\end{equation}

\noindent which splits into 4 terms. The first term is:

\begin{equation}{\label{firstterm}}
\sum_{{\mathbf{r}}_{A},{\mathbf{r}}_{B}}\, \sum_{{\mathbf{Q}}^{*}\in 
{\mathbb{L}}}\,
- \frac{w_{0}({\mathbf{Q}}^{*}{\mathbf{\cdot e}}_{\perp})} 
{2\,({\mathbf{Q-Q^{*}}}){\mathbf{\cdot e}}_{\parallel}}
e^{\imath ([ \,({\mathbf{Q-Q}}^{*})_{\parallel} + q(1,1)/\tau\,]
{\mathbf{\cdot r}}_{B}
+ [\,(0,1)\psi_{k} - q(1,1)\,]{\mathbf{\cdot r}}_{A}/\tau )}.
\end{equation}

\noindent For this term it is convenient to perform first the sum over
${\mathbf{r}}_{B}$. Let us first calculate this sum
without the condition  $x_{A} < x_{B}$.
This unconditional sum over ${\mathbf{r}}_{B}$ is the 
{\em two-dimensional} Fourier transform
 for a ${\mathbf{Q}}$-vector $({\mathbf{Q-Q}}^{*})_{\parallel} + q(1,1)/\tau$
 of the part of the two-dimensional superlattice that lies in the strip 
${\cal{W}}_{B}\times E_{\parallel}$ that is used to define
${\mathbb{QC}}_{B}$. (Here $E_{\parallel}$ is parallel space). 
The two-dimensional Fourier transform of this strip-restricted
superlattice
is different 
from zero only for ${\mathbf{Q}}$-vectors 
${\mathbf{G}} \in {\mathbb{L}} \oplus  E_{\perp}$,
that belong to the set composed of all straight lines,
that are perpendicular to the cut and contain a 
two-dimensional Bragg peak. (Here $E_{\perp}$ is 
perpendicular space).
The intensity is thus non-zero
if $(\exists {\mathbf{Q}}^{**} \in  {\mathbb{L}})\,(\exists x \in {\mathbb{R}})
(({\mathbf{Q-Q}}^{*})_{\parallel} + q(1,1)/\tau =
 {\mathbf{Q}}^{**} + x {\mathbf{e}}_{\perp}$.
The parallel component of $q(1,1)/\tau$ is $q\tau/\sqrt{2+\tau}$.
Hence the intensity will be non-zero when $({\mathbf{Q-Q}}^{*})_{\parallel}
+ q\tau/\sqrt{2+\tau} = {\mathbf{Q}}^{**}_{\parallel}$,
such that ${\mathbf{Q}}_{\parallel}  = 
{\mathbf{Q}}^{**}_{\parallel} + 
{\mathbf{Q}}^{*}_{\parallel} - q\tau/\sqrt{2+\tau}$.
Let us first keep ${\mathbf{Q}}^{*}$ fixed.
Then for each ${\mathbf{Q}}^{**}$-value a corresponding 
${\mathbf{Q}}_{\parallel}$-value
can be found that satisfies the condition.
This defines thus a set of satellites at a 
distance $-q\tau/\sqrt{2+\tau}$
from each QC Bragg peak. All these satellites are 
valid ${\mathbf{Q}}_{\parallel}$-values.
The term $1/({\mathbf{Q-Q^{*}}}){\mathbf{\cdot e}}_{\parallel}$ 
reduces to 
$1/({\mathbf{Q}}^{**}_{\parallel} - q\tau/\sqrt{2+\tau})$.
The perpendicular component of $q(1,1)/\tau$ is 
$-q/(\tau^{2}\sqrt{2+\tau})$,
such that $[ \,({\mathbf{Q-Q}}^{*})_{\parallel} + q(1,1)/\tau\,]$
will be at this $Q_{\perp}$ value on the line 
through ${\mathbf{Q}}^{**}$.
The  amplitude in the Fourier transform for the satellites is thus 
given by $w_{B}(Q^{**}_{\perp} + q/(\tau^{2}\sqrt{2+\tau}))/ 
({\mathbf{Q}}^{**}_{\parallel} - q\tau/\sqrt{2+\tau})$.
As ${\mathbf{Q}}^{*}$ also runs through ${\mathbb{L}}$, 
each particular
satellite is realized once for every value of 
${\mathbf{Q}}^{**}$.
The dominant amplitude contribution to a satellite 
will come from its realization 
with ${\mathbf{Q}}^{**} =0$, such that in a first
approximation the intensity of a satellite is 
proportional to $1/q^{2}$.
${\mathbf{Q}}^{**} =0$ implies that the satellite position 
${\mathbf{Q}}_{\parallel}$ lies close to the Bragg peak at
${\mathbf{Q}}^{*}_{\parallel}$, and the factor
$w_{0}({\mathbf{-Q}}^{*}{\mathbf{\cdot e}}_{\perp})$ makes sure
that the intensities of the dominant realizations scale with the 
intensities of their associated Bragg peaks at 
${\mathbf{Q}}^{*}_{\parallel}$.

In reality we must perform the sum with the condition $x_{B} > x_{A}$.
The condition can be introduced by multiplying the 
acceptance strip further with a step function along $E_{\parallel}$
at $x_{A}$. Hence we must replace ${\mathbb{L}} \oplus  E_{\perp}$
by its convolution with the Fourier transform of this step function,
which is $\delta({\mathbf{G}}_{\perp}) \,
e^{
{\imath\mathbf{G}}_{\parallel}
{\mathbf{\cdot r}}_{A}
}\,
{1\over{2}} 
[\,\delta({\mathbf{G}}_{\parallel}) -
 \imath /(\pi {\mathbf{G}}_{\parallel} )\,]$.
We must thus convolute our spectrum of satellites with
this function. 
The term $e^{\imath{\mathbf{G}}_{\parallel}{\mathbf{\cdot r}}_{A}}$
leads to $e^{\imath[\,({\mathbf{Q-Q}}^{*})_{\parallel} + 
q(1,1)/\tau\,]{\mathbf{\cdot r}}_{A}}$,
and this enters into the final sum, which becomes
a {\em two-dimensional} Fourier transform
 for a $Q$-vector $({\mathbf{Q-Q}}^{*})_{\parallel} + 
 (0,1)\psi_{k}/\tau$
of the part of the superlattice that lies in 
the full strip
that is used to define
${\mathbb{QC}}^{+}_{A}$. 
This will again lead to satellites. 
However, as a whole continuum of $\psi_{k}$-values lead 
to the same value of $\lambda$, 
this will only multiply the $Q$-dependence of the spectrum obtained
so far with a constant.

{\em Description of the further calculations}. We have 
now accomplished 
one small step of the 
total calculation. We must also
perform the sums of the three other terms in Eq. (\ref{4terms}}). 
Then we must
also perform the calculations for the cases that the tile
diffuses to the left. 
The contributions must be weighted by
$e^{\pm\imath Q\sigma} - 1$.
We must
repeat all this {\em mutatis mutandis}
for ${\mathbf{p}}^{(q,-,k)}$ (for which there will be
no parameter $\psi$). 
And finally,
we must add terms that correspond to
the first, second and fourth terms
of Eq. (\ref{generalform}). 
In principle
those that correspond to
this first term will correspond only to 
elastic scattering: The sum of the components
of an eigenvector that does not correspond to $\lambda = 0$
is zero, as this eigenvector
is orthogonal to the vector 
$(1,1,1,\cdots 1,\cdots)$ that corresponds
to $\lambda =0$. Those 
that correspond to the third and fourth terms can be
expected to be vanishingly small
with respect to the other terms.
If we assume that all configurations have equal
starting probabilities, we must then
square the sums obtained this way,
and add up all these squares that
go with a same value of $\lambda$.

In principle, it should thus be possible
to derive the completely rigorous mathematical
expression for the coherent quasielastic scattering
signal that corresponds to the model.
But the complete calculation would be extremely tedious.
We will therefore not perform it,
because what we have obtained so far
is sufficient for the point that we want to make,
viz. that the general response function
shares many characteristics of the signal
observed by Francoual {\em et al.}
Let us just for the sake of completeness
give the three other terms from Eq. (\ref{4terms}) that 
occur at the same level 
as the ``first'' term given by Eq. (\ref{firstterm}).
These terms are:

\begin{equation}
\sum_{{\mathbf{r}}_{A},{\mathbf{r}}_{B}}\, \sum_{{\mathbf{Q}}^{*}\in 
{\mathbb{L}}}\,
\frac{w_{0}({\mathbf{Q}}^{*}{\mathbf{\cdot e}}_{\perp})} 
{2\,({\mathbf{Q-Q^{*}}}){\mathbf{\cdot e}}_{\parallel}}
e^{\imath ([ \,({\mathbf{Q-Q}}^{*})_{\parallel} - q(1,1)/\tau\,]
{\mathbf{\cdot r}}_{B}
+ [\,(0,1)\psi_{k} + q(1,1)\,]{\mathbf{\cdot r}}_{A}/\tau )},
\end{equation}

\begin{equation}
\sum_{{\mathbf{r}}_{A},{\mathbf{r}}_{B}}\, \sum_{{\mathbf{Q}}^{*}\in 
{\mathbb{L}}}\,
\frac{w_{0}({\mathbf{Q}}^{*}{\mathbf{\cdot e}}_{\perp})} 
{2\,({\mathbf{Q-Q^{*}}}){\mathbf{\cdot e}}_{\parallel}}
e^{\imath ([\,q(1,1)\,]{\mathbf{\cdot r}}_{B}/\tau
+ [\,({\mathbf{Q-Q}}^{*})_{\parallel} +((0,1)\psi_{k} 
- q(1,1))/\tau \,]{\mathbf{\cdot r}}_{A} )},
\end{equation}

\begin{equation}
\sum_{{\mathbf{r}}_{A},{\mathbf{r}}_{B}}\, \sum_{{\mathbf{Q}}^{*}\in 
{\mathbb{L}}}\,
-\frac{w_{0}({\mathbf{Q}}^{*}{\mathbf{\cdot e}}_{\perp})} 
{2\,({\mathbf{Q-Q^{*}}}){\mathbf{\cdot e}}_{\parallel}}
e^{\imath ([\,-q(1,1)\,]{\mathbf{\cdot r}}_{B}/\tau
+ [\,({\mathbf{Q-Q}}^{*})_{\parallel} +((0,1)\psi_{k} 
+ q(1,1))/\tau \,]{\mathbf{\cdot r}}_{A} )}.
\end{equation}

\noindent In the third and fourth term we must first sum over $x_{A}$
rather than over $x_{B}$. 
In these terms the fact that $\psi_{k}$ runs through a dense set
will turn the total contributions from such terms into
a flat (dynamical) background.

{\em Important Conclusions}. 
When we take into account all eigenvalues
$\lambda$, we will obtain
an intensity with a $1/q^{2}$ shape close to each Bragg peak,
where $q$ is the distance from the Bragg peak.
The intensity will scale with the intensity of the Bragg peak.
And in the energy domain, the signal will consist
of Lorentzians  with a width that is of a
$Dq^{2}$ type.

We must insist with emphasis on the fact
that the signals for different $q$-values
are all obtained by {\em a summing over the
very same configurations of the QC}, viz.
the set of configurations  we described above.
The satellites with wave vector
$q$ are not at all caused by
some wave with wavelength $2\pi/q$ in the QC,
but by a weighting of QC configurations
that are devoid of any wave structure
with eigenvectors of the dynamical
jump matrix. It is thus not at all
some correlation in the atomic positions
in the QC that brings about the satellites,
but the special character of the eigenvectors
that are used as weighting functions. 
These eigenvectors have a
wavelike structure due to the fact
that the configuration space
is the discrete analogon of a manifold assembled from
parts that have a topology that
is {\em almost} translationally invariant.
The reader can check that the eigenvectors of the jump model
calculated in \cite{Coddens} also show such a wavelike
dependence due to the translational invariance of the topology.
It is this wavelike behaviour of the eigenvectors that
is responsible for the possibility of having satellites, but
it depends on the way such eigenvectors couple to the Fourier
transforms of the configurations 
if a trace of this wavelike character 
will survive and show up in the final result 
under the form of satellites.

It is generally believed that a wavelike
Fourier component of random-tiling-like structural disorder
is needed to provoke the kind of diffuse scattering
observed in QCs. In other words that the Fourier components
of the diffuse
scattering are structural cut ondulations. The diffusing-tile 
model shows that also this interpretation may
not be unique, especially as the diffuse scattering
has now been proved to be the signature of 
very slow kinetics. The fact that the signal is dynamical
opens the way for alternative readings of the diffuse scattering.

Our calculation further confirms
a point that we already stressed ever so often,
viz. that an interpretation
of the satellites in terms
of phasons is precluded by
the fact that the satellites
would have to follow the same temperature
dependence as the elementary jumps:
All characteristic times of 
our model have the temperature dependence of
$\tau_{0}$.
If the signal observed in the speckle experiments
is due to atomic displacements, then it
can at the very best be due to
a secondary relaxation of the lattice in the form
of small atomic displacements, e.g. in response to the
freezing of the phason dynamics.
It is quite possible that both contributions
(small atomic displacements with a Debye-Waller
factor that shows an anomalous temperature behaviour
due to a response to the freezing of the phason jumps and
a smaller tile diffusion contribution with a  Debye-Waller factor
which follows the normal temperature behaviour)
are simultaneously present in the data.

We may finally note that
it is not obvious that in our model
the quasielastic intensity pattern would comply with
a phason elasticity {\em Ansatz}.
(But it is possible to think of variants of our model,
e.g. by introducing different jump times between
${\cal{C}}_{0}$ and ${\cal{C}}_{m,1}$, and this
way modulating the probabilities of the branches of the star
with the $x_{\perp}$ coordinate of ${\mathbf{r}}_{A}$. 
The eigenvalue
problem would still remain exactly solvable).

For all these reasons, it is not
necessary to pursue our calculations of this
diffusing-tile model
any further. 
In fact, in the third paragraph of subsection III.F about 
conflicting time scales 
we pointed out that the kind of model we have developed here is
physically unrealistic: If one wanted to impose such a model, 
one would need to introduce
a notion that the QC is highly stable with respect to tile flips, in 
order to validate the assumption that there would be 
only one flipped tile at a time within the whole QC. 
A real system will contain many tile flips simultaneously.
The prolificity of these flips, with their
high hopping rates, raises the question if the 
long-time persistence of correlations 
between configurations as observed in our model 
can survive in a more realistic setting.

The aim of our model is therefore not to give an alternative 
interpretation of the
data but rather to give a counter example to show
that the interpretation of the data according to reference \cite{Francoual}
is not unique and that it contains unjustified tacit assumptions:
(1) that the presence of satellites would be a proof for the
presence of a wavelike modulation
within the structure, (2) that phason elasticity 
would be in 1-1-correspondence
with phason dynamics, and (3) that one could ignore the fact
that the temperature dependence observed in the quasielastic neutron 
scattering results precludes
any interpretation of the diffuse scattering data 
in terms of tile flip dynamics.
Note that the unjustified introduction
of correlations in the atomic positions on the basis of (1)
is only a first ingredient for the introduction of
the wrong paradigm of dynamical phason waves,
a second one being the erroneous identification of these
correlations in the atomic positions with unphysical
long-distance correlations between atomic
jumps.\\

{\subsubsection{The presence of satellites at wave vector $q$ is 
not a proof for a presence
of a modulation wave with wavelegth $2\pi/q$. }}

This conclusion from model 3 about  the
invalidity of tacit assumption (1) can be generalized to all jump models.
The algorithm to calculate a coherent quasielastic signal for
a model of atomic jump dynamics is based on a set of coupled
differential equations expressed in terms of a jump matrix,
that transcribes the original problem into a model of 
jumps between configurations.
To solve the differential equations one must find the eigenvalues and
eigenvectors of the jump matrix. These eigenvectors have a component for each
configuration. The components of an eigenvector that corresponds
to an eigenvalue $\lambda$
must be used to weight the Fourier transforms of the corresponding configurations.
These weighted Fourier transforms are 
added up such as to yield the amplitude of the structure factor that
goes with the eigenvalue $\lambda$.
As the very same set of configurations comes into play for different
eigenvalues, it is meaningless to want to associate a wavelength
or a particular wave pattern in the QC structure
with the structure factor of some eigenvalue: The coupling of a 
wavelength or wave pattern
with an eigenvalue
would mean that each eigenvalue is associated with its own
typical configuration, while the truth is that each eigenvalue is coupled
to the same (full) set of configurations, with a different
weighting procedure, defined by the components of 
the corresponding
eigenvector. Hence any relationship between a 
measured ``wave vector'' $q$ and a 
special relaxation time $\tau_{q}$, such as e.g. 
$1/\tau_{q} \propto q^{2}$,
can only originate in the weighting procedure, not in 
a modulation wave pattern
with wave vector $q$,
that would exist within the structure. The calculation of model 3 may 
serve as a worked-out illustration of
this argument, rendering it less abstract, but the argument 
is truly generally valid 
for all jump models. The weighting factors correspond to 
a probability formalism for the configurations. We can thus 
summarize the situation
by stating that the signal for a value of $q$ does not 
correspond to a configuration
of the QC
but to an invariant, equilibrium probability distribution on 
the set of all possible 
configurations of the QC. (As such an equilibrium probability 
distribution never resumes to a single
configuration with probability 1, the system can act up to
the probability distribution only by fluctuating indefinitely).
Therefore the ``wave vector'' $q$
does not correspond to a single configuration
with a wavelike pattern  on the physical structure of the QC.
At the very best it can  
correspond to a probability wave
on the set of configurations, like it is the case in model 3, 
where the presence of these waves is due to some
translational invariance on the configuration manifold. 
In more complicated
models even this abstract kind of translational invariance 
can be absent.
The {\em dynamical} signal is thus even not due to a density 
wave as discussed in Section II for
{\em static} diffuse scattering: It is only a probabilistic 
Van Hove correlation between
configurations at time 0 and  configurations at a later time $t$.
We will now illustrate the point that the presence of satellites 
does not prove
a presence of correlations in the even more salient example 
of single-vacancy diffusion.\\

{\subsubsection{Single-vacancy diffusion }}

Let us consider the problem of the diffusion of a single 
vacancy on the Fibonacci chain.
The probabilities $p_{j}$ that the vacancy is located at 
site $j$, are governed by
the jump equations
$(\forall j \in {\mathbb{Z}})\, 
({\frac{dp_{j}}{dt}} = - {\frac{1}{\tau_{j-1,j}}} \,(p_{j}-p_{j-1}) 
                       + {\frac{1}{\tau_{j,j+1}}} \,(p_{j+1}-p_{j}))$,
where the relaxation times $\tau_{j,j+1} = \tau_{L}$ 
if the bond between atoms $j$ and $j+1$
is long, and $\tau_{j,j+1} = \tau_{S}$ if it is short. We can 
compare this with
the dynamical equations for the atomic displacements $u_{j}$ in 
the lattice dynamics,
when only first-neighbour intercations are considered:
$(\forall j \in {\mathbb{Z}})\, 
({\frac{d^{2}u_{j}}{dt^{2}}} = - k_{j-1,j} \,(u_{j}-u_{j-1})  
                               + k_{j,j+1} \,(u_{j+1}-u_{j}))$,
where the spring constants $k_{j,j+1} = k_{L}$ 
if the bond between atoms $j$ and $j+1$
is long, and $k_{j,j+1} = k_{S}$ if it is short.
We see that apart from the specific values of the 
constants $\tau_{L}, \tau_{S}$,
and $k_{L}, k_{S}$, the sets of equations have the same structure,
such that the eigenvalue problems of the jump matrix and the dynamical
matrix are identical.

In the case of the jump matrix the eigenvalue 
$\lambda$ and the corresponding
eigenvector ${\mathbf{v}}^{(\lambda)}$ will be used to find a solution $
{\mathbf{p}} = e^{-t/\tau_{\lambda}} {\mathbf{v}}^{(\lambda)}$,
such that $- 1/\tau_{\lambda} = \lambda$.
In the case of the dynamical matrix the eigenvalue $\lambda$ 
and the corresponding
eigenvector ${\mathbf{v}}^{(\lambda)}$ will be used to find a solution
${\mathbf{u}} = e^{i\omega_{\lambda} t} {\mathbf{v}}^{(\lambda)}$,
such that $- \omega^{2}_{\lambda} = \lambda$. The solution of such eigenvalue
problems on the Fibonacci chain is horrendously difficult. 
Apart for $\lambda = 0$,
the solutions are in general not quasiperiodic. We should thus not expect
eigenvector solutions in the form of Bloch waves 
$v^{(\lambda)}_{j} = e^{\imath q j}$. 
At the best, we can attempt a Fourier decomposition of 
the eigenvectors into waves,
$v^{(\lambda)}_{j} = \sum_{q}\,a^{(\lambda)}_{q}\,e^{\imath q j}$.
Nevertheless,  in the
long-wavelength limit for the phonon problem
the distribution of $q$-values will be extremely narrow
and  the dominant Fourier component 
of $v^{(\lambda)}$ will be
$w^{(\lambda)}_{j} = e^{\imath q j}$, with a linear 
relationship $\omega = c q$, 
that defines   the speed of sound $c$. This is at least 
what is observed experimentally
in all real QCs.
From the analogy it will then follow that in the long-wavelength limit for
the vacancy diffusion problem the dominant Fourier component
of the eigenvector will also follow a linear relationship
$1/\tau = D q^{2}$, that defines the vacancy diffusion constant.
When $k_{L} = k_{S}$
and $\tau_{L} = \tau_{S}$ the eigenvectors 
are true waves and these results become exact.

In the problem of the lattice dynamics we will 
calculate the amplitude of the 
coherent neutron scattering
signal for a neutron momentum transfer $Q$ as
$\sum_{j \in {\mathbb{Z}}}\, e^{\imath Q (x_{j} + u_{j})}$,
where $x_{j}$ is the equilibrium position of atom $j$. To first order
of a Taylor series expansion this can be developed as
$\sum_{j \in {\mathbb{Z}}}\, e^{\imath Q x_{j}} 
( 1 + \imath Q u_{j}) = {\mathbb{L}}_{Q} 
+ \imath Q \sum_{j \in {\mathbb{Z}}}\,v^{(\lambda)}_{j}\, 
e^{\imath Q x_{j}} e^{\imath \omega_{\lambda} t}$.
Here the first term ${\mathbb{L}}_{Q}$ is the elastic signal in the form 
of the spectrum of Bragg peaks
as a function of $Q$. The second term is the inelastic contribution.
Of course for $u_{j}$ we have taken 
$v^{(\lambda)}_{j} e^{\imath \omega_{\lambda} t}$.

In the problem of single-vacancy diffusion  the configuration space
can be put into 1-1-correspondence with the sites 
of the lattice. The configuration
${\cal{C}}_{j}$ will be the one where the vacancy is situated 
at the lattice site
$j$.  Its Fourier transform ${\cal{F}}_{Q} ({\cal{C}}_{j})$ 
will be given by
${\mathbb{L}}_{Q} - e^{\imath Q x_{j}}$.
The amplitude of the 
coherent neutron scattering
signal for a neutron momentum transfer $Q$ is then given by
$\sum_{j \in {\mathbb{Z}}}\, p_{j} \,{\cal{F}}_{Q} ({\cal{C}}_{j}) = 
\sum_{j \in {\mathbb{Z}}}\, v^{(\lambda)}_{j} 
\,({\mathbb{L}}_{Q} - e^{\imath Q x_{j}})\, e^{- t/\tau_{\lambda} }$.
The first term is of the form 
${\mathbf{v}}^{(\lambda)} {\mathbf{\cdot v}}^{(0)} 
{\mathbb{L}} = \delta(\lambda) {\mathbb{L}}$,
as for $\lambda = 0$ the eigenvector is of the form 
$[1,1,1 \cdots 1,1, \cdots ]$.
When $\lambda = 0$ we recover the elastic signal as
$ {\mathbf{\delta}} \cdot {\mathbb{L}} - {\mathbb{L}}$, 
where weight of the second part is negligible
with respect to that of the Dirac measure $ {\mathbf{\delta}}$.
For $\lambda \neq 0$ we recover the inelastic signal. 
We can thus compare the $Q$ dependences
for the dynamical parts of the amplitudes of the signals:

\begin{equation}
\imath Q \,(\,\sum_{j \in {\mathbb{Z}}}\,v^{(\lambda)}_{j}\, 
e^{\imath Q x_{j}}\,)\, 
e^{\imath \omega_{\lambda} t} \,\,\,\, [\,phonons\,\,]\,
 \Leftrightarrow \,\,
- (\,\sum_{j \in {\mathbb{Z}}}\,v^{(\lambda)}_{j}\, e^{\imath Q x_{j}}\,)\, 
e^{- t/\tau_{\lambda} } \,\,\,\, [\,vacancy\,\,].
\end{equation}

Hence, apart from a factor $Q^{2}$, the $Q$-dependences 
of the inelastic signals 
for the phonon dynamics 
and for the vacancy diffusion
have for a given eigenvalue $\lambda$ the same structure. 
In the long-wavelength limit
this Q-dependence is coupled to phonon waves $e^{\imath c \hbar q t}$
and to relaxation times $e^{-Dq^{2} t}$, where $D$ is the 
vacancy diffusion constant.
In the energy domain $e^{\imath c \hbar q t}$ will lead to 
an inelastic delta peak at non-zero 
energy $c \hbar q$ and of zero width, while $e^{-Dq^{2} t}$ will lead 
to an inelastic Lorentzian peak
centered at zero energy and of width $\hbar Dq^{2}$.
(Because the inelastic signal is centered at zero energy it is quasielastic).
In the long-wavelength approximation 
$v^{(\lambda)}_{j} \approx w^{(\lambda)}_{j} = e^{\imath q j}$,
and through $qj = q(1,1)\cdot {\mathbf{r}}_{j}$ in superspace,
the $Q$-dependence leads to satellites at distance 
$q/{\sqrt{2+\tau}}$ from each Bragg peak.
It is interesting to note that when $k_{L} = k_{S}$
and $\tau_{L} = \tau_{S}$ these results become exact, 
even for large
wave vectors. 

Single-vacancy mediated atomic jump diffusion is a school 
example by exellence 
of  what in folk lore is being considered as an
``incoherent process'', and the calculation of the incoherent 
signal for this process
is presented in many text books.
Our calculation illustrates very clearly 
the falsehood of the wide-spread belief that 
the incoherence or coherence of a scattering process or signal
would stand in 1-1-correspondence with the incoherence or
coherence of the physical process in the sample
that is being probed by it:
The calculation yields a coherent signal in the form of a set of
satellites at a distance $q$ from the Bragg peaks, while there
is no correlation between the atomic jumps whatsoever.
When we limit ourselves to first-neighbour models, 
the previous argument
can be easily generalized to two- and three-dimensional QCs.

{\subsubsection{Fluctuating Fourier components}} {\label{fluctuation}}

Imagine that we have a lattice where a single atom has been displaced over a
distance ${\mathbf{u}}$ from its initial place ${\mathbf{r}}_{j}$.
Call the Fourier transform of the perfect lattice ${\cal{F}}({\mathbf{Q}})$.
The Fourier transform of the lattice with the single defect will
then be ${\cal{F}}({\mathbf{Q}}) + 
e^{\imath{\mathbf{Q\cdot}}({\mathbf{r}}_{j}+{\mathbf{u}})} 
- e^{\imath{\mathbf{Q\cdot r}}_{j}}$. 
This can be expanded in a Taylor series as
${\cal{F}}({\mathbf{Q}}) + e^{\imath{\mathbf{Q\cdot r}}_{j}}\,
[\, \imath ({\mathbf{Q\cdot u}}\,)\,
 - {\frac{1}{2}}({\mathbf{Q\cdot u}})^{2}\, \ldots\,]$. 
 This signal would be detected by coherent scattering.
Hence it follows that for all wave vectors ${\mathbf{Q}}$ 
(except those in the plane defined by ${\mathbf{Q\cdot u}} = 0$)
the intensity changes. At every reciprocal-lattice point ${\mathbf{Q}}$ 
(that does not belong to the plane) there is
a change of intensity. This idea can be generalized 
to a kind of pertubation theory that permits to 
get some idea about the changes of the static signal,
when a change 
of the position of many atoms is at stake.
Does the observation of 
such changes (on a characteristic time scale $\tau_{0}$) mean that 
there are dynamical modes with
wave vector ${\mathbf{Q}}$ on the characteristic time scale $\tau_{0}$ 
within the system? Certainly not.
We must emphasize that what we just described
is not what is being measured by photon correlation spectroscopy.
What is being measured involves correlations between 
two such intensities at two different times:
The measurement of an intermediate scattering function
 is more subtle than just a measurement of the fluctuating
static structure facor.  Photon correlation spectroscopy 
is a correct measurement of the dynamics and the intermediate scattering function.
But the paradigm we just described is exactly what has been used
to analyze the data: The fact that the diffuse scattering
can be Fourier analyzed through a superspace description 
(in terms of physically meaningless density waves) and that it exihibits
slow kinetics (on a time scale that forbids any confrontation with
dynamical data)
has been used to claim that dynamical 
phason modes have been
observed in QCs. 
From the calculation one can see that
the major changes in intensity will be roughly speaking 
proportional to ${\cal{F}}({\mathbf{Q}})$, due to the cross products
between ${\cal{F}}({\mathbf{Q}})$ (and ${\cal{F}}({\mathbf{Q}})^{*}$)
and the other, smaller terms. The terms that do not
contain ${\cal{F}}({\mathbf{Q}})$ (and ${\cal{F}}({\mathbf{Q}})^{*}$)
can be neglected in this approach.
Hence the intensities which will fluctuate the most under any
change of the system that is small enough to
warrant such a pertubational approach will be
the diffuse scattering and the Bragg peaks. They will thus no doubt
lead to the most prominent features in the measured correlation function.
Even in the wrong paradigm of the static structure factor approach
the fluctuating speckle would 
not prove that it is the  phason modulation that is (hypothetically) observed
in the static signal that fluctuates. In reality, the signal must be analyzed
in a completely different fashion as we have shown on the basis
of our model calculations in the preceding subsubsections.
The fact that the speckle signal fluctuates is therefore not at all
a proof that a phason modulation field, claimed to cause
the diffuse scattering, would exist and fluctuate, as has been claimed.
The few additional arguments that have been advanced to strengthen
the credibility of this claim have been analyzed above. They are not convincing.
One cannot study dynamics on the basis of a few selected points ${\mathbf{Q}}$,
in reciprocal space and interpret the intensities in these points
as though they were produced by the same physical causes as the
intensity  of the elastic structure factor at the same values ${\mathbf{Q}}$.
To identify the dynamical process one must measure 
the dynamical structure factor over the whole of reciprocal space
and confront it with an appropriate calculation.
Simultaneously to this unjustified claim that ``dynamical phason modes''
would have been observed, the 
doctrine has been 
published
that the definition of phason dynamics must be changed in such a way 
that our work on phason hopping can be dismissed as
a true observation of phason dynamics (see below). 
\\

{\section{Lack of proof for the random tiling model}}

\subsection{Introduction}

In reference\cite{ISIS} the authors stated
that the temperature dependence of the data are 
``in contradiction 
with the hypothesis
of a simple random tiling model''.
They added that the tile flip interpretation 
can be maintained by 
introducing a more complex random 
tiling model.\cite{Widom}
The random tiling model  and 
the tile flip scenario can thus be claimed
on the basis completely
opposite  temperature behaviours.
This is thus not very convincing
evidence for the random tiling model.
To cite Sir Karl Popper:
{\em ``Falsifiability should be the criterion
 of demarcation in science''}.\cite{Popper}
 Moreover, as already pointed out,
all the temperature data might show is
that there is some softening of the elastic constants.
 What the temperature dependence settles {\em within} the
context of the random tiling model, is that the
pristine model should be abandoned in favour of Widom's 
model.\cite{Widom}

The diffuse scattering data, without a mention 
of their temperature
dependence, were eventually reported 
 in reference \cite{wrong1}.
Nonetheless, in the final discussion of that paper,
the authors already introduce the arguments to 
thwart the criticism that can be formulated based
on the temperature behaviour  observed. As the reader
had no access to the information about the  temperature
dependence, he could not possibly understand 
the issues at stake.
Rather than the mention of a 
crucial problem raising serious doubts\cite{IJMPB}
about the validity of the interpretation,
it looked like a very puzzling 
digression. 

In that paper, the authors concluded 
that their data were ``compatible'' with the 
random tiling
model. The problem with this statement does 
not reside in what it
literally states, but in its
tacit {\em tertium non datur}.  It  does not
say that the data may also be compatible 
with other models.
Due to this formulation the reader may pick 
 up the totally
unjustified belief that the data would have 
conclusively proved the random tiling model.
As a matter of fact, such overinterpretations have indeed
made their way to the literature.\cite{Elser}.
This is of course very harmful.
It has installed a strong oral tradition, that
made it very difficult to point out
the absence of convincing proof for the random tiling model.
The unassailable character
of the literal interpretation of the formulation 
may have contributed to this situation.
It is not by qualifying this problem 
as an old story, 
 that the validity of these claims 
can be settled. \\

\subsection{The data claimed to be evidence are too unspecific}

At this point it is perhaps good to 
review the evidence
the authors proposed in favor of their 
interpretation
in terms of tile flips. 

(a) An $1/q_{\parallel}^{2}$ 
dependence
of the intensity. This is unspecific 
and occurs in many
other cases of Huang scattering in 
crystals.\cite{Salje} 
 
(b) A certain shape
of the diffuse scattering intensity contours.
Again, this is unspecific, and purely 
due to symmetry
in analogy with the situation in 
conventional crystals.\cite{Salje}
Such symmetry-based arguments do not 
contain any information
about the underlying mechanisms or 
interactions. 

(c) The shapes of the contours depend only 
on the phason-phason
elastic  constants.
This is the only feature that is not 
unspecific. As such,
it may rule out a number of alternative 
models.
But this does not mean  that
it rules out all possible alternative 
models. We have clearly shown
that it is still not unique,  even 
if it has been  presented as self-evident
that it would be unique.

As the temperature dependence of the neutron scattering data
does not agree
with an interpretation in terms of tile flips, we 
actually know that
the correct model must be different. 

We should
insist on the fact that an interpretation 
in terms of tile flips
is a {\em derived} application of the 
elasticity theory,
which is formulated in terms of a 
continuum
of small atomic displacements rather 
than on a discrete set 
devoid of infinitesimals. The very 
definition
of an elastic constant cannot be written down 
if we cannot assume that the atomic 
displacements
explore a continuum. The validity of the derived 
application is not
obvious, as on the microscopic level
tile flips do not explore a continuum
of atomic displacements. This is very different
from the continuum limit for phonons.
There is thus {\em a priori} no good
theoretical rationale to explain the results
of Tang et al.\cite{Tang,Shaw} except the {\em post facto} observation
that it works despite such theoretical objections.

To improve on this situation,
Henley\cite{Henley} has proposed an argument
of coarse graining.
Unfortunately, this may rather have contributed
to the fact that it was not realized
that the interpretation of
the data is not unique.
Without this argument  
 of coarse graining,
 the results of Tang et al.\cite{Tang,Shaw}
 could have stood out as a
 clearly different alternative in the form of 
 a discrete displacement field
 rather than a continuous one.
 
 In any case, it remains
 a cracking pass to generalize 
the finding by Tang et al.\cite{Tang,Shaw} in the sense that one
takes it for granted that the only exclusive way
to obtain such a dependence on the phason
elastic constants would be the tile flips
that are so improper for the 
first-hand application
of the theory. Similarly, in Widom's\cite{Widom} 
Landau-type theory
it is not granted that the phason elastic constants
can only correspond to tile flips: 
Even if some elastic
instability were observed in diffraction 
experiments that 
completely tallied with
his calculations, it would not yet 
prove the random tiling scenario. 

We may note that the phason elasticity constants obtained by
the authors of their diffuse scattering data do not
agree with calculations based on canonical cell tilings.\cite{Henleyprivate}

Finally, we must point out that the alternative that the 
diffuse scattering could be due to
chemical disorder has not been properly ruled out.
If the data really were due to chemical disorder, than they 
would not contain any proof
for the random tiling model.

In conclusion, we think that the experimental evidence is too narrow
to claim that it would be a proof for the validity of the 
random tiling model.
It could be misleading to want to harness the observation
of diffuse scattering into a stability issue in the form
of an opposition between random tiling models
and a perfect quasicristal model.\\

{\subsection{Absence of proof for a phase transition}} 
 
We may note that the QC-crystal transition postulated by the random tiling model
is of first order, while in the example of NaNO$_{2}$ cited 
in reference \cite{Francoual},
as a paradigm for the observations,
the transition is second order.

We may also note that there is no proof for the existence
of the phase transition postulated by the random tiling model.
In fact, one must introduce 
{\em ad hoc} assumptions
in order to deny the clear experimental evidence 
that there is no phase transition.
One might well invoke
the superstructure reported by 
Ishimasa,\cite{Ishimasa} which shows both
satellite peaks and diffuse scattering. 
However, for this
exceptional observation there are many 
other ones where
there is no phase transition at all.\cite{wrong2}
But even in the favorable case reported by 
Ishimasa, the dominant 
diffuse scattering maxima are centered
on the Bragg positions of the QC, not on 
the satellites.
The scenario is thus not one of diffuse scattering
progressively building up at the future 
satellite positions, and
eventually turning into satellite Bragg peaks.

Let us finally point out that in a discussion
about the presence of a phase transition or otherwise,
only those who claim such a transition exists
can solve the issue by  providing the proof.
It is in fact virtually impossible to prove that 
there is no phase transition: absence of proof can simply not be taken
as  proof of absence.\\

{\section{``Phason jumps'' is by all criteria an appropriate terminology}}

It has been claimed that the terminology 
``phason jumps'' would not be appropriate,
and that the terminology ``phason'' should be reserved to the modes
postulated in Eq. [42] in the paper by Janssen et al.\cite{Janssen}
There are several problems with this restriction or change of definitions:

(1) Before defining ``phason  modes'' as is done in reference\cite{Janssen},
one could wish to wait until convincing proof of their existence has been given.
The ``phason modes''
are just a conjecture. We have seen that they are introduced on the basis
of an argument that is not rigorous.
It has to be checked that the definition is meaningful. 
It rather seems as though this is not the case.

(2) To motivate the claim mentioned,
it has been argued that phason jumps are not characteristic
of quasicrystals because they also occur in 
1/1 approximants.\cite{Ames1} This capitalizes
on undue remarks in references \cite{Dubois} where it was
stated that phason jumps are ``not a special property for QCs''. These
remarks focused on the fact
that extremely fast jumps also occur in a B2-based phase that can contain
up to 12 \% vacancies.  But phason jumps are not only ``special'' 
in just being fast,
and it is doubtful that the presence of atomic jumps in a very unusual sample
with an exceptional amount of vacancies could shed light on such  issues.
What is obvious is, that if atomic jumps occur in a system,
they will forcedly produce diffuse scattering, and that diffuse scattering
can occur for many other reasons.
Diffuse scattering or speckle are certainly even 
less characteristic of quasicrystals.
And the modes that are supposed to replace the phason jumps
as the unique observation of phason dynamics 
most probably just do not exist.

(3) The terminology ``phason'' was 
introduced by Overhauser\cite{Overhauser}
in the field of dynamics of charge-density waves. 
It did not rely on elasticity considerations.
It is thus the terminology ``phason elasticity'' that is a derived
concept, rather than the phason jumps.

(4) The argument that ``phason jumps'' is not a good terminology
has been pushed 
to the extreme. One has even proposed a sarcastic, belittling terminology
``flipon'' for them,\cite{Ames2,Avignon}  but e.g. Edegawa et al. have 
clearly pointed out that the terminology is perfectly 
appropriate.\cite{Ames2}\\

{\section{Conclusion}}

In conclusion, we can resume the situation as follows.
The diffraction pattern of a QC allows for a Fourier decomposition
in superspace based on (physically meaningless) static density waves.
The Fourier components that describe the diffuse scattering in this Fourier decomposition
can be interpreted as static modulation waves in superspace.
When there are fluctuations in the QC, automatically
the intensities of all these Fourier components will change, i.e.
the intensities of all Fourier components will fluctuate.
Such fluctuations of the intensity of a Fourier component do of course not imply
that the Fourier component would be a dynamical wave, even not in superspace.
It is just that the contribution of the Fourier component to
the diffraction pattern fluctuates, whenever the diffraction pattern fluctuates,
whatever the kind of fluctuation might be.
What the authors measure is not such a fluctuation
of the Fourier components, but a component of the intermediate scattering function.
Nevertheless, their interpretation of the ${\mathbf{Q}}$-dependence of the
intermediate
scattering function follows exactly
the mental image of such a fluctuation of a Fourier component 
of the static structure factor, as it interprets
the intermediate scattering function
using the modulation waves from the analysis of the static structure factor. 
Although a dynamical structure factor
cannot be analyzed
as  a static structure factor,
the authors start from 
an {\em Ansatz} of dynamical waves on the basis of the 
superspace modulation picture which is only valid
for the static structure factor analysis.
As the topology and the translational invariance of the 
network of the bonds in the superspace crystal
(which are part of the mechanism needed to justify
such an assumption of dynamical waves) are destroyed
by the introduction of the cut,
it is physically wrong to carry over the {\em Ansatz} 
of dynamical waves from
the virtual crystal in superspace to the real QC in physical space.
This error has profound physical implications.
Even if one were to allow for this error, there is 
a whole series of secondary objections
against such an interpretation: (1) Even within
the context of a wave {\em Ansatz} it is not 
unique, e.g. chemical disorder or 
a field of small atomic displacements can
lead to the same result. These alternatives are 
more physical and they have not 
been properly excluded.
(2) A closer look at them reveals that the hypothetical waves are 
all but waves, and their hypothetical wavelengths
all but wavelengths as can clearly be seen in
the limit of small amplitudes.
(3) The idea of a phason wave is conceptually wrong 
because it amalgamates 
without any discussion two conflicting
physical pictures (propagating phonons and diffusing atomic jumps) 
and introduces 
unphysical long-distance correlations between atomic jumps
by confusing correlations between atomic positions with correlations
between atomic jumps. It uses an interpretation
of a static structure factor to analyze a dynamical structure factor.
(4) In order to justify certain approximations,
the model of the authors requires that one assumes that the phason 
wave has a small amplitude
while it is not mentioned that in the limit of very small amplitudes 
long-range diffusion is locked. 
(5) The model associates a phonon-like wavelength with a $q$-vector
while the discrete character of these $q$-values is not observed 
in the form of clear isolated {\em inelastic}  peaks as for phonons,
 and $q$ can have a completely different meaning than a wave-length.
(6) The presence of such discrete satellites
does not require any presence of a physical wave,
which further emphasizes how the interpretation
that has been proposed is not at all unique.
(7) The {\em Ansatz} is contradicted by the 
temperature behaviour observed in
neutron scattering studies whose interpretation is much more reliable
and open to cross-checking by other methods, in 
sharp contrast with
the data of the authors which are unassailable.
(8) In passing,
the authors have confused  (a) coherent scattering signals 
as evidence for the presence of correlations and
(b) mathematical Fourier components as evidence of physical waves.

Hence, it is not true that the experimental data would be evidence 
for the existence of phason waves in QCs, while from a conceptual
and theoretical viewpoint
 the introduction of such a notion of phason waves 
is awkward.


\begin{thebibliography}{99}
\bibitem{note} We calculate the Fourier transform
within the cut formalism, which is more general
than the projection formalism.
\end{thebibliography}
\end{document}